\documentclass[11pt,reqno]{amsart}

\usepackage{amsfonts,amssymb,amsbsy,amsmath,amsthm}
\def\qed{\hfill $\blacksquare$}

\newtheorem{definition}{\underline{Definition}}[section]
\newtheorem{proposition}[definition]{Proposition}
\newtheorem{theorem}[definition]{Theorem}
\newtheorem{lema}[definition]{Lemma}
\newtheorem{corollary}[definition]{Corollary}

\theoremstyle{definition}
\newtheorem{remark}[definition]{Remark}

\numberwithin{equation}{section}

\begin{document}
\begin{titlepage}
\begin{center}

\vskip 1cm

\renewcommand{\thefootnote}{\fnsymbol{footnote}}

\section*{A rigorous approach  to  the magnetic response in disordered systems.}

\vskip 7mm

\noindent
Philippe Briet, Baptiste Savoie.

\vskip 7mm

{\em
\medskip
\noindent
Centre de Physique Th\'eorique\footnote{Unit\'e Mixte de Recherche
(UMR 6207) du CNRS et de l'Universit\'e d'Aix-Marseille et
de l'Universit\'e du Sud Toulon-Var. Unit\'e affili\'ee \`a la
FRUMAM F\'ed\'eration de Recherche 2291 du CNRS.}, Case postale 907,\\
F--13288 Marseille Cedex 9, France.\\[2mm]
}

\vskip 1cm

\noindent
{\bf Abstract:}
\end{center}

This paper is a part of an ongoing study on the diamagnetic behavior of a 3-dimensional quantum gas of non-interacting charged particles subjected to an external uniform magnetic field together with a random electric potential.  We prove the existence of an almost-sure non-random thermodynamic limit for the  grand-canonical pressure,  magnetization and    zero-field orbital magnetic susceptibility. We also give an explicit  formulation of these thermodynamic limits. Our results cover a wide class of physically relevant random potentials which model not only \textit{crystalline disordered solids}, but also \textit{amorphous solids}.

\vfill

\noindent
PACS-2010 number: 05.30.-d, 51.30.+i, 51.60.+a, 02.30.Tb, 75.20.En, 75.20.Ck

\medskip

\noindent
MSC-2010 number: 82B10, 82B44, 81Q10, 81V99.

\medskip

\noindent
Keywords: Operator theory, Random Scrhr\" odinger operators,  quantum statistical mechanics, diamagnetism,  disordered solids.

\end{titlepage}


\section{Introduction and main results}

\subsection{Introduction}
In this paper we study  rigorously the diamagnetic response of a quantum gas of non-interacting charged particles trapped in an amorphous medium and subjected to an uniform magnetic field of intensity $B \geq 0$. Under the grand-canonical conditions  and in the weak-field regime, this response is completely characterized by the pressure and its first two derivatives w.r.t. $B$, i.e. the  magnetization and the susceptibility. Especially we are  focusing  on the bulk response which  is of great interest since it is independent of the boundary effects. This is  obtained  by proving  the existence of the thermodynamic limit for these quantities firstly  defined at finite volume. From a mathematical point of view, this consists of showing  that the derivatives of the pressure w.r.t. $B$ (performed at finite volume) commute with the thermodynamic limit.\\
\indent Our paper is an extension  of the works of Briet \textit{et al.} \cite{BCL1,BrCoLo1,BrCoLo3} where the case of  a perfect quantum gas has been  treated. All these papers  are in fact  in the  continuation of a study initiated by Angelescu \textit{et al.} \cite{ABN, AC}; for a brief review see also \cite{C1, BrCoLo1}. In the  regime of  positive temperature $T$ and small fugacity $z$, it is  proved  in \cite{BCL1, BrCoLo1}  the existence of the thermodynamic limit for the pressure and all its derivatives  w.r.t. $B$ for any positive value of  the cyclotron frequency, $b:= qB/c$. This  proof is based on a main ingredient:   the application of the so-called gauge invariant magnetic perturbation theory to  the corresponding Gibbs semigroups, see also \cite{C1}. Afterwards  in \cite{BrCoLo3}, they extend  these  results  to the   $z$-complex domain   of analyticity of the pressure, through the  Vitali convergence theorem.  Consequently  they obtained    the existence but no explicit formula  of the limits for all admissible values of $z$. \\
\indent Recently, all these results have been improved in \cite{BCS1}  covering the case of   periodic  interactions with  local singularities; basically of the Kato class.   But in \cite{BCS1}, only the limit of the  pressure has been considered. The proof  is essentially  based on the Pastur-Shubin formula for the integrated density of states \cite{If}. Later on the generalized susceptibilities  were studied in \cite{Sa}.  These results have been used in \cite{BCS2}   to get  a zero-field orbital susceptibility  formula  for a Bloch electron gas and the justification of  the Landau-Peierls  approximation  at small density and  zero temperature. \\
\indent In this paper, the background  electric potential is assumed to be a $\mathbb{G}^{3}$-ergodic ($\mathbb{G}= \mathbb{Z}$ or $\mathbb{R}$) random field having two types of singularities: local singularities and a polynomial growth at infinity, see assumptions (R1)-(R2) below. These assumptions cover  most of the electric potentials widely used in the quantum theory of solids, see Section 1.3 for examples.  Our main results  prove generically  the  existence of an almost-sure non-random thermodynamic limit for the  pressure, magnetization and zero-field orbital susceptibility. Furthermore we  give an explicit expression of these  limits on the maximal (independent of $b$) $z$-complex domains without resorting to the Vitali theorem. These significant advances  are made possible by employing the  gauge invariant magnetic perturbation theory to control the perturbed  resolvent operator, see \cite{CN3} for further  applications.

\subsection{The setting and the main result}

Consider a 3-dimensional quantum gas composed of non-relativistic identical charged particles, obeying either the  Bose-Einstein or the Fermi-Dirac statistics, and subjected to an external constant magnetic field. Since we are only   interested in orbital diamagnetic effects, we do not consider the spin of particles. Besides each particle interacts with a random electric potential (the sense will make precise hereafter) modeling a disordered medium. The interactions between particles are neglected (strongly diluted gas assumption) and the gas is at   thermal equilibrium.\\
\indent Let us precise our assumptions. The gas is confined in a cubic box centered at the origin given by $\Lambda_{L} = \Lambda_{L}(\bold{0}) := (-L/2,L/2)^{3}$, $L>0$. The magnetic field is defined by $\bold{B} := (0,0,B)$ with $B \geq 0$, and we use the symmetric gauge, i.e. the magnetic  vector potential is defined  by $B\bold{a}(\mathbf{x}) :=\frac{B}{2} \bold{e}_{3} \wedge \bold{x} = \frac{B}{2}(-x_{2},x_{1},0)$. In the following we denote $ b := \frac{qB}{c} \in \mathbb{R}$.\\
Let $(\Omega,\mathcal{F},\mathbb{P})$ be a complete probability space and $\mathbb{E}[\cdot\,] := \int_{\Omega} \mathbb{P}(\mathrm{d}\omega)(\cdot\,) $ be the  associated expectation. We consider random electric potentials,  i.e. scalar random fields $V:\Omega \times \mathbb{R}^{3} \rightarrow \mathbb{R}$, $(\omega,\bold{x}) \mapsto V^{(\omega)}(\bold{x})$ which are assumed to be jointly measurable with respect to the product of the $\sigma$-algebra $\mathcal{F}$ on $\Omega$ and the Borel-algebra $\mathcal{B}(\mathbb{R}^{3})$. In the whole of the paper we suppose:
\begin{itemize}
\item[(E)] $V^{(\omega)}$ is a $\mathbb{R}^{3}$-ergodic random field.
\end{itemize}
Recall that this assumption (see e.g. \cite{Ki2}) requires the existence of an ergodic group $\{\tau_{\bold{k}}\}_{\bold{k} \in \mathbb{R}^{3}}$ of measure-preserving automorphisms on $\Omega$ s.t. $V^{(\omega)}$ is $\mathbb{R}^{3}$-stationary in the sense that:
\begin{equation}
\label{Vstation}
V^{(\tau_{\bold{k}}\omega)}(\bold{x}) = V^{(\omega)}(\bold{x} - \bold{k})\quad \forall\bold{x} \in \mathbb{R}^{3},\,\forall\bold{k} \in \mathbb{R}^{3},\,\forall\omega \in \Omega.
\end{equation}
\begin{itemize}
\item[(R)] {The realizations of $V^{(\omega)}$ are given by:}
\begin{equation}
\label{decV}
V^{(\omega)}(\bold{x}) = V_{1}^{(\omega)}(\bold{x}) + V_{2}^{(\omega)}(\bold{x}) \quad \bold{x} \in \mathbb{R}^{3},\,\omega \in \Omega,
\end{equation}
\noindent where $\mathbb{P}$-a.s.  on $\Omega$:
\begin{itemize}
\item[(R1)] $V_{1}^{(\omega)}$ is an uniformly locally integrable function,   i.e.
$V_{1}^{(\omega)} \in L_{\mathrm{uloc}}^{p}(\mathbb{R}^{3})$ with $p>3$.
\item[(R2)] $V_{2}^{(\omega)}$ obeys the conditions:
\begin{equation}
\label{V2cond1}
0\leq V_{2}^{(\omega)}({\bold x}) \leq c_{\alpha}(\omega) (1+\vert \bold{x}\vert^{\alpha}) \quad \textrm{with $\alpha \in (0, \frac{1}{3})$ and $c_{\alpha}(\omega) >0$}.
\end{equation}
\end{itemize}
\end{itemize}
Recall that the space $L_{\mathrm{uloc}}^{p}(\mathbb{R}^{3})$ consists of measurable functions $f: \mathbb{R}^{3} \rightarrow \mathbb{C}$ satisfying:
\begin{equation}
\label{defuloc}
\Vert f\Vert_{1\leq p<\infty, \mathrm{uloc}} :=
\sup_{\bold{x} \in \mathbb{R}^{3}} \bigg(\int_{\vert \bold{x} - \bold{y}\vert < 1} \mathrm{d}\bold{y}\,\vert f(\bold{y})\vert^{p}\bigg)^{\frac{1}{p}}< \infty,\,\,\, \Vert f\Vert_{\infty, \mathrm{uloc}} := \mathrm{ess} \sup_{\bold{x} \in \mathbb{R}^{3}} \vert f(\bold{x}) \vert < \infty.
\end{equation}
We  discuss below about the choice of these assumptions, see  Section 1.3.\\

Introduce the 'one-particle' Hamiltonian in $L^{2}(\Lambda_{L})$. On $\mathcal{C}_0^\infty (\Lambda_L)$ consider the operator:
\begin{equation}
\label{HL}
H_{L}(b, \omega) := \frac{1}{2} (-i\nabla - b\bold{a})^{2} + V^{(\omega)} , \quad b  \in \mathbb{R}.
\end{equation}
It is well-known (see e.g. \cite[Prop. 2.1]{HLMW2}) that $\mathbb{P}$-a.s. on $ \Omega$, $\forall b\in\mathbb{R}$, \eqref{HL} defines a family of  self-adjoint and bounded below operators for any $L\in(0,\infty)$, denoted again by $H_L(b, \omega)$, with domain $D(H_L(b, \omega))= \mathcal{H}_{0}^{1}(\Lambda_{L}) \cap \mathcal{H}^{2}(\Lambda_{L})$. Obviously this definition corresponds to choose Dirichlet boundary conditions on $\partial\Lambda_{L}$. Moreover $H_{L}(b,\omega)$ has purely discrete spectrum; we denote the set of eigenvalues (counting multiplicities and in increasing order) by $\{e_{j}^{(\omega)}(b,L)\}_{j \geq 1}$. Besides by \cite[Prop. 2.2]{BCS1}, $\mathbb{P}$-a.s. on $ \Omega$, $\{H_L(b, \omega), b \in \mathbb C \}$ is a type (A)-entire family of operators.\\
\indent When $L=\infty$, define on  $\mathcal{C}_{0}^{\infty}(\mathbb{R}^{3})$ the operator:
\begin{equation}
\label{Hinfini}
H_{\infty}(b, \omega):=  \frac{1}{2}(-i\nabla - b\mathbf{a})^{2} + V^{(\omega)}, \quad  b \in \mathbb R.
\end{equation}
Then $\mathbb{P}$-a.s. on $ \Omega$,  $\forall b \in \mathbb R$, $H_{\infty}(b, \omega)$ is   essentially self-adjoint   and  its self-adjoint extension is bounded below, see \cite[Thm. B.13.4]{Si1}. Furthermore $H_{\infty}(b, \omega)$ is a family of $\mathbb{R}^{3}$-ergodic self-adjoint operators. This comes from the measurability of the mapping $\omega \in \Omega \mapsto H_{\infty}(b,\omega)$, see \cite[Coro. 3]{KM}, associated to the assumption (E) which leads to the covariance relation $T_{\bold{k},b} H_{\infty}(b, \omega)T_{-\bold{k},b}=H_{\infty}(b, \tau_{\bold{k}} \omega)$, $\forall\bold{k} \in \mathbb{R}^{3}$. Here $\{T_{\bold{k},b}\}_{\bold{k} \in \mathbb{R}^{3}}$ stands for the family of the usual real magnetic translations, see \eqref{trm}. We denote by $\Sigma$ the  $\mathbb{P}$-a.s.  spectrum of  $ H_{\infty}(0,\omega) $, \cite{PF}.

Our analysis is  based on  the  fact  that due to   assumption (R),  the variational principle  and the diamagnetic inequality \cite{Si1}, imply:
\begin{equation}
\label{minmax}
\forall\, b\in \mathbb{R},\quad \inf \sigma(H_{L}(b,\omega)) \geq \inf \sigma(H_{\infty}(b,\omega) )\geq  E_0,\quad E_0 :=  \inf \Sigma, \end{equation}
 as soon as the corresponding   self-adjoint operators are well-defined.\\
\indent Let us recall the basic points of the  grand-canonical formalism of the quantum statistical mechanics. Let $\beta := (k_{B}T)^{-1} > 0$ be the 'inverse' temperature ($k_{B}$ is the Boltzmann constant). Define the domains $\mathcal{D}_{\epsilon}=  \mathcal{D}_{\epsilon}(E_0)$, $\epsilon= \pm 1$, by:
\begin{equation}
\label{domz}
\mathcal{D}_{-1}:= \mathbb{C} \setminus [\mathrm{e}^{\beta E_{0}},+\infty),\quad  \mathcal{D}_{+1} := \mathbb{C} \setminus (-\infty, -\mathrm{e}^{\beta E_{0}}].
\end{equation}
In the following the parameter $\epsilon=-1$ refers to the bosonic case, $\epsilon=+1$ to the fermionic case.\\
For  $\beta>0$, $b \in \mathbb{R}$ and $z \in \mathcal{D}_{\epsilon} \cap \mathbb{R}_{+}^{*}$, the finite-volume  pressure and density are defined as \cite{ABN, AC}:
\begin{gather}
\label{PL}
P_{L}^{(\omega)}(\beta,b,z,\epsilon) := \frac{\epsilon}{\beta \vert \Lambda_{L}\vert} \mathrm{Tr}_{L^{2}(\Lambda_{L})} \ln(\mathbb{I} + \epsilon z \mathrm{e}^{-\beta H_{L}(b,\omega)}) = \frac{\epsilon}{\beta \vert \Lambda_{L}\vert} \sum_{j=1}^{\infty} \ln(1 + \epsilon z \mathrm{e}^{-\beta e_{j}^{(\omega)}(b,L)}), \\
\label{rhoL}
\rho_{L}^{(\omega)}(\beta,b,z,\epsilon) := \beta z \frac{\partial P_{L}^{(\omega)}}{\partial z}(\beta,b,z,\epsilon) = \frac{1}{\vert \Lambda_{L}\vert} \sum_{j=1}^{\infty} \frac{z \mathrm{e}^{-\beta e_{j}^{(\omega)}(b,L)}}{1 + \epsilon z \mathrm{e}^{-\beta e_{j}^{(\omega)}(b,L)}}.
\end{gather}
The relations \eqref{PL}-\eqref{rhoL} are well-defined since $\mathbb{P}$-a.s. on $\Omega$, $ \forall b \in \mathbb R$, the semigroup $\{\mathrm{e}^{-\beta H_{L}(b, \omega)}, \beta >0\}$ is trace class, see \cite[Eq. (2.12)]{BCS1}. Moreover from \cite[Thm. 1.1]{BCS1}, then $\mathbb{P}$-a.s. on $\Omega$,  $\forall \beta>0$, $P_{L}^{(\omega)}(\beta,\cdot, \cdot,\epsilon)$ is an analytic function in $(z,b)   \in \mathcal{D}_{\epsilon}\times \mathbb R$. This allows us to define the finite-volume magnetization and orbital susceptibility at $\beta>0$, $b \in \mathbb{R}$ and $z\in \mathcal{D}_{\epsilon} \cap \mathbb{R}_{+}^{*}$ as \cite{ABN, BCL1, BrCoLo1}:
\begin{equation}
\label{gensuscepti}
\mathcal{X}_{L,n}^{(\omega)}(\beta,b,z,\epsilon) :=  \bigg(\frac{q}{c}\bigg)^n  \frac{\partial^{n} P_{L}^{(\omega)}}{\partial b^{n}}(\beta, b,z,\epsilon) \quad n=1,2.
\end{equation}
Hereafter we will sometimes use the notation $\mathcal{X}_{L,0}^{(\omega)}(\beta,b,z,\epsilon) = P_{L}(\beta,b, z,\epsilon)$.\\
\indent We now want to formulate our  main results. We first introduce some notations.\\
\noindent By \cite[Thm. B.7.2]{Si1}, $\mathbb{P}$-a.s. on $\Omega$,  $\forall b \in \mathbb R$ and  $ \forall \xi \in \mathbb{C} \setminus [E_{0},\infty)$, $R_{\infty}(b,\omega,\xi):= (H_{\infty}(b,\omega) - \xi)^{-1} $  has an integral  kernel   $R_{\infty}^{(1)}(\cdot\,,\cdot\,;b,\omega,\xi)$  jointly continuous on $\mathbb{R}^{6} \setminus D_{\infty}$, where $D_{\infty} := \{(\bold{x},\bold{y})\in \mathbb{R}^{6}:\bold{x} = \bold{y}\}$. Under the same conditions,  let   $T_{j,\infty}(b,\omega,\xi)$, $j=1,2$, be the operators on $L^{2}(\mathbb{R}^{3})$  defined via  their integral kernel:
\begin{align}
\label{T1inftyk}
T_{1,\infty}(\bold{x},\bold{y};b, \omega,\xi) &:= \bold{a}(\bold{x} - \bold{y}) \cdot (i\nabla_{\bold{x}} + b \bold{a}(\bold{x})) R_{\infty}^{(1)}(\bold{x},\bold{y};b,\omega,\xi), \\
\label{T2inftyk}
T_{2,\infty}(\bold{x},\bold{y};b, \omega,\xi) &:= \frac{1}{2} \bold{a}^{2}(\bold{x} - \bold{y})R_{\infty}^{(1)}(\bold{x},\bold{y};b,\omega,\xi),\quad (\bold{x},\bold{y}) \in  \mathbb{R}^{6}\setminus D_{\infty}.
\end{align}
Let $\beta>0$, $b \in \mathbb{R}$, $z \in  \mathcal{D}_{\epsilon}$, and $ K \subset \mathcal{D}_{\epsilon}$  be a compact set containing $z$. Let $\Gamma_{K}$ be the positively oriented contour around $[E_{0},\infty)$ defined in \eqref{GammaK}. Introduce the following operators on $L^{2}(\mathbb{R}^{3})$:
\begin{gather}
\label{opIinfty0}
\mathcal{L}_{\infty,0}^{(\omega)}(\beta,b,z,\epsilon) := \frac{i}{2\pi} \int_{\Gamma_{K}} \mathrm{d}\xi\,  \mathfrak{f}_{\epsilon}(\beta,z;\xi)  R_{\infty}(b,\omega,\xi), \\
\label{opIinfty1}
\mathcal{L}_{\infty,1}^{(\omega)}(\beta,b,z,\epsilon) := -\frac{i}{2\pi} \int_{\Gamma_{K}} \mathrm{d}\xi\, \ \mathfrak{f}_{\epsilon}(\beta,z;\xi)  R_{\infty}(b,\omega,\xi) T_{1,\infty}(b,\omega,\xi), \\
\label{opIinfty2}
\mathcal{L}_{\infty,2}^{(\omega)}(\beta,b,z,\epsilon) := \frac{i}{\pi} \int_{\Gamma_{K}} \mathrm{d}\xi\,  \mathfrak{f}_{\epsilon}(\beta,z;\xi)  R_{\infty}(b,\omega,\xi)
\big((T_{1,\infty}(b,\omega,\xi))^{2} - T_{2,\infty}(b,\omega,\xi)\big),
\end{gather}
where $\mathfrak{f}_{\epsilon}(\beta,z; \cdot ) := \ln(1 + \epsilon z \mathrm{e}^{-\beta \cdot })$. We   will  prove   that  generically these operators  admit   a jointly continuous integral kernel on $\mathbb{R}^{6}$   denoted by $\mathcal{L}_{\infty,n}^{(\omega)}(\cdot\,,\cdot\,;\beta,b,z,\epsilon)$, and  are locally trace class.\\

Our  results concerning the pressure and the magnetization are the following:

\begin{theorem}
\label{maintheorem}
Suppose (E) and (R). Then:\\
i) $\mathbb{P}$-a.s. on $\Omega$, $\forall b \in \mathbb{R}$, $\forall 0<\beta_1<\beta_2$ and for any compact subset $K$ of $\mathcal{D}_{\epsilon}$, the thermodynamic limit of the pressure and magnetization exist.  These  limits are non-random, and:
\begin{equation}
\label{limitthermok}
\mathcal{X}_{\infty,n}(\beta,b, z,\epsilon) := \lim_{L \rightarrow \infty} \mathcal{X}_{L,n}^{(\omega)}(\beta,b, z,\epsilon)=  \bigg(\frac{q}{c}\bigg)^n \frac{\epsilon}{\beta} \mathbb{E}\big[\mathcal{L}_{\infty,n}^{(\omega)}(\bold{0},\bold{0};\beta,b,z,\epsilon)\big]\quad n=0,1,
\end{equation}
uniformly in $ (\beta, z) \in  [\beta_1, \beta_2]\times K$.\\
ii) $\mathbb{P}$-a.s. on $\Omega$, $ \forall b \in \mathbb R$ and $\forall \beta>0$, $P_{\infty}(\beta,b,\cdot\,,\epsilon):= \frac{\epsilon}{\beta} \mathbb{E}[\mathcal{L}_{\infty,0}^{(\omega)}(\bold{0},\bold{0};\beta,b,\cdot\,,\epsilon)]$ is an analytic function on $\mathcal{D}_{\epsilon}$. Moreover, $\forall 0<\beta_1<\beta_2$ and for any compact subset $K$ of $\mathcal{D}_{\epsilon}$, one has:
\begin{equation}
\label{limdens}
\rho_{\infty}(\beta,b,z,\epsilon) := \lim_{L \rightarrow \infty} \rho_{L}^{(\omega)}(\beta,b,z,\epsilon) = \beta z \frac{\partial  P_{\infty}}{\partial z}(\beta,b,z,\epsilon),
\end{equation}
uniformly in $ (\beta, z) \in  [\beta_1, \beta_2]\times K$.\\
iii) $\mathbb{P}$-a.s. on $\Omega$, $ \forall \beta>0$, $ \forall z\in\ \mathcal{D}_{\epsilon}$, $P_{\infty}(\beta,\cdot\,,z,\epsilon) \in  \mathcal{C}^{1}( \mathbb{R} )$, and
$\mathcal{X}_{\infty,1}(\beta,b,z,\epsilon) =  (\frac{q}{c}) \frac{\partial P_{\infty}}{\partial b} (\beta,b,z,\epsilon)$.
\end{theorem}

For  the zero-field orbital susceptibility, we have:

\begin{theorem}
\label{maintheorem1}
Suppose also (E) and (R).  Then:\\
i) $\mathbb{P}$-a.s. on $\Omega$, $\forall 0<\beta_1<\beta_2$ and for any compact subset $K$ of $\mathcal{D}_{\epsilon}$, the thermodynamic limit of the zero-field orbital susceptibility exists.  The limit is non-random and it is given by:
\begin{equation}
\label{limitthermok2}
\mathcal{X}_{\infty,2}(\beta,0, z,\epsilon) := \lim_{L \rightarrow \infty} \mathcal{X}_{L,2}^{(\omega)}(\beta,0,z,\epsilon)=  \bigg(\frac{q}{c}\bigg)^2 \frac{\epsilon}{\beta} \mathbb{E}\big[\mathcal{L}_{\infty,2}^{(\omega)}(\bold{0},\bold{0};\beta,0,z,\epsilon)\big],
\end{equation}
uniformly in $(\beta,z) \in [\beta_1, \beta_2]\times K$.\\
ii) Let $\alpha \in [0, \frac{1}{4})$. Then  $\mathbb{P}$-a.s. on $\Omega$, $ \forall \beta>0$ and $ \forall z\in\ \mathcal{D}_{\epsilon}$, $P_{\infty}(\beta,\cdot\,,z,\epsilon)$ is a $\mathcal{C}^{2}$-function near $b=0$, and
$ \mathcal{X}_{\infty,2}(\beta,0,z,\epsilon) =  (\frac{q}{c})^2 \frac{\partial^2P_{\infty}}{\partial b^{2}}(\beta,0,z,\epsilon)$.
\end{theorem}

\begin {remark}    $i)$ Theorems \ref{maintheorem} and \ref{maintheorem1}  define  the pressure, the magnetization  and the susceptibility of the system, although the corresponding  physical quantities, strictly speaking, require   the strict positivity of the  fugacity $z$. 

 $ii)$ By using  simple  arguments, $ P_{\infty}(\beta,-b,z,\epsilon) = \overline{P_{\infty}(\beta,b,z,\epsilon)} = P_{\infty}(\beta,b,z,\epsilon).$
Then Theorem \ref{maintheorem} $iii)$  implies that  $\mathbb{P}$-a.s., $
\forall\, \beta >0,\, \forall\,z \in  \mathcal{D}_{\epsilon}$, $  \mathcal{X}_{\infty,1}(\beta,0,z,\epsilon) = 0$.
\end{remark}
\subsection{Discussions and examples} Let us comment (R1). Our approach is based on the representation of the finite-volume pressure by using the  Dunford-Schwartz integral formula, see Section 4. However this requires the use of bounded below Schr\"odinger operators, see \cite[Sect. VII.9]{DS}. As we allow realizations of $V_{1}^{(\omega)}$ to be negative with local singularities, this condition is fulfilled if  $\mathbb{P}$-a.s. $V_{1}^{(\omega)} \in L^{p}_{\mathrm{uloc}}(\mathbb{R}^{3})$ with $p>\frac{3}{2}$, see \cite[Eq. (A21)]{Si1}. The additional condition $p>3$ will appear when estimating the derivative of the infinite-volume resolvent's integral kernel in \eqref{T1inftyk}, see the proof of Lemma \ref{prokdres} $ii)$. Notice that (R1) does not cover  Coulomb-type singularities and implies that $V_{1}^{(\omega)}$  roughly behaves locally   like $\mathcal{O}(\vert \bold{x}\vert^{-(1- \epsilon)})$, $\epsilon>0$. Finally  we mention that under the conditions on $V_{1}^{(\omega)}$ given in Section 1.2, there  exists a $\omega$-independent constant $C >0$ s.t. $\mathbb{P}$-a.s. on $\Omega$, \cite{Ki2}:
\begin{equation}
\label{constv}
\Vert V_1^{(\omega)}  \Vert_{p, \mathrm{uloc}} \leq C.
\end{equation}

Let us discuss  \eqref{V2cond1}. Unlike $V_{1}^{(\omega)},$  we allow   $V_{2}^{(\omega)}$ to have a  polynomial growth at infinity. Notice that $V_{2}^{(\omega)}$ is not supposed to be monotone; indeed in that case the operator $ H_\infty (b, \omega)$ may  have only discrete spectrum, and then the problem becomes irrelevant. On the contrary,  $V_{2}^{(\omega)}$  is  basically of sparse barrier potential type,  so that
  the corresponding operator  has  $[E_0, \infty)$ in its   spectrum and    with  a non-trivial   spectral type. By considering an arbitrary  polynomial growth: $\vert \bold{x}\vert^{\alpha}$, $\alpha>0$, the condition $(n+1)\alpha< 1$ with $n=1,2$ will appear when proving the existence of the thermodynamic limits \eqref{limitthermok}-\eqref{limitthermok2}, see Proposition \ref{secondresult}.  However the relations given in Theorems \ref{maintheorem} $iii)$ and \ref{maintheorem1} $ii)$ require  the additional condition  $(n+2)\alpha< 1$. Notice that if we only are interested in the case of the pressure or density, the assumption (R) can be relaxed to the ones ensuring the validity of the Pastur-Shubin formula for the integrated density of states, see \cite[Sect. 2]{Vs} and the method in \cite[Sect. 3.4]{BCS1}.\\
\indent Let us discuss  the assumption (E). Since the proof of Theorems \ref{maintheorem} $i)$ and \ref{maintheorem1} $i)$ is based on the Birkhoff-Khintchine theorem in \cite[Prop. 1.13]{PF}, the use of $\mathbb{R}^{3}$-ergodic random field is crucial. Nevertheless we can replace the assumption (E) with:
\begin{itemize}
\item [(E')] $V^{(\omega)}$ is a $\mathbb{Z}^{3}$-ergodic random field,
\end{itemize}
since  this reduces  to a $\mathbb{R}^{3}$-ergodic random field with  the suspension technique, see \cite{PF}.\\

\indent We now give physically relevant examples covered by the assumptions (E)-(R) or (E')-(R).
\begin{itemize}
\item [\textbf{0}.] \textit{The non-negative Poisson random field}.
\end{itemize}
In \eqref{decV} set $V_{1}^{(\omega)} = 0$ and choose for $V_{2}^{(\omega)}$ the random field with realizations given by:
\begin{equation*}
\label{amorphous}
V_{P}^{(\omega)}(\bold{x}) = \int_{\mathbb{R}^{3}} \boldsymbol{\mu}_{\lambda}^{(\omega)}(\mathrm{d}\bold{y})\,u(\bold{y} - \bold{x})\quad \bold{x} \in \mathbb{R}^{3},\,\, \omega \in \Omega,
\end{equation*}
where $\boldsymbol{\mu}_{\lambda}^{(\omega)}$ denotes the random Poisson measure on $\mathbb{R}^{3}$ with parameter $\lambda>0$ and $u(\cdot\,) : \mathbb{R}^{3} \rightarrow [0,\infty)$ is the single-site potential, see e.g. \cite{PF, Ki1}. By assuming that $u$ is compactly supported and  $u \in L^{\infty}(\mathbb{R}^{3})$, then (E) is satisfied as well as (R2), since we have $\mathbb{P}$-a.s., $\forall \bold{x} \in \mathbb{R}^{3}$, $0 \leq V_{P}^{(\omega)}(\bold{x}) \leq c(\omega) \ln(1 + \vert \bold{x} \vert)$, see \cite[Lem. 2.2]{GHK}.


\begin{itemize}
\item [\textbf{1}.] \textit{The alloy-type random field (the so-called 'Anderson potential')}.
\end{itemize}
In \eqref{decV} set $V_{2}^{(\omega)} = 0$ and choose for $V_{1}^{(\omega)}$ the random field with realizations given by:
\begin{equation*}
\label{anderson}
V_{A}^{(\omega)}(\bold{x}) = g\sum_{\bold{j} \in \mathbb{Z}^{3}} \lambda_{\bold{j}}(\omega) u(\bold{x} - \bold{x}_{\bold{j}})\quad \bold{x} \in \mathbb{R}^{3},\,\, \omega \in \Omega, \quad g \in \mathbb R.
\end{equation*}
Here  $\{\lambda_{\bold{j}}\}_{\bold{j} \in \mathbb{Z}^{3}}$ is a family of i.i.d. random variables with a  common  distribution  what ensures (E'), see e.g. \cite{PF,Ki1}. Besides we suppose that $\forall\, \bold{j} \in \mathbb{Z}^{3}$, $\vert \lambda_{\bold{j}}(\omega)\vert \leq 1 $ and $u$ satisfies the Birman-Solomyak condition: $\sum_{\bold{j} \in \mathbb{Z}^{3}} (\int_{\Lambda_{1}(\bold{j})} \mathrm{d}\bold{x}\, \vert u(\bold{x})\vert^{p})^{\frac{1}{p}} < \infty$, $p > 3$, where $\Lambda_{1}(\bold{j})$ denotes the unit cube centered on site $\bold{j}$.
Then $V_{A}^{(\omega)} \in L_{\mathrm{uloc}}^{p}(\mathbb{R}^{3})$ with $p>3$, and we have (see also \cite{Sa}):
 \begin{equation} \label{and}
\mathcal{X}_{\infty,n}(\beta,b,z,\epsilon) =   \bigg(\frac{q}{c}\bigg)^n \frac{\epsilon}{\beta \vert \Lambda_{1}(\bold{0})\vert}  \int_{\Lambda_{1}(\bold{0})} \mathrm{d}\bold{x}\, \mathbb{E}\big[\mathcal{L}_{\infty,n}^{(\omega)}(\bold{x},\bold{x};\beta,b,z,\epsilon)\big]\quad n=0,1,2.
\end{equation}

\begin{itemize}
\item [\textbf{2}.] \textit{Further models}.
\end{itemize}
{\it The periodic case}. Set $V_2=0$.  We assume that $V_1 \in L_{\mathrm{uloc}}^{p}(\mathbb{R}^{3}),p> 3$ and  $\mathbb Z^3$-periodic. The suspension method can be applied to this case leading to define from $V_1$, a $\mathbb R^3$-ergodic random field $V_1^{(\omega)}$. So  \eqref{and} holds true but in that case,
$ \mathbb{E}\big[\mathcal{L}_{\infty,n}^{(\omega)}(\cdot\,,\cdot\,;\beta,b,z,\epsilon)\big]= \mathcal{L}_{\infty,n}^{(\omega)}(\cdot\,,\cdot\,;\beta,z,\epsilon)$, see also \cite{BCS1,Sa}. The  {\it  random displacements model} on $\mathbb{R}^{3}$ (see e.g. \cite{Ki1}) or {\it the quasi-periodic case} (see e.g. \cite{PF,Ki2}) are also  covered by our results.
\subsection{The content}  In Section 2 we investigate the $\mathbb{P}$-a.s. analyticity (in the Hilbert-Schmidt topology) of the finite-volume resolvent and the analyticity of its integral kernel w.r.t. $b$.\\
\indent In Section 3 we apply the gauge invariant magnetic perturbation theory (see e.g. \cite{CN3, N1}) to the finite-volume  perturbed resolvent. This allows us to get an expression of the partial derivatives w.r.t. $b$ of its integral kernel by keeping a good control over the linear growth of the magnetic vector potential.  Although taking the resolvent as 'central object' leads to further technical difficulties, but it allows us to get more powerful results.\\
\indent In Section 4 we formulate the finite-volume quantities  by using the expressions of the derivatives w.r.t. $b$ of the  kernel of the finite-volume resolvent  obtained in the previous section. \\
\indent In Section 5 we investigate the main properties on the infinite-volume operators involved in the thermodynamic limits. Note that here the gauge invariant perturbation theory gives $\mathbb{R}^{3}$-stationary
quantities, see (1.1). This is necessary to apply the limit ergodic theorem.\\
\indent In Section 6 we prove the almost-sure non-random thermodynamic limits by the Birkhoff-Khintchine theorem and we investigate the bulk properties. This section also contains the proof of Theorems \ref{maintheorem} and \ref{maintheorem1}.  
\section{Regularity of the finite-volume resolvent in the $b$-parameter}

\subsection{Analyticity in the Hilbert-Schmidt topology}

Hereafter we will denote respectively by $\Vert \cdot \Vert_{\mathfrak{I}_{1}}$, $\Vert \cdot\Vert_{\mathfrak{I}_{2}}$ and $\Vert \cdot\Vert$, the trace norm in $\mathfrak{I}_{1}(L^{2}(\Lambda_{L}))$, the Hilbert-Schmidt (H-S) norm in $\mathfrak{I}_{2}(L^{2}(\Lambda_{L}))$, and the operator norm in $\mathfrak{B}(L^{2}(\Lambda_{L}))$.\\
\indent From \cite[Prop. 3.1]{BCS1},  $\mathbb{P}$-a.s.,
 $ \forall b_{0} \in \mathbb{R}$, $ \forall 0< L< \infty$ and $ \forall \xi \in \mathbb{C} \setminus [E_{0},\infty)$,
there exists a complex neighborhood $\mathcal{V}_{\xi,L}(b_{0})$ of $b_{0}$ s.t. the operator-valued function $\mathcal{V}_{\xi,L}(b_{0}) \owns b \mapsto R_{L}(b,\omega,\xi):= (H_{L}(b,\omega)-\xi)^{-1}$ is analytic in the H-S topology. We now precise this result.\\
\indent Consider the following  operators on $L^2(\Lambda_L)$:
\begin{equation}
\label{opSiL}
S_{1,L}(b_0,\omega,\xi) := \bold{a} \cdot (i \nabla + b_0\bold{a})R_{L}(b_0,\omega,\xi),\quad
S_{2,L}(b_0,\omega,\xi) := \frac{1}{2} \bold{a}^{2} R_{L}(b_0,\omega,\xi).
\end{equation}
Again from \cite[Sect. 2]{BCS1},  $\mathbb{P}$-a.s.,  $\forall b_0 \in \mathbb R$, $ \forall 0< L< \infty$ and $ \forall \xi \in \mathbb{C} \setminus [E_{0},\infty)$, they are bounded and,   denote by   $d(\xi):= \mathrm{dist}(\xi, [E_{0},\infty)) $, then there exists a constant $c>0$ s.t.:
\begin{equation}
\label{normSiL1}
\Vert S_{1,L}(b_0,\omega,\xi) \Vert \leq c (1+ \frac{1}{d(\xi)})(1 + \vert \xi\vert) L \quad \textrm{and} \quad \Vert S_{2,L}(b_0,\omega,\xi) \Vert \leq  \frac{c}{d(\xi)} L^{2}.
\end{equation}
For all integer $k\geq 1$, introduce the following family of H-S  operators on $L^{2}(\Lambda_{L})$:
\begin{equation}
\label{opJkLi}
J_{k,L}({\bf i})(b_0, \omega,\xi) := R_{L}(b_0,\omega,\xi) \prod_{m=1}^{k} S_{i_{m},L}(b_0,\omega,\xi),\quad {\bf i}= \{i_{1},\ldots,i_{k}\} \in \{1,2\}^k,
\end{equation}
and for  $n\geq k \geq 1$, $\chi_{k}^{n}$,   the characteristic function :
\begin{equation*}
\chi_{k}^{n}(\bf{i}):= \left\{\begin{array}{ll}
1\quad \textrm{if $i_{1} + \dotsb + i_{k} = n$} \\
0\quad \textrm{otherwise.}
\end{array}\right.
\end{equation*}
\begin{proposition}
\label{thm1}
$\mathbb{P}$-a.s. on $ \Omega$, $ \forall b_{0} \in \mathbb{R}$, $ \forall 0< L< \infty$  and $ \forall \xi \in \mathbb{C} \setminus [E_{0},\infty)$, then there exists a complex neighborhood $\mathcal{V}_{\xi,L}(b_{0})$ of $b_{0}$ s.t.   in the H-S operators sense:
\begin{equation}
\label{opresserie}
\forall b \in \mathcal{V}_{\xi,L}(b_{0}),\quad R_{L}(b,\omega,\xi) =  R_{L}(b_{0},\omega,\xi) +  \sum_{n=1}^{\infty} \frac{(b - b_{0})^{n}}{n!} \frac{\partial^{n} R_{L}}{\partial b^{n}}(b_{0},\omega,\xi),
\end{equation}
where for $n \geq 1$:
\begin{equation}
\label{opdnres}
\frac{\partial^{n} R_{L}}{\partial b^{n}}(b_{0},\omega,\xi)
:=
n! \sum_{k=1}^{n} (-1)^{k} \sum_{{\bf i}\in \{1,2\}^{k}} \chi_{k}^{n}({\bf i}) J_{k,L}({\bf i})(b_{0},\omega,\xi).
\end{equation}
\end{proposition}

\noindent\textit{Proof.} Let $\xi \in \mathbb{C} \setminus [E_{0},\infty)$, $0< L< \infty$  and $(b,b_0) \in  {\mathbb C}\times{\mathbb R}$. Set: \begin{equation}
\label{opSL}
S_{L}(b,b_{0},\omega,\xi) := \delta b\, S_{1,L}(b_{0},\omega,\xi) + (\delta b)^{2} S_{2,L}(b_{0},\omega,\xi), \quad \delta b:=b-b_{0}.
\end{equation}
From the $n$-th iterated second resolvent equation, one has:
\begin{equation*}
R_{L}(b,\omega,\xi) = R_{L}(b_{0},\omega,\xi)\bigg[\mathbb{I}+ \sum_{k=1}^{n} (-1)^{k} \big( S_{L}(b,b_{0},\omega,\xi)\big)^{k}\bigg]
 + (-1)^{n+1} R_{L}(b,\omega,\xi) \big(S_{L}(b,b_{0},\omega,\xi)\big)^{n+1}.
\end{equation*}
Then by \eqref{opSL},  one gets after some rearranging:
\begin{equation}
\label{rearange}
R_{L}(b,\omega,\xi) = R_{L}(b_{0},\omega,\xi)
+ \sum_{k=1}^{n} (\delta b)^{k} \sum_{l=1}^{k} (-1)^{l} \sum_{{\bf i} \in \{1,2\}^{l}} \chi_{l}^{k}({\bf i}) J_{l,L}({\bf i})(b_{0},\omega,\xi) + \mathcal{S}_{n+1,L}(b,b_{0},\omega,\xi),
\end{equation}
\begin{multline}
\label{Sn+1}
\textrm{with:}\quad \mathcal{S}_{n+1,L}(b,b_{0},\omega,\xi) :=
(\delta b)^{n+1} \bigg\{\sum_{k=0}^{n-1} (\delta b)^{k} \sum_{l=1}^{n} (-1)^{l} \sum_{{\bf i}\in \{1,2\}^{l}} \chi_{l}^{k+n+1}({\bf i}) J_{l,L}({\bf i})(b_{0},\omega,\xi) + \\
+ (-1)^{n+1} \sum_{k=0}^{n+1} (\delta b)^{k}   \sum_{{\bf i} \in \{1,2\}^{n+1}} \chi_{n+1}^{k+n+1}({\bf i}) R_{L}(b,\omega,\xi) \prod_{m=1}^{n+1} S_{i_{m},L}(b_{0},\omega,\xi)\bigg\}.
\end{multline}
By using the analyticity properties of the resolvent in the H-S topology given above, the proposition follows from  \eqref{rearange} and \eqref{Sn+1}.\qed



\subsection{Regularity of the integral kernel} We know from \cite[Thm. B.7.2]{Si1} that $\mathbb{P}$-a.s., $\forall b \in \mathbb{R}$,  $\forall L \in (0, \infty]$,  $\forall \eta>0$ and $ \forall \xi \in \mathbb{C}$, $ d(\xi) \geq \eta$, $R_{L}(b,\omega,\xi)$ has an integral kernel $R_{L}^{(1)}(\cdot\,,\cdot\,;b,\omega,\xi)$ jointly continuous on $\Lambda_{L}^2\setminus D_{L}$,  $D_{L}:=\{(\bold{x},\bold{y})\in \Lambda_{L}^2:\bold{x}=\bold{y}\}$. Moreover $\mathbb{P}$-a.s., $\forall \eta>0$ there exists a constant $\gamma = \gamma(\eta) >0$ and a polynomial $p(\cdot\,)$ s.t. $\forall b \in \mathbb{R}$, $\forall L \in (0, \infty]$   and $ \forall \xi \in \mathbb{C}$, $ d(\xi)\geq \eta$:
\begin{equation}
\label{eskres}
\forall\, (\bold{x},\bold{y}) \in \Lambda_{L}^2\setminus D_{L},\quad \vert R_{L}^{(1)}(\bold{x},\bold{y};b,\omega,\xi)\vert \leq \vert p( \xi) \vert \frac{\mathrm{e}^{- \gamma_\xi \vert \bold{x} - \bold{y}\vert}}{\vert \bold{x} - \bold{y}\vert}\quad \textrm{with $\gamma_\xi  := \frac{ \gamma}{1 + \vert \xi\vert}$}.
\end{equation}
Notice that $\gamma_\xi $ can  be more explicit w.r.t. the energy parameter. The one given in \eqref{eskres}, valid for $\xi \in \mathbb{C}, d(\xi) \geq \eta>0$, contains the $\xi$-dependence we need in the whole of this work.

\remark \label{2.2}     Consequently    the product $\Pi_{l=1}^m R_{L}(b,\omega,\xi_l)$, $m\geq 2$ has an integral kernel jointly continuous on $\Lambda_{L}^2$, and moreover, $\mathbb{P}$-a.s.,     $\forall \eta>0$,  there exists a constant $\gamma = \gamma(\eta,m) >0$ and a polynomial $p(\cdot\,)$ s.t. $\forall b \in \mathbb{R}$, $\forall L \in (0, \infty]$ and $ \forall \xi_{l}\in \mathbb{C},  l=1,\ldots,m $, $ d(\xi_{l})\geq \eta$:
\begin{equation}
\label{eskrespow}
\forall\, (\bold{x},\bold{y}) \in \Lambda_{L}^2,\quad \vert \Big(  \Pi_{l=1}^m R_{L} (b,\omega,\xi_l)\Big)(\bold{x},\bold{y})\vert \leq  ( \Pi_{l=1}^m  \vert p( \xi_l )\vert )\mathrm{e}^{-\ \gamma_\xi  \vert \bold{x} - \bold{y}\vert}, \quad \gamma_\xi  := \frac{ \gamma}{1 + \vert \xi\vert}.
\end{equation}
This follows by induction on $m$ from the case of $m=2$ with:
\begin{equation*}
 \Big(  \Pi_{l=1}^m R_{L} (b,\omega,\xi_l)\Big)(\bold{x},\bold{y}) = \int_{\Lambda_{L}} \mathrm{d}\bold{z}\,
  \Big(  \Pi_{l=1}^{m-1} R_{L} (b,\omega,\xi_l)\Big)(\bold{x},\bold{z}) R_{L}^{(1)}(\bold{z},\bold{y};b,\omega,\xi_m).
\end{equation*}
When $m=2$ the continuity property holds true since the kernel $R_{L}^{(1)}(\cdot\,,\cdot\,;b,\omega,\xi)$ fulfills the assumptions of Lemma \ref{continuity}. Furthermore from \eqref{eskres} and \eqref{esintk}, we get   \eqref{eskrespow}.
\medskip


From Proposition \ref{thm1} together with these  results, we now  prove:

\begin{proposition}
\label{thm2}
$\mathbb{P}$-a.s. on $\Omega$, $\forall b_{0} \in \mathbb{R}$, $ \forall 0< L< \infty$, $ \forall \eta >0$ and  $ \forall \xi \in \mathbb{C}$, $d(\xi) \geq \eta $, then there exists a complex neighborhood $\nu_{\xi,L,\omega}(b_{0})$ of $b_{0}$ s.t.   $ \forall (\bold{x},\bold{y}) \in \Lambda_{L}^2\setminus D_{L}$, $b \mapsto R_{L}^{(1)}(\bold{x},\bold{y}; b ,\omega,\xi) $ is an analytic function on $\nu_{\xi,L,\omega}(b_0)$ .
\end{proposition}

An important point for the proof of Proposition \ref{thm2} is the following estimate. We choose to give its proof in the appendix of the paper, see Section 7.
\begin{lema}
\label{prokdres}
i) $\mathbb{P}$-a.s. on $\Omega$,  $ \forall b \in \mathbb R$, $\forall L \in (0, \infty] $, $\forall  \eta>0$ and  $\forall \xi \in \mathbb{C}$,  $d(\xi) \geq \eta$, then $(i \nabla_{\bold{x}} + b \bold{a}(\bold{x}))  R_{L}(b,\omega,\xi)$  has an integral  kernel  jointly continuous on $\Lambda_{L}^2\setminus D_{L}$.\\
ii) $\mathbb{P}$-a.s. on $\Omega$, $\forall  \eta>0$, there exists a constant $\gamma=\gamma(\eta) >0$ and a polynomial $p(\cdot\,)$ s.t. $\forall L \in (0, \infty]$, $\forall b \in \mathbb{R}$, $\forall \xi \in \mathbb{C}$, $d(\xi) \geq \eta$, one has on $ \Lambda_{L}^2 \setminus D_{L}$:
\begin{gather}
\label{eskdres}
\vert  (i \nabla_{\bold{x}} + b \bold{a}(\bold{x})) R_{L}^{(1)}(\bold{x},\bold{y};b,\omega,\xi)\vert \leq(1+ \vert b\vert)^3 \vert p( \xi) \vert (1 + \vert \bold{x} \vert^{\alpha} + \vert \bold{y} \vert^{\alpha}) \frac{\mathrm{e}^{- \gamma_\xi  \vert \bold{x} - \bold{y}\vert}}{\vert \bold{x} - \bold{y}\vert^{2}}, \quad \gamma_\xi  := \frac{ \gamma}{1 + \vert \xi\vert}.
\end{gather}
\end{lema}
\noindent\textit{Proof of Proposition \ref{thm2}.} Under the conditions of the proposition, from  \eqref{eskres} and Lemma \ref{prokdres}, the operators in  \eqref{opSiL} have an integral kernel  jointly continuous on $ \Lambda_{L}^2\setminus D_{L} $, given by:
\begin{equation*}
\begin{split}
S_{1,L}(\bold{x},\bold{y};b_0, \omega,\xi) &:= \bold{a}(\bold{x})\cdot(i\nabla_{\bold{x}} + b_0 \bold{a}(\bold{x})) R_{L}^{(1)}(\bold{x},\bold{y};b_0, \omega,\xi),\\
S_{2,L}(\bold{x},\bold{y};b_0, \omega,\xi) &:= \frac{1}{2} \bold{a}^{2}(\bold{x}) R_{L}^{(1)}(\bold{x},\bold{y};b_0,\omega,\xi),
\end{split}
\end{equation*}
and moreover there exists a constant $\gamma=\gamma(\eta)>0$ and a polynomial $p(\cdot\,)$ s.t. on $ \Lambda_{L}^2\setminus D_{L}$:
\begin{equation}
\label{eskSiL}
\vert S_{j,L}(\bold{x},\bold{y};b_0,\omega,\xi)\vert \leq \vert p(\xi) \vert (1+L^{\alpha}) L^{j} \frac{\mathrm{e}^{- \gamma_{\xi}\vert \bold{x} - \bold{y}\vert}}{\vert \bold{x} - \bold{y}\vert^{2}}, \quad j=1,2.
\end{equation}
Then, considering \eqref{opdnres}, $\mathbb{P}$-a.s., $\forall  b_{0} \in \mathbb{R}$,  $ \forall 0< L< \infty$,  $\forall \eta >0$ and  $ \forall \xi \in \mathbb{C}$, $d(\xi) \geq \eta $,  the operator $\frac{\partial^{n} R_{L}}{\partial b^{n}}(b_{0},\omega,\xi)$, $n\geq 1 $ has an integral kernel given on  $ \Lambda_{L}^2\setminus D_{L}$ by:
\begin{equation}
\label{kdnres}
\frac{\partial^{n} R_{L}}{\partial b^{n}}(\bold{x},\bold{y};b_{0},\omega,\xi) := n! \sum_{k=1}^{n} (-1)^{k} \sum_{{\bf i} \in \{1,2\}^{k}} \chi_{k}^{n}({\bf i}) J_{k,L}({\bf i})(\bold{x},\bold{y};b_{0},\omega,\xi),
\end{equation}
where $J_{k,L}({\bold i})(\cdot\,,\cdot\,;b_{0},\omega,\xi)$ stands for the integral kernel of the operator in \eqref{opJkLi}. It reads as:
\begin{multline*}
J_{k,L}({\bold i})(\bold{x},\bold{y};b_{0},\omega,\xi) :=
\int_{\Lambda_{L}} \mathrm{d}\bold{z}_{1}\dotsb \int_{\Lambda_{L}}\mathrm{d}\bold{z}_{k}\, R_{L}^{(1)}(\bold{x},\bold{z}_{1};b_{0},\omega,\xi) \times \\
\times S_{i_{1},L}(\bold{z}_{1},\bold{z}_{2};b_{0},\omega,\xi)\dotsb S_{i_{k},L}(\bold{z}_{k},\bold{y};b_{0},\omega,\xi).
\end{multline*}
  Furthermore, from estimates \eqref{eskres} and \eqref{eskSiL}, by applying $k$-times successively  Lemma \ref{continuity} $i)$ combined with \eqref{secestie}, we obtain that    $J_{k,L}({\bold i})(\cdot\,,\cdot\,;b_{0},\omega,\xi)$ is jointly continuous on $\Lambda_{L}^2 \setminus D_{L}$. By using \eqref{eskres}, \eqref{eskSiL} with  Lemma \ref{proestim} $ii)$, $\mathbb{P}$-a.s., $\forall b_0 \in \mathbb{R}$, $\forall \eta >0 $, there exists a constant $\gamma=\gamma(\eta)>0$ and a polynomial $p(\cdot\,)$ s.t.  $\forall 0< L< \infty$, $\forall \xi \in \mathbb{C}$, $d(\xi) \geq \eta $ and $\forall (\bold{x},\bold{y}) \in \Lambda_{L}^{2}\setminus D_{L}$:
\begin{equation*}
\vert J_{k,L}({\bold i})(\bold{x},\bold{y};b_{0},\omega,\xi) \vert \leq \vert p(\xi) \vert^{k} (1+L^{\alpha})^{k} L^{i_{1} + \dotsb + i_{k}} \frac{\mathrm{e}^{- \frac{\gamma_{\xi}}{2^{k}} \vert \bold{x} - \bold{y} \vert}}{\vert \bold{x} - \bold{y} \vert}, \quad \gamma_\xi  := \frac{ \gamma}{1 + \vert \xi\vert}.
\end{equation*}
This estimate then  imply the following rough estimate which holds on $\Lambda_{L}^{2}\setminus D_{L}$:
\begin{equation}
\label{eskdnres}
\forall n \in \mathbb{N}^{*},\quad \frac{1}{n!} \bigg\vert \frac{\partial^{n} R_{L}}{\partial b^{n}}(\bold{x},\bold{y};b_{0},\omega,\xi)\bigg\vert \leq  c^n\vert p( \xi) \vert^n  (1+ L^{\alpha})^{n} L^{n} \frac{\mathrm{e}^{- \frac{{\gamma_{\xi}}}{2^{n}} \vert \bold{x} - \bold{y}\vert}}{\vert \bold{x} - \bold{y}\vert},
\end{equation}
for some constant $c>0$. So the analyticity property follows from \eqref{opresserie}, \eqref{eskdnres}  since for $\vert b -b_{0}\vert$ sufficiently small, the corresponding Taylor expansion converges. Here we use, 
\begin{equation} \label{CF}
 \sup_{\bold{x} \in \Lambda_{L}} \int_{\mathbb{R}^{3}} \mathrm{d}\bold{y}\, \frac{\mathrm{e}^{- \frac{\varsigma}{2^{n}} \vert \bold{x} - \bold{y} \vert}}{\vert \bold{x} - \bold{y} \vert} = \Big(\frac{2^{n}}{\varsigma}\Big)^{2},\quad  \varsigma>0.
 \end{equation} \qed



\section{A new expression for the first and second derivatives w.r.t. $ b$}

In order  to determine the thermodynamic limits, we want to isolate in the expression \eqref{kdnres} the term giving rise to the growth  w.r.t. $L$ when $L \to \infty$, see \eqref{eskdnres}. \\
\indent Let $\bold{x},\bold{y}\in \Lambda_{L}$. Define the magnetic phase $\phi$ as:
\begin{equation}
\label{phase}
\phi(\bold{x},\bold{y}) := \frac{1}{2} \bold{e}_{3} \cdot (\bold{y} \wedge \bold{x}) = - \phi (\bold{y},\bold{x})  \quad \textrm{with $\bold{e}_{3}:=(0,0,1)$}.
\end{equation}
Introduce on $L^{2}(\Lambda_{L})$ the operators $T_{j,L}(b,\omega,\xi)$, $j=1,2$ defined via their integral kernel:
\begin{align}
\label{kT1L}
\forall (\bold{x},\bold{y}) \in \Lambda_{L}^{2}\setminus D_{L},\quad T_{1,L}(\bold{x},\bold{y};b,\omega,\xi) &:= \bold{a}(\bold{x} - \bold{y})\cdot(i \nabla_{\bold{x}} + b\bold{a}(\bold{x}))R_{L}^{(1)}(\bold{x},\bold{y};b,\omega,\xi), \\
\label{kT2L}
T_{2,L}(\bold{x},\bold{y};b,\omega,\xi) &:= \frac{1}{2} \bold{a}^{2}(\bold{x} - \bold{y}) R_{L}^{(1)}(\bold{x},\bold{y};b,\omega,\xi).
\end{align}
Obviously $\vert \bold{a}(\bold{x} - \bold{y}) \vert \leq \vert \bold{x} - \bold{y} \vert$, then from  \eqref{eskres} and \eqref{eskdres},  $\mathbb{P}$-a.s., $\forall b \in \mathbb{R}$, $\forall \eta >0 $, there exists  $\gamma=\gamma(\eta)>0$ and a polynomial $p(\cdot\,) $ s.t.   $\forall 0< L< \infty$ and  $\forall \xi \in \mathbb{C}$, $d(\xi) \geq \eta $:
\begin{equation}
\label{eskTiL}
 \vert T_{j,L}(\bold{x},\bold{y};b,\omega,\xi)\vert \leq \vert p(\xi) \vert  (1 + L^{\alpha}) \frac{\mathrm{e}^{- \gamma_{\xi}{\vert \bold{x} - \bold{y}} \vert}}{\vert \bold{x} - \bold{y} \vert}\quad  j=1,2.
\end{equation}
Hence $T_{j,L}(b,\omega,\xi)$, $j=1,2$ are bounded operators and
\begin{equation}
\label{nTiLB}
 \Vert T_{j,L}(b,\omega,\xi) \Vert \leq \vert p(\xi) \vert  (1 + L^{\alpha})\quad  j=1,2,
\end{equation}
for  some polynomial $p(\cdot\,)$. For any $k \in \{1,2\}$ and $m \in \{0,1\}$, define on $\Lambda_{L}^{2}$:
\begin{multline}
\label{kTkLm}
\mathcal{T}_{k,L}^{m}(\bold{x},\bold{y};b,\omega,\xi) :=
\sum_{j=1}^{k} (-1)^{j} \sum_{{\bf i} \in \{1,2\}^{j}} \chi_{j}^{k}({\bf i}) \int_{\Lambda_{L}} \mathrm{d}\bold{z}_{1}\dotsb \int_{\Lambda_{L}} \mathrm{d}\bold{z}_{j}\,
\big(i \phi(\bold{z}_{j},\bold{y}) - i\phi(\bold{z}_{j},\bold{x}) \big)^{m} \times \\ \times R_{L}^{(1)}(\bold{x},\bold{z}_{1};b,\omega,\xi)T_{i_{1},L}(\bold{z}_{1},\bold{z}_{2};b,\omega,\xi) \dotsb T_{i_{j},L}(\bold{z}_{j},\bold{y};b,\omega,\xi).
\end{multline}
Here we set $0^0=1$. Notice that for $\bold{x} = \bold{y}$, the terms  in the r.h.s. of \eqref{kTkLm}   containing    the magnetic phase  vanish.  Clearly $\mathbb{P}$-a.s., $ \forall  b \in \mathbb{R}$,   $\forall 0< L< \infty$, $\forall \eta >0$ and  $\forall \xi \in \mathbb{C}$, $d(\xi) \geq \eta$, ${\mathcal{T}}^{m}_{k,L}(\cdot\,, \cdot\, ;b,\omega,\xi)$ is jointly continuous on $\Lambda_{L}^2$. To see that, we  apply $j$-times Lemma \ref{continuity} considering \eqref{eskres}, \eqref{eskTiL} and \eqref{fiestie}. This  also gives:
\begin{equation}
\label{eskTkLm2}
\forall( \bold{x}, \bold{y})  \in  \Lambda_{L}^2,\quad \vert \mathcal{T}_{k,L}^{m}(\bold{x},\bold{y};b,\omega,\xi)\vert  \leq \vert p(\xi)\vert^{k} L^m(1+ L^{\alpha})^{k},\quad k \in \{1,2\},\,m \in \{0,1\},
\end{equation}
for some polynomial $p(\cdot\,) $. Note also that  when $\bold{x}=\bold{y}$, the r.h.s. of \eqref{eskTkLm2} behaves like $L^{\alpha k}$.\\

We now formulate the main result of this section; its proof is given  in the next subsections:

\begin{proposition}
\label{thm3}
$\mathbb{P}$-a.s. on $\Omega$, $\forall  b \in \mathbb{R}$,  $\forall 0< L< \infty$, $ \forall  \eta>0$ and $ \forall \xi \in \mathbb{C}$, $d(\xi) \geq \eta $, then $\forall (\bold{x},\bold{y}) \in \Lambda_{L}^2 \setminus D_{L}$ and for $n=1,2$:
\begin{equation}
\label{nkdnres}
\frac{1}{n!} \frac{\partial^{n} R_{L}^{(1)}}{\partial b^{n}}(\bold{x},\bold{y};b,\omega,\xi) = \frac{\big(i \phi(\bold{x},\bold{y})\big)^{n}}{n!} R_{L}^{(1)}(\bold{x},\bold{y};b,\omega,\xi) +
\sum_{k=1}^{n} \mathcal{T}_{k,L}^{n-k}(\bold{x},\bold{y};b,\omega,\xi).
\end{equation}
\end{proposition}


\subsection{Some preliminary results}

\indent Let $(b,b_{0})\in \mathbb{R}^2$ and set $\delta b := b - b_{0}$. Let $\eta>0$ and $\xi \in \mathbb{C}$, $d(\xi) \geq \eta$. Introduce on $L^{2}(\Lambda_{L})$ the operators $\tilde{R}_{L}(b,b_{0},\omega,\xi)$ and $\tilde{T}_{j,L}(b,b_{0},\omega,\xi)$, $j=1,2$ through their integral kernel which are respectively defined by:
\begin{align}
\label{regkres}
\forall\, (\bold{x},\bold{y}) \in \Lambda_{L}^2 \setminus D_{L},\quad \tilde{R}_{L}^{(1)}(\bold{x},\bold{y};b,b_{0},\omega,\xi) &:= \mathrm{e}^{i \delta b \phi(\bold{x},\bold{y})}
R_{L}^{(1)}(\bold{x},\bold{y};b_{0},\omega,\xi), \\
\label{regkTiL}
\tilde{T}_{j,L}(\bold{x},\bold{y};b,b_{0},\omega,\xi) &:= \mathrm{e}^{i \delta b \phi(\bold{x},\bold{y})} T_{j,L}(\bold{x},\bold{y};b_{0},\omega,\xi).
\end{align}																						
Set also:
\begin{equation}
\label{regopTL}
\tilde{T}_{L}(b,b_{0},\omega,\xi) := \delta b\, \tilde{T}_{1,L}(b,b_{0},\omega,\xi) + (\delta b)^{2} \tilde{T}_{2,L}(b,b_{0},\omega,\xi).
\end{equation}															
Except  a gauge phase factor, the integral kernel of $\tilde{R}_{L}(b,b_{0},\omega,\xi)$ and
$\tilde{T}_{j,L}(b,b_{0},\omega,\xi)$ is the same as the one of $R_{L}(b_{0},\omega,\xi)$ and $T_{j,L}(b_{0},\omega,\xi)$ respectively. Therefore, $\mathbb{P}$-a.s., $\forall (b_0,b) \in \mathbb{R}^2$,  $\forall 0< L< \infty$, $ \forall  \eta>0$ and $ \forall \xi \in \mathbb{C}$, $d(\xi) \geq \eta $,  they are  bounded operators with  a  norm satisfying \eqref{nTiLB}. Besides they are  eventually H-S operators on $L^{2}(\Lambda_{L})$, and:
\begin{equation}
\label{regI2}
\Vert \tilde{R}_{L}(b,b_{0},\omega,\xi)\Vert_{\mathfrak{I}_{2}}\leq  \vert p( \xi ) \vert  L^{\frac{3}{2}}, \quad \Vert \tilde{T}_{j,L}(b,b_{0},\omega,\xi) \Vert_{\mathfrak{I}_{2}} \leq   \vert p( \xi ) \vert  (1 + L^{\alpha}) L^{\frac{3}{2}}.
\end{equation}
Under the same conditions as above, introduce the following bounded operators on $L^{2}(\Lambda_{L})$:
\begin{gather}
\label{regopTkL}
\tilde{\mathcal{T}}_{1,L}(b,b_{0},\omega,\xi) := -\tilde{R}_{L}(b, b_{0},\omega,\xi) \tilde{T}_{1,L}(b,b_{0},\omega,\xi),\\
\label{regopTkL1}
\tilde{\mathcal{T}}_{2,L}(b,b_{0},\omega,\xi) :=  \tilde{R}_{L}(b, b_{0},\omega,\xi) \Big( \big(\tilde{T}_{1,L}(b,b_{0},\omega,\xi)\big)^{2}- \tilde{T}_{2,L}(b,b_{0},\omega,\xi)\Big),
\end{gather}
\begin{multline}\label{regopTkL2}
\tilde{\mathcal{T}}_{3,L}(b,b_{0},\omega,\xi) := (\delta b)^3 \sum_{k=0}^1(\delta b)^k \sum_{{\bold i} \in \{1,2\}^2 }  \chi _2^{3+k }({\bold i} ) \tilde{R}_{L}(b, b_{0},\omega,\xi)  \tilde{T}_{i_1,L}(b,b_{0},\omega,\xi) \tilde{T}_{i_2,L}(b,b_{0},\omega,\xi)   + \\
- R_{L}(b,\omega,\xi)\big(\tilde {T}_{L}(b,b_{0},\omega,\xi)\big)^{3}.
\end{multline}
Then we prove (see also \cite[Lem. 3.2]{BrCoLo1} and \cite[proof of Prop. 3.2]{CN3}):
\begin{lema}
\label{lema3.1}  $\mathbb{P}$-a.s. on $\Omega$, $ \forall (b,b_0) \in \mathbb{R}^{2}$, $\forall 0<L<\infty$, $ \forall  \eta>0$ and $\ \forall \xi \in \mathbb{C}$, $d(\xi) \geq \eta$, then one has in the H-S operators sense:
\begin{equation}
\label{regres}
R_{L}(b,\omega,\xi) = \tilde{R}_{L}(b,b_{0},\omega,\xi) + \sum_{k=1}^{2} (\delta b)^{k}
\tilde{\mathcal{T}}_{k,L}(b,b_{0},\omega,\xi) +  \tilde{\mathcal{T}}_{3,L}(b,b_{0},\omega,\xi).
\end{equation}
\end{lema}

\noindent\textit{Proof.} From \cite[Sect. 2]{C1}, $\mathbb{P}$-a.s, $\forall b \in \mathbb R$, $\tilde{\mathrm{D}}:=\{ \varphi \in \mathcal{C}^{1}(\overline{\Lambda_{L}})\cap \mathcal{C}^{2}(\Lambda_{L}), \varphi\vert_{\partial \Lambda_{L}}=0 \}$ is a core for $H_{L}(b,\omega )$. Since $\bold{a}(\cdot\,)$ is the symmetric gauge, one has in the form sense on $\tilde{\mathrm{D}}\times  \tilde{\mathrm{D}}$:
\begin{equation}
\label{comute}
(- i \nabla_{\bold{x}} - b \bold{a}(\bold{x})) \mathrm{e}^{i \delta b \phi(\bold{x},\bold{y})} = \mathrm{e}^{i \delta b \phi(\bold{x},\bold{y})}(- i \nabla_{\bold{x}} - b_{0} \bold{a}(\bold{x}) - \delta b \bold{a}(\bold{x} - \bold{y})).
\end{equation}
In view of \eqref{regkres}-\eqref{regopTL} and   \eqref{comute}, we get for any $(\varphi,\psi) \in \tilde{\mathrm{D}} \times \mathcal{C}_{0}^{\infty}(\Lambda_{L})$:
\begin{equation*}
l_{L}(\varphi,\psi) := \langle (H_{L}(b,\omega) - \overline{\xi}) \varphi, \tilde{R}_{L}(b,b_{0},\omega,\xi) \psi \rangle = \langle \varphi, \psi \rangle + \langle \varphi, \tilde{T}_{L}(b,b_{0},\omega,\xi) \psi \rangle.
\end{equation*}
By standard  arguments, $l_{L}$ can be extended in a bounded form on $D(H_L (b, \omega)) \times L^{2}(\Lambda_{L})$. Let
$ \varphi= R_{L}(b,\omega, \bar\xi)\hat{\varphi}$, with $\hat{\varphi} \in L^{2}(\Lambda_{L}) $. Then  the following identity  holds in $\mathfrak B(L^{2}(\Lambda_{L}))$, and eventually in the H-S operators sense (see \eqref{regI2}):
\begin{equation}
\label{1eqresmod}
R_{L}(b,\omega,\xi) = \tilde{R}_{L}(b,b_{0},\omega,\xi) - R_{L}(b,\omega,\xi)\tilde{T}_{L}(b,b_{0},\omega,\xi).
\end{equation}
Now  we iterate twice     \eqref{1eqresmod},   in view of   \eqref{regopTL} and  \eqref{regopTkL}-\eqref{regopTkL2}  the lemma  follows.\qed


\subsection{Proof of Proposition \ref{thm3}}

\indent  Following  Lemma \ref{lema3.1}, and by rewriting \eqref{regres} in terms of corresponding integral kernels, we get on $\Lambda_{L}^{2} \setminus D_{L}$:
\begin{equation}
\label{kregres}
R_{L}^{(1)}(\bold{x},\bold{y};b,\omega,\xi) = \tilde{R}_{L}^{(1)}(\bold{x},\bold{y};b,b_{0},\omega,\xi)
+ \sum_{k=1}^{2} (\delta b)^{k} \tilde{\mathcal{T}}_{k,L}(\bold{x},\bold{y};b,b_{0},\omega,\xi) +
\tilde{\mathcal{T}}_{3,L}(\bold{x},\bold{y};b,b_{0},\omega,\xi),
\end{equation}
where, for all integer $k \in \{1,2\}$ and for any $(\bold{x},\bold{y}) \in \Lambda_{L}^2$:
\begin{multline}
\label{kregTkL}
\tilde{\mathcal{T}}_{k,L}(\bold{x},\bold{y};b,b_{0},\omega,\xi) := \sum_{j=1}^{k} (-1)^{j} \sum_{\bold{i} \in \{1,2\}^{j}}
\chi_{j}^{k}(\bold{i}) \int_{\Lambda_{L}^{j}}\mathrm{d}\bold{z}_{1}\dotsb \mathrm{d}\bold{z}_{j}\,
\mathrm{e}^{i \delta b( \phi(\bold{x},\bold{z}_{1}) + \dotsb + \phi(\bold{z}_{j},\bold{y}))} \times\\ \times R_{L}^{(1)}(\bold{x},\bold{z}_{1};b_{0},\omega,\xi) T_{i_{1},L}(\bold{z}_{1},\bold{z}_{2};b_{0},\omega,\xi)\dotsb T_{i_{j},L}(\bold{z}_{j},\bold{y};b_{0},\omega,\xi),
\end{multline}
and $\tilde{\mathcal{T}}_{3,L}(\cdot\,,\cdot\,;b,b_{0},\omega,\xi)$ stands for the kernel of  $\tilde{\mathcal{T}}_{3,L}(b,b_{0},\omega,\xi)$, see  \eqref{regopTkL2}.
We now remove the $b$-dependence of the coefficient of $(\delta b)^{k}$ in the  sum  \eqref{kregres} by expanding in Taylor series the exponential phase factor in \eqref{regkres} and in \eqref{kregTkL} up to the second order. Thus on $\Lambda_{L}^{2}\setminus D_{L}$:
\begin{multline*}
\tilde{R}_{L}^{(1)}(\bold{x},\bold{y};b,b_{0},\omega,\xi)
+ \sum_{k=1}^{2} (\delta b)^{k} \tilde{\mathcal{T}}_{k,L}(\bold{x},\bold{y};b,b_{0},\omega,\xi) =
\sum_{k=0}^{2} (\delta b)^{k} \frac{\big(i \phi(\bold{x},\bold{y})\big)^{k}}{k!} R_{L}^{(1)}(\bold{x},\bold{y};b_{0},\omega,\xi) + \\
+ \sum_{k=1}^{2} (\delta b)^{k} \sum_{m=1}^{k} \mathcal{T}_{m,L}^{k-m}(\bold{x},\bold{y};b_{0},\omega,\xi) + \mathcal{T}_{3,L}(\bold{x},\bold{y};b,b_{0},\omega,\xi),
\end{multline*}
where by construction the remainder term $\mathcal{T}_{3,L}(\bold{x},\bold{y};\cdot\,,b_{0},\omega,\xi)$ satisfies the property that its first   and second  derivatives at $b_{0}$ are all zero. It remains to use the definitions \eqref{regkres}-\eqref{regopTL} combined with  \eqref{eskres}, \eqref{eskTiL} and Lemma \ref{proestim},  this shows  that $\mathbb{P}$-a.s., $ \forall (b,b_0) \in \mathbb{R}^{2}$, $\forall 0<L<\infty$, $ \forall  \eta>0$ and $\ \forall \xi \in \mathbb{C}$, $d(\xi) \geq \eta$, $\vert \tilde{\mathcal{T}}_{3,L}(\bold{x},\bold{y};b,b_{0},\omega,\xi)\vert = \mathcal{O}(\vert \delta b\vert^{3})$ when $\vert \delta b\vert \rightarrow 0$ uniformly in $\bold{x},\bold{y} \in \Lambda_{L}$. Then the proposition follows from  Proposition \ref{thm2}. \qed
\section{The finite-volume diamagnetic response}

Here, by using the results of Section 3, we  want to get a new expression  for   the magnetization and  the susceptibility.  Recall that by applying \cite[Thm 1.1 $i)$]{BCS1}, $\mathbb{P}$-a.s.,  $\forall \beta >0$ and $\forall b \in \mathbb{R}$, the  pressure defined in \eqref{PL} has an analytic extension in $ z \in \mathcal{D}_{\epsilon}$   (see \eqref{domz}).  This analytic continuation  is defined as the following.\\
\indent Let $\beta >0$ and $K \subset \mathcal{D}_{\epsilon}$ be a compact subset. Let $\Gamma_{K}$ be the positively oriented contour around the interval $ [E_0, \infty)$ defined by:
 \begin{multline}
\label{GammaK}
\Gamma_{K} := \{ \Re \xi = E_K ,\,\, \Im \xi \in [\frac{- \vartheta_{K}}{2 \beta}, \frac{\vartheta_{K}}{2 \beta}]\} \cup \{\Re \xi \in [E_K,\xi_{K}),\,\, \Im \xi = \pm \frac{\vartheta_{K}}{2\beta}\} \cup \\
\cup \{ \Re \xi \geq \xi_{K},\,\, \mathrm{arg}(\xi - \xi_{K} \mp i \frac{\vartheta_{K}}{2\beta}) = \pm \varsigma\}.
\end{multline}
The constants $\vartheta_{K}>0$, $E_K < E_0$, $\varsigma \in (0, \pi/2)$
and $\xi_{K}> E_0$  are chosen so that for all $z\in K$, the  closed subset surrounding by $\Gamma_K$ is a strict subset of  the holomorphic domain of the map
$\xi \in \mathbb C\mapsto \mathfrak{f}_{\epsilon}(\beta,z;\xi) := \ln(1 + \epsilon z \mathrm{e}^{-\beta \xi})$, see \cite[Lem. 3.4]{BCS1}.
 Besides $\mathfrak{f}_{\epsilon}(\beta,z;\cdot\,)$ admits an exponentially decreasing estimate on $\Gamma_{K}$, i.e. there exists  a constant $c=c(\beta,K)>0$ s.t.:
\begin{equation}
\label{expdecr}
\forall  z \in K,\,\,\forall \xi\in\Gamma_{K},\quad \vert \mathfrak{f}_{\epsilon}(\beta,z;\xi)\vert \leq c \mathrm{e}^{-\beta \Re\xi}.
\end{equation}
Let $\beta>0$, $b \in \mathbb{R}$, $ 0<L<\infty$, $ z \in \mathcal{D}_{\epsilon}$,  $K \subset \mathcal{D}_{\epsilon}$ be a compact neighborhood of $z $ and  $\Gamma_K$ given  in \eqref{GammaK}.  Introduce on $L^{2}(\Lambda_{L})$:
\begin{equation}
\label{opLNDS}
\mathcal{L}_{L}^{(\omega)}(\beta,b,z,\epsilon) := \frac{i}{2\pi} \int_{\Gamma_{K}} \mathrm{d}\xi\, \mathfrak{f}_{\epsilon}(\beta,z;\xi) R_{L}(b,\omega,\xi).
\end{equation}
Then $\mathbb{P}$-a.s.,  $ \forall  \beta>0$, $\forall  b \in \mathbb R$, $\forall 0<L< \infty$ and  $ \forall z \in  \mathcal{D}_{\epsilon} \cap \mathbb R$, \eqref{opLNDS}   defines a trace class operator on $L^{2}(\Lambda_{L})$ (see \cite[Sect. 3.2]{BCS1}), and  via the standard  functional calculus  we have,
$ \mathcal{L}_{L}^{(\omega)}(\beta,b,z,\epsilon)=\ln(\mathbb{I} + \epsilon z \mathrm{e}^{-\beta H_{L}(b,\omega  )}).$
Hence,   this allows  us  to define    the  finite-volume pressure  as:
\begin{equation}
\label{panaPL}
P_{L}^{(\omega)}(\beta,b,z,\epsilon) =\frac{\epsilon}{\beta \vert \Lambda_{L}\vert}  \mathrm{Tr}_{L^{2}(\Lambda_{L})}\Big(\mathcal{L}_{L}^{(\omega)}(\beta,b,z,\epsilon)\Big).
\end{equation}
It is shown in  \cite[Sect. 3.3]{BCS1} that \eqref{panaPL} can be analytically extended to any  $z \in   \mathcal{D}_{\epsilon}$, and on the other hand,  the definition  \eqref{panaPL} is independent of the choice of the compact subset $K$.

 \begin{proposition}
\label{propochiLn}
$\mathbb{P}$-a.s. on $\Omega$, $\forall 0< L < \infty$, $\forall \beta > 0$, $\forall b \in \mathbb{R}$, $\forall z \in \mathcal{D}_{\epsilon}$ and for any compact subset $K$ of $\mathcal{D}_{\epsilon}$ s.t. $z \in K$, then one has for $n=1,2$:
\begin{multline}
\label{opochiLn}
\mathcal{X}_{L,n}^{(\omega)}(\beta,b,z,\epsilon) = \bigg(\frac{q}{c}\bigg)^n \frac{\epsilon}{\beta \vert \Lambda_{L} \vert} \mathrm{Tr}_{L^{2}(\Lambda_{L})}\Big( \frac{\partial^{n} \mathcal{L}_{L}^{(\omega)}}{\partial b^{n}}(\beta,b,z,\epsilon)\Big)
 =  \\  \bigg(\frac{q}{c}\bigg)^n \frac{\epsilon}{\beta \vert \Lambda_{L} \vert} \frac{i}{2\pi} \mathrm{Tr}_{L^{2}(\Lambda_{L})} \bigg(\int_{\Gamma_{K}} \mathrm{d}\xi\, \mathfrak{f}_{\epsilon}(\beta,z;\xi) \frac{\partial^{n} R_{L}}{\partial b^{n}}(b,\omega,\xi)\bigg).
\end{multline}
\end{proposition}

\noindent \textit{Proof.} Let $\eta := \min\{E_{0}-E_{K}, \frac{\vartheta_{K}}{2\beta}\}>0$. From \eqref{rearange} and  \eqref{opdnres}, $\mathbb{P}$-a.s., $\forall(b,b_0) \in \mathbb{R}^2$, $\forall 0< L< \infty $ and $\forall \xi \in \mathbb{C}$, $d(\xi)\geq \eta$:
\begin{equation}
\label{devres2}
R_{L}(b,\omega,\xi)= R_{L}(b_0,\omega,\xi) + \sum_{k=1}^{2} \frac{(\delta b)^{k}}{k!} \frac{\partial^{k} R_{L}}{\partial b^{k}}(b_{0},\omega,\xi)  +  \mathcal{S}_{3,L}(b,b_{0},\omega,\xi).
\end{equation}
\eqref{devres2} holds in the bounded operators sense. Under the conditions of Proposition \ref{propochiLn} and  in view of \eqref{GammaK}, we get from \eqref{opLNDS} followed by \eqref{devres2}:
\begin{multline}
\label{exlnserie1}
\mathcal{L}_{L}^{(\omega)}(\beta,b,z,\epsilon) =  \sum_{k=0}^{2} \frac{(\delta b)^{k}}{k!} \bigg\{\frac{i}{2\pi} \int_{\Gamma_{K}} \mathrm{d}\xi\, \mathfrak{f}_{\epsilon}(\beta,z;\xi) \frac{\partial^{k} R_{L}}{\partial b^{k}}(b_{0},\omega,\xi) \bigg\} + \\ \frac{i}{2\pi} \int_{\Gamma_{K}} \mathrm{d}\xi\, \mathfrak{f}_{\epsilon}(\beta,z;\xi) \mathcal{S}_{3,L}(b,b_{0},\omega,\xi).
\end{multline}
Now from  \eqref{normSiL1}, \eqref{opJkLi}, \eqref{opdnres}  and the estimate $\Vert R_L (b_0,\omega,\xi) \Vert \leq \eta^{-1}$,  $ \xi \in \Gamma_{K}$, then $\mathbb{P}$-a.s., $\forall 0<L < \infty$ there exists a polynomial $ p(\cdot\,)$ s.t. $\forall b_{0} \in \mathbb{R}$, $\forall \xi \in \Gamma_K$ and $k=0,1,2$,  $\Vert \frac{\partial^{k} R_{L}}{\partial b^{k}}(b_{0},\omega,\xi)\Vert \leq  \vert p(\xi )\vert$. Hence from  \eqref{expdecr}, all the operators in the sum of the r.h.s. of \eqref{exlnserie1} are  bounded operators. Moreover in view of \eqref{Sn+1} with $n=2$, and since  $b \in \mathbb R$, the same arguments as above applied to the last term of the r.h.s. of \eqref{exlnserie1}, show that this term behaves like $\mathcal{O}(\vert \delta b \vert^3 )$ in the $\mathfrak{B}(L^{2}(\Lambda_{L}))$-sense. The proposition follows  since $\mathbb{P}$-a.s., $\forall \beta > 0$, $\forall z \in K$ and $\forall 0<L<\infty $, $ \mathcal{L}_{L}^{(\omega)}(\beta,\cdot,z,\epsilon)$ is an $ \mathfrak{I}_{1}(L^{2}(\Lambda_{L}))$-real analytic operator-valued function, see \cite[proof of Prop. 3.5]{BCS1}.\qed\\

We now give the main result of this section:
\begin{theorem}
\label{thm5}
$\mathbb{P}$-a.s. on $\Omega$, $\forall 0<L< \infty$, $\forall \beta > 0$, $\forall b \in \mathbb{R}$, $\forall z \in \mathcal{D}_{\epsilon}$ and for any compact subset $K$ of $\mathcal{D}_{\epsilon}$ s.t. $z \in K$, we have for $n=1,2$:
\begin{gather}
\label{PLk}
P_{L}^{(\omega)}(\beta,b,z,\epsilon) = \frac{\epsilon}{\beta \vert \Lambda_{L} \vert} \frac{i}{2\pi}\int_{\Lambda_{L}} \mathrm{d}\bold{x}\, \bigg(\int_{\Gamma_{K}} \mathrm{d}\xi\, \mathfrak{f}_{\epsilon}(\beta,z;\xi) R_{L}^{(1)}(\bold{x},\bold{y};b,\omega,\xi) \bigg)\bigg\vert_{\bold{y} = \bold{x}}, \\
\label{chiLk}
\mathcal{X}_{L,n}^{(\omega)}(\beta,b,z,\epsilon) =\bigg(\frac{q}{c}\bigg)^n \frac{n! \epsilon}{\beta \vert \Lambda_{L} \vert} \frac{i}{2\pi} \int_{\Gamma_{K}} \mathrm{d}\xi\, \mathfrak{f}_{\epsilon}(\beta,z;\xi) \int_{\Lambda_{L}} \mathrm{d}\bold{x}\,
\sum_{k=1}^{n} \mathcal{T}_{k,L}^{n-k}(\bold{x},\bold{x};b,\omega,\xi),
\end{gather}
where $\mathcal{T}_{k,L}^{m}(\cdot\,,\cdot\,;b,\omega,\xi)$ with $k\in\{1,2\}$, $m \in \{0,1\}$ are  given in \eqref{kTkLm}.
\end{theorem}


The above result together with the joint continuity of $\mathcal{T}_{k,L}^{m}(\cdot\,,\cdot\,;b,\omega,\xi)$ on $\Lambda_L^2 $ lead to:

\begin{corollary}
\label{corro5} Under the same conditions as in Theorem \ref{thm5}, we have:
\begin{gather}
\label{magnz}
\mathcal{X}_{L,1}^{(\omega)}(\beta,b,z,\epsilon) =-\bigg(\frac{q}{c}\bigg) \frac{\epsilon}{\beta \vert \Lambda_{L} \vert} \frac{i}{2\pi} \int_{\Gamma_{K}} \mathrm{d}\xi\, \mathfrak{f}_{\epsilon}(\beta,z;\xi) \mathrm{Tr}_{L^{2}(\Lambda_{L})} \Big (R_{L}(b,\omega,\xi) T_{1,L}(b,\omega,\xi)\Big),
\end{gather}
\begin{multline}
\label{mos}
\mathcal{X}_{L,2}^{(\omega)}(\beta,b,z,\epsilon) =
\bigg(\frac{q}{c}\bigg)^2 \frac{\epsilon}{\beta \vert \Lambda_{L} \vert} \frac{i}{\pi} \int_{\Gamma_{K}} \mathrm{d}\xi\, \mathfrak{f}_{\epsilon}(\beta,z;\xi) \times \\
 \times \mathrm{Tr}_{L^{2}(\Lambda_{L})} \bigg (R_{L}(b,\omega,\xi) \Big(\big(T_{1,L}(b,\omega,\xi)\big)^{2} - T_{2,L}(b,\omega,\xi)\Big)\bigg),
\end{multline}
where ${T}_{j,L}(b,\omega,\xi)$, $j=1,2$ are defined via their kernel in \eqref{kT1L} and \eqref{kT2L} respectively.
\end{corollary}

\noindent \textit{Proof of Theorem \ref{thm5}.}  Under the conditions of Proposition \ref{propochiLn} and for a fixed $\xi_0 < \min\{0,E_0\}$ and large enough, the first resolvent equation followed by the Cauchy integral formula lead to:
\begin{equation*}
\mathcal{L}_{L}^{(\omega)}(\beta,b,z,\epsilon) =   \frac{i}{2\pi} \bigg(\int_{\Gamma_{K}} \mathrm{d}\xi\, (\xi-\xi_0) \mathfrak{f}_{\epsilon}(\beta,z;\xi) R_{L}^{}(b,\omega,\xi) \bigg)R_{L}^{}(b,\omega,\xi_0).
\end{equation*}
From Remark \ref{2.2} and in view of \eqref{GammaK}, then $\mathbb{P}$-a.s., $\forall b \in \mathbb R$, $\forall 0< L< \infty$ and $\forall \xi \in \Gamma_{K}$, the operator $R_{L}(b,\omega,\xi) R_{L}(b,\omega,\xi_0)$ has a jointly continuous kernel. Moreover $\mathbb{P}$-a.s., there exists a polynomial $p(\cdot\,)$ s.t. $\forall b \in \mathbb{R}$, $\forall \xi \in \Gamma_{K}$ and $\forall(\bold{x},\bold{y})\in \Lambda_{L}^{2}$, $\vert (R_{L}(b,\omega,\xi) R_{L}(b,\omega,\xi_0))(\bold{x},\bold{y})\vert \leq \vert p(\xi )\vert$. By using \eqref{expdecr}, the joint continuity of the integral kernel of $\mathcal{L}_{L}^{(\omega)}(\beta,b,z,\epsilon)$ follows by standard arguments. This proves \eqref{PLk}.
Let $n=1,2$.  Clearly as  in the proof of Proposition \ref{propochiLn}, we have for any $\varphi \in \mathcal{C}_{0}^{\infty}(\Lambda_{L})$ and $\bold{x} \in \Lambda_{L}$:
\begin{equation}
\label{identkILn}
\Big(\frac{\partial^{n} \mathcal{L}_{L}^{(\omega)}}{\partial b^{n}}(\beta,b,z,\epsilon)\varphi\Big)(\bold{x}) =
\frac{i}{2 \pi} \int_{\Gamma_{K}} \mathrm{d}\xi\, \mathfrak{f}_{\epsilon}(\beta,z;\xi) \int_{\Lambda_{L}} \mathrm{d}\bold{y}\, \frac{\partial^{n}R_{L}^{(1)}}{\partial b^{n}}(\bold{x},\bold{y};b,\omega,\xi) \varphi(\bold{y}).
\end{equation}
The estimates \eqref{eskdnres}, \eqref{CF},  \eqref{expdecr}  and standard arguments  then  imply:
\begin{equation}
\label{iif}
 \forall  (\bold{x},\bold{y})\in\Lambda_{L}^{2}, \quad   \frac{\partial^{n} \mathcal{L}_{L}^{(\omega)}}{\partial b^{n}}(\bold{x},\bold{y};\beta,b,z,\epsilon) =  \frac{i}{2\pi} \int_{\Gamma_{K}} \mathrm{d}\xi\, \mathfrak{f}_{\epsilon}(\beta,z;\xi) \frac{\partial^{n} R_{L}^{(1)}}{\partial b^{n}}(\bold{x},\bold{y};b, \omega,\xi).
\end{equation}
Now use \eqref{nkdnres}.  Then   from  \eqref{iif}, we have on $\Lambda_{L}^2$:
\begin{equation} \label{besoin0}
\begin{split}
\frac{\partial^{n} \mathcal{L}_{L}^{(\omega)}}{\partial b^{n}}(\bold{x},\bold{y};\beta,b,z,\epsilon) &=
n! \sum_{k=1}^{n} \frac{i}{2\pi} \int_{\Gamma_{K}} \mathrm{d}\xi\, \mathfrak{f}_{\epsilon}(\beta,z;\xi) \mathcal{T}_{k,L}^{n-k}(\bold{x},\bold{y};b,\omega,\xi) + \\
&+ (i\phi(\bold{x},\bold{y}))^{n} \mathcal{L}_{L}^{(\omega)}(\bold{x},\bold{y};\beta,b,z,\epsilon).
\end{split}
\end{equation}
We have already proved that the second term in the r.h.s. of \eqref{besoin0} is jointly continuous on $\Lambda_{L}^2$. Besides we know from Section 3 that $\mathbb{P}$-a.s., $\forall b \in \mathbb R$, $ \forall 0< L< \infty $, $ \forall \xi \in \Gamma_K$,
$\mathcal{T}_{k,L}^{m}(\cdot\,,\cdot\,;b,\omega,\xi)$, $k=1,2$, $m=0,1$, are jointly continuous on $\Lambda_{L}^2$. Moreover we have the bound \eqref{eskTkLm2}. Again by standard arguments, we conclude that each term in the sum of the r.h.s. of  \eqref{besoin0} is jointly continuous on $\Lambda_{L}^2$. So we perform the trace in \eqref{opochiLn} as the integral on $\Lambda_{L}$ of the diagonal part of the integral kernel  \eqref{besoin0}. This together with $\phi(\bold{x},\bold{x})=0$ imply \eqref{chiLk}.\qed


\remark \label{pour la fin1} Due to \eqref{opdnres}, \eqref{opochiLn} can be extended for any integer $n \geq 3$ to  define the  generalized susceptibilities at finite volume, see \cite{BCL1, BrCoLo1, BrCoLo3}. If $n=3$:
\begin{multline}
\label{suscep3}
\mathcal{X}_{L,3}^{(\omega)}(\beta,b,z,\epsilon) :=  \bigg(\frac{q}{c}\bigg)^3 \frac{\partial^{3} P_{L}^{(\omega)}}{\partial b^{3}}(\beta, b,z,\epsilon) =\\
\bigg(\frac{q}{c}\bigg)^3 \frac{\epsilon}{\beta \vert \Lambda_{L}\vert} \frac{3i}{\pi} \int_{\Gamma_{K}} \mathrm{d}\xi\, \mathfrak{f}_{\epsilon}(\beta,z;\xi)
\bigg\{ \int_{\Lambda_{L}} \mathrm{d}\bold{x}\, \mathcal{U}^{(\omega)}_{L,1}(\bold{x},\bold{x};b, \xi) +
\int_{\Lambda_{L}} \mathrm{d}\bold{x}\, \mathcal{U}^{(\omega)}_{L,2}(\bold{x},\bold{x};b, \xi) \bigg\},
\end{multline}
\begin{multline}
\label{Ujk}
\textrm{where:} \quad \mathcal{U}^{(\omega)}_{L,1}(\bold{x},\bold{y};b,\xi):=\\ \bigg(R_{L}(b,\omega,\xi)\Big(T_{1,L}(b,\omega,\xi)T_{2,L}(b,\omega,\xi) + T_{2,L}(b,\omega,\xi)T_{1,L}(b,\omega,\xi) - \big(T_{1,L}(b,\omega,\xi)\big)^{3}\Big)\bigg)(\bold{x},\bold{y}),\\
\mathcal{U}^{(\omega)}_{L,2}(\bold{x},\bold{y};b,\xi) := \int_{\Lambda_{L}} \mathrm{d}\bold{z}_{1}\int_{\Lambda_{L}} \mathrm{d}\bold{z}_{2}\int_{\Lambda_{L}} \mathrm{d}\bold{z}_{3}\, \big(i(\phi(\bold{x},\bold{z}_{1})+\phi(\bold{z}_{1},\bold{z}_{2}) + \phi(\bold{z}_{2},\bold{y}))\big) \times \\ R_{L}^{(1)}(\bold{x},\bold{z}_{1};b,\omega,\xi) T_{1,L}(\bold{z}_{1},\bold{z}_{2};b,\omega,\xi) T_{1,L}(\bold{z}_{2},\bold{y};b,\omega,\xi).
\end{multline}
Due to \eqref{eskres}, \eqref{eskTiL} and the estimate $\vert\phi(\bold{x},\bold{z}_{1})+\phi(\bold{z}_{1},\bold{z}_{2}) + \phi(\bold{z}_{2},\bold{x})\vert \leq \vert \bold{x}-\bold{z}_{1}\vert\vert\bold{z}_{1} - \bold{z}_{2}\vert$ (see \eqref{phase}), $\mathbb{P}$-a.s., $\forall b \in \mathbb R$, $ \forall 0< L< \infty $, $\forall \xi \in \Gamma_K$, the  integral kernels $\mathcal{U}^{(\omega)}_{L,j}(\cdot\,,\cdot\,;b,\xi)$, $j=1,2$ are well-defined on $\Lambda_{L}^{2}$ and eventually jointly continuous.


\section{The bulk operators}

\subsection{Preliminaries}

In the following we denote by  $P_{l}(b):= \bold{P}(b)\cdot \bold{e}_{l}$, $l=1,2$ where $\bold{P}(b) := (i\nabla_{\bold{x}} + b \bold{a}(\bold{x}))$, $\bold{e}_{1}:= (1,0,0) $ and $\bold{e}_{2}:= (0,1,0)$.

\begin{lema}
\label{Tiinftybounded} $\mathbb{P}$-a.s. on $\Omega$, $\forall b \in \mathbb{R}$, $\forall \eta >0$, $\forall\xi \in \mathbb{C}$, $d(\xi)\geq \eta$, then $P_k(b)   R_{\infty}(b,\omega,\xi)P_l(b)$, $k,l=1,2$ are bounded operators and there exists a polynomial $p(\cdot\,)$ independent of $(\omega, b)$ s.t.:
\begin{equation}
\label{pRp}
\Vert P_k(b)   R_{\infty}(b,\omega,\xi)P_l(b)\Vert \leq \vert p(\xi) \vert.
\end{equation}
 \end{lema}

\noindent \textit{Proof.}  By using \eqref{constv}, the analysis of  \cite[Sect. A2]{Si1}  and  the diamagnetic inequality, we know that 
for any $\varepsilon >0$,  then  if  $ \xi < E_0 $ and large enough, $\mathbb{P}$-a.s., $\forall b \in \mathbb{R}$:
\begin{equation*}
\label{pVp}
\Vert      \vert  V_1^{(\omega)} \vert ^{1/2 } R_{\infty}(b,\omega,\xi)\vert V_1^{(\omega)} \vert ^{1/2 } \Vert \leq \varepsilon.
\end{equation*}
This implies that $\forall\varepsilon >0$, there exists $a(\varepsilon)$ independent of $\omega$ and $b$ s.t. $\forall\varphi \in D(H_\infty(b, \omega))$:
\begin{equation}
\label{V}
(\varphi, \vert V_1^{(\omega)}\vert \varphi)  \leq \varepsilon \vert (\varphi, H_\infty(b, \omega)  \varphi) \vert  + a(\varepsilon)\Vert \varphi\Vert_{2}^{2}.
\end{equation}
Besides $\mathbb{P}$-a.s., $ \forall b \in \mathbb R$, $D(H_{\infty}(b,\omega)) \subset \{\phi \in L^{2}(\mathbb{R}^{3}): (i\nabla + b \bold{a})\phi \in (L^{2}(\mathbb{R}^{3}))^{3},\, (V_{2}^{(\omega)})^{\frac{1}{2}} \phi \in L^{2}(\mathbb{R}^{3})\}$, see \cite[Sect. B13]{Si1}. Under these  conditions we conclude the following.  From \eqref{V}, for any   $\varphi \in D(H_{\infty}(b,\omega))$:
\begin{equation}
\label{V1}   \frac{1}{2} \Vert P_k(b) \varphi \Vert_{2}^2  \leq (1+ \varepsilon)\Re(\varphi, (H_\infty(b, \omega)-\xi)  \varphi)  +  ((1+ \varepsilon)  \vert \Re\xi  \vert  +a(\varepsilon) ) \Vert \varphi \Vert_{2}^2.
\end{equation}
Choose $ \varphi =  R_{\infty}(b,\omega,\xi) \psi$ with $ \psi \in L^2(\mathbb{R}^{3})$, $\Vert  \psi \Vert_{2} =1$ and $\xi \in \mathbb{C}$, $d(\xi)\geq \eta$,  then \eqref{V1} shows that  there exists a constant $c_1>0$ independent of $\omega$ and $b$ s.t.:
\begin{equation}
\label{V2}
\Vert P_k(b) R_{\infty}(b,\omega,\xi) \psi \Vert_{2} \leq c_1( 1+  \vert  \xi \vert)^{1/2}.
 \end{equation}
Therefore  $P_k(b)  R_{\infty}(b,\omega,\xi) $ is a bounded operator and  $  \Vert P_k(b) R_{\infty}(b,\omega,\xi) \Vert \leq c_1( 1+  \vert  \xi \vert)^{1/2}$.
Now choose $ \varphi = R_{\infty}(b,\omega,\xi) P_k(b) \psi$ with $ \psi \in D(H_{\infty}(b,\omega))$, $\Vert  \psi \Vert_{2}=1$, and $\xi \in \mathbb{C}$, $d(\xi) \geq \eta$, then \eqref{V1} and \eqref{V2} imply that there exists a constant $c_2 >0$ independent of $(\omega, b)$ s.t.:
\begin{equation}
\label{V3}
 \Vert P_k(b) R_{\infty}(b,\omega,\xi) P_k(b) \psi \Vert_{2}^2 \leq  c_2(  \Vert P_k(b) R_{\infty}(b,\omega,\xi) P_k(b) \psi \Vert_{2} + (1 +
 \vert  \xi \vert)^2).
 \end{equation}
Hence  $P_k(b) R_{\infty}(b,\omega,\xi) P_k(b) $ is bounded and $ \Vert P_k(b) R_{\infty}(b,\omega,\xi) P_k(b)\Vert \leq   c'_2  ( 1+  \vert  \xi \vert)$ for another constant $c'_2 >0 $ independent  of $ (\omega,b)$.
Let $ \varphi = R_{\infty}(b,\omega,\xi) P_l(b) \psi$, with $ \psi \in D(H_{\infty}(b,\omega))$, $\Vert  \psi \Vert_{2}=1$, then \eqref{V1}-\eqref{V3} together imply:
 $$ \Vert P_k(b) R_{\infty}(b,\omega,\xi) P_l(b) \psi \Vert_{2}^2 \leq  c_2(  \Vert P_l(b) R_{\infty}(b,\omega,\xi) P_l(b) \psi \Vert_{2} + (1 +
 \vert  \xi \vert)^2). $$
 Again  $P_k(b) R_{\infty}(b,\omega,\xi) P_l(b), k \not= l $ is bounded and $ \Vert P_k(b) R_{\infty}(b,\omega,\xi) P_l(b)\Vert \leq   c_3 ( 1+  \vert  \xi \vert)$  for some   constant $c_3>0$  independent of $(\omega,b)$. \qed\\

Let $\{{ \bf T}_{\bold{k},b}\}_{\bold{k} \in \mathbb{R}^{3}}$ be the family of the usual real magnetic translations defined as the following. Let $\phi$ be the phase defined as in \eqref{phase}, and:
\begin{equation}
\label{trm}
\forall \bold{k} \in \mathbb{R}^{3},\quad (\bold{T}_{\bold{k},b}\psi)(\bold{x}):= e^{ib \phi(\bold{x},\bold{k}) } \psi( \bold{x}-\bold{k}) \quad \psi \in L^{2}(\mathbb{R}^{3}).
\end{equation}

\begin{lema}
\label{Tiinftybounded1}$\mathbb{P}$-a.s. on $\Omega$, $ \forall b \in \mathbb{R}$, $\forall\eta>0$ and  $ \forall \xi \in \mathbb{C}$, $d(\xi) \geq \eta$, then for $j=1,2 $:\\
$i)$ $T_{j,\infty}(b,\omega,\xi)$ is a bounded operator and  there exists a polynomial $p(\cdot)$ independent of $(\omega,b)$ s.t.:
\begin{equation}
\label{esopTiinfty}
\Vert T_{j,\infty}(b,\omega,\xi)\Vert \leq \vert p( \xi) \vert.
\end{equation}
$ii)$ $T_{j,\infty}(b,\omega,\xi)$ satisfies the covariance relation:
\begin{equation}
\label{covrelato}
\forall\bold{k} \in \mathbb{R}^{3},\quad { \bf T}_{\bold{k},b} T_{j,\infty}(b,\omega,\xi) { \bf T}_{-\bold{k},b} = T_{j,\infty}(b, \tau_{\bold{k}} \omega,\xi).
\end{equation}
Here  $T_{j,\infty}(b,\omega,\xi)$, $j=1,2$ are defined via their integral kernel  in  \eqref{T1inftyk}, \eqref{T2inftyk} respectively.
\end{lema}

\noindent\textit{Proof.} Let us  consider the operator $ T_{1,\infty}(b,\omega,\xi)$. In view of its integral  kernel \eqref{T1inftyk}, the definition of the symmetric gauge and under the conditions of Lemma \ref{Tiinftybounded1}, we have on $ \mathbb{R}^{6}\setminus D_{\infty}$:
\begin{equation*}
T_{1,\infty}(\bold{x},\bold{y};b,\omega,\xi) = \frac{1}{2} (i\nabla_{\bold{x}} + b\bold{a}(\bold{x}))\cdot (-(x_{2}-y_{2}) \bold{e}_{1} + (x_{1}-y_{1})\bold{e}_{2})R_{\infty}^{(1)}(\bold{x},\bold{y};b,\omega,\xi).
\end{equation*}
By using the same arguments as the ones in the proof of \cite[Prop. 3.2]{BCS2}, we get:
\begin{equation}\label {T1C}
T_{1,\infty}(b,\omega,\xi) = \frac{i}{2}
\bigg(P_{1}(b) R_{\infty}(b,\omega,\xi) P_{2}(b)-
P_{2}(b) R_{\infty}(b,\omega,\xi) P_{1}(b)\bigg)R_{\infty}(b,\omega,\xi),
\end{equation}
which is valid in the form sense on
$ \mathcal{C}_{0}^{\infty}(\mathbb{R}^{3}) \times  \mathcal{C}_{0}^{\infty}(\mathbb{R}^{3})$. By Lemma \ref{Tiinftybounded}, $\mathbb{P}$-a.s., $ \forall b \in \mathbb{R}$,  $\forall \eta >0$, $\forall \xi \in \mathbb{C}$, $d(\xi) \geq \eta$, the r.h.s of \eqref{T1C} defines a bounded operator on $L^2(\mathbb{R}^{3})$. So  this holds for the operator $T_{1,\infty}(b,\omega,\xi) $. The same arguments can be applied to the operator $T_{2,\infty}(b,\omega,\xi)$, but from the equality valid in the form sense on
$ L^2(\mathbb{R}^{3})$:
\begin{multline}\label {T2C}
T_{2,\infty}(b,\omega,\xi) = -\frac{1}{4}\bigg(R_{\infty}(b,\omega,\xi)P_{1}(b) R_{\infty}(b,\omega,\xi) P_{1}(b) + \\
+ R_{\infty}(b,\omega,\xi)P_{2}(b) R_{\infty}(b,\omega,\xi) P_{2}(b) - R_{\infty}(b,\omega,\xi)\bigg)R_{\infty}(b,\omega,\xi).
\end{multline}
This proves $i)$. Let us show $ii)$. The measurability of $\omega \mapsto H_{\infty}(b,\omega)$ combined with the assumption (E) lead to the covariance relation:
\begin{equation}
\label{covrela}
\forall \bold{k} \in \mathbb{R}^{3},\quad {\bf T}_{\bold{k},b} R_{\infty}(b,\omega,\xi) {\bf T}_{-\bold{k},b} = R_{\infty}(b, \tau_{\bold{k}} \omega,\xi).
\end{equation}
This  implies the following identity  on $\mathbb{R}^{6}\setminus D_{\infty}$:
\begin{equation*}
\forall \bold{k} \in \mathbb{R}^{3},\quad R_{\infty}^{(1)}(\bold{x}-\bold{k},\bold{y}-\bold{k};b,\omega,\xi) \mathrm{e}^{-i b \phi(\bold{y},\bold{k})} = \mathrm{e}^{-i b \phi(\bold{x},\bold{k})} R_{\infty}^{(1)}(\bold{x},\bold{y};b,\tau_{\bold{k}}\omega,\xi),
\end{equation*}
 and from  \eqref{comute},
$ \mathrm{e}^{-i b \phi(\bold{x},\bold{k})}  \bold{a}(\bold{x}-\bold{y})(i\nabla_{\bold{x}} + b\bold{a}(\bold{x})) =
    \bold{a}(\bold{x}-\bold{y})(i\nabla_{\bold{x}} + b\bold{a}(\bold{x}-\bold{k} )) \mathrm{e}^{-i b \phi(\bold{x},\bold{k})} $.
Then   \eqref{covrelato} follows from these two relations together with \eqref{T1inftyk}-\eqref{T2inftyk}. \qed\\

We now use these   results to investigate the $\mathbb{P}$-a.s. properties of the operators involved  in the definition \eqref{opIinfty1} and \eqref{opIinfty2}. Introduce the notation:
\begin{gather}
\label{calIj}
\mathcal{I}_{j}(b,\omega,\xi) := R_{\infty}(b,\omega,\xi) T_{j,\infty}(b,\omega,\xi) \quad  j\in \{1,2\},\\
\label{calI3}
\mathcal{I}_{3}(b,\omega,\xi) := R_{\infty}(b,\omega,\xi) T^2_{1,\infty}(b,\omega,\xi), \end{gather}
and  also set:
\begin{equation}
\label{calI0}
\mathcal{I}_0(b,\omega,\xi)= \mathcal{I}_0(b,\omega,\xi,\xi_{0}) := R_{\infty}(b,\omega,\xi)R^2_{\infty}(b,\omega,\xi_{0}),
\end{equation}
for a fixed $ \xi_0 < \min\{0, E_0\} $ (see \eqref{minmax}) and large enough.
Notice that from Lemma \ref{Tiinftybounded1},  $\mathbb{P}$-a.s., $\forall b \in \mathbb{R}$, $\forall \eta >0$ and $\forall \xi \in \mathbb{C}$, $d(\xi) \geq \eta$, these operators are bounded on  $L^{2}(\mathbb{R}^{3})$. Below  we denote by $\chi_{U}$ the characteristic function of a given $U \subset \mathbb R^3$.

\begin{proposition}
\label{ffithm} $\mathbb{P}$-a.s. on $ \Omega$, $\forall b \in \mathbb{R}$, $\forall \eta >0$, $\forall \xi \in \mathbb{C}$, $d(\xi) \geq \eta$, one has for $j=0,1,2,3$:\\
$i)$ For any open and bounded set $U \subset \mathbb{R}^{3}$,   $\chi_{U} \mathcal{I}_{j}(b,\omega,\xi)\mathcal{\chi}_{U}$ is trace class on $L^{2}(\mathbb{R}^{3})$ and
\begin{equation*}
 \Vert \chi_{U} \mathcal{I}_{j}(b,\omega,\xi)\mathcal{\chi}_{U} \Vert_{\mathfrak{I}_{1}} \leq \vert p(\xi)\vert,
\end{equation*}
for some polynomial $p(\cdot\,)$.\\
$ii)$ $\mathcal{I}_{j}(b,\omega,\xi)$ has an  integral kernel $\mathcal{I}_{j}(\cdot\,,\cdot\,;b,\omega,\xi)$ jointly continuous on $\mathbb{R}^{6}$, and moreover there exists a polynomial $p(\cdot\,)$ independent of $(\omega,b)$ s.t.:
\begin{equation}
\label{majpinom}
\forall (\bold{x},\bold{y}) \in \mathbb{R}^{6}, \quad  \vert \mathcal{I}_{j}(\bold{x},\bold{y};b,\omega,\xi)\vert \leq \vert p(\xi)\vert.
\end{equation}
$iii)$ $(\omega,\bold{x}) \mapsto \mathcal{I}_{j}(\bold{x},\bold{x};b,\omega,\xi)$ is a $\mathbb{R}^{3}$-ergodic random field with a finite expectation:
\begin{equation}
\label{expectfi}
\forall \bold{x} \in \mathbb{R}^{3},\quad  \mathbb{E}[\vert \mathcal{I}_{j}(\bold{x},\bold{x};b,\omega,\xi)\vert] < \infty.
\end{equation}
\end{proposition}

\proof From \eqref{eskres} and \eqref{eskdres}, $\mathbb{P}$-a.s., $ \forall b \in \mathbb R$, $ \forall \eta >0$  there exists $\gamma=\gamma(\eta)>0$ and a polynomial $p(\cdot\,)$ s.t. $\forall L \in (0, \infty]$ and $ \forall \xi \in \mathbb{C}$, $d(\xi) \geq \eta$, on $\Lambda_{L}^{2}\setminus D_{L}$:
\begin{gather}
\label{essecbul1}
\vert R_{L}^{(1)}(\bold{x},\bold{y};b,\omega,\xi)\vert, \vert T_{2,L}(\bold{x},\bold{y};b,\omega,\xi)\vert
\leq \vert p( \xi)\vert \frac{\mathrm{e}^{-\frac{\gamma}{1 + \vert \xi\vert}\vert \bold{x} - \bold{y}\vert}}{\vert \bold{x} - \bold{y}\vert},\\
\label{essecbul2}
\vert T_{1, L}(\bold{x},\bold{y};b,\omega,\xi) \vert \leq \vert p(\xi)\vert (1+ \vert \bold{x} \vert^{\alpha} +
\vert \bold{y} \vert^{\alpha}) \frac{\mathrm{e}^{-\frac{\gamma}{1 + \vert \xi\vert}\vert \bold{x} - \bold{y}\vert}}{\vert \bold{x} - \bold{y}\vert}.
\end{gather}
Then we have  the following estimates \cite{K}:
\begin{equation*}
\label{normI2res}
\Vert \chi_{U} R_{\infty}(b,\omega,\xi)\Vert_{\mathfrak{I}_{2}}, \Vert \chi_{U} T_{1,\infty}(b,\omega,\xi) \Vert_{\mathfrak{I}_{2}}, \;  \Vert \chi_{U} T_{2,\infty}(b,\omega,\xi)\Vert_{\mathfrak{I}_{2}} \leq \vert p( \xi)\vert,
\end{equation*}
for some polynomial $p(\cdot\,)$. Besides we have  \eqref{esopTiinfty}. By standard operator estimates, $i)$ follows.\\ Let us show $ii)$. From \eqref{essecbul1}, \eqref{essecbul2} and Lemma \ref{continuity} $ii)$,   the integral kernels
of $\mathcal{I}_{j}(b,\omega,\xi)$, $j=0,1,2$ are  jointly continuous on $\mathbb{R}^{6}$. This also holds for $\mathcal{I}_{3}(\cdot\,,\cdot\,;b,\omega,\xi)$ from the identity:
\begin{equation*}
\label{calIjmk2}
\mathcal{I}_{3}(\bold{x},\bold{y};b,\omega,\xi) = \int_{\mathbb{R}^{3}} \mathrm{d}\bold{z}\, \mathcal{I}_{1}(\bold{x},\bold{z};b,\omega,\xi) T_{1,\infty}(\bold{z},\bold{y};b,\omega,\xi),
\end{equation*}
followed by Lemma \ref{continuity} $ii)$ together with \eqref{essecbul1} and the rough estimate:
\begin{equation*}
\forall\, (\bold{x},\bold{y}) \in \mathbb{R}^{6},\quad
\vert \mathcal{I}_{1}(\bold{x},\bold{y};b,\omega,\xi) \vert \leq  \vert p(\xi)\vert  (1+\vert \bold{x}\vert^{\alpha} + \vert \bold{y}\vert^{\alpha}),
\end{equation*}
for some polynomial $p(\cdot\,)$.  We now prove  \eqref{majpinom}. Although the case of $j=0$ is straightforward, the method given below covers all cases $j=0,1,2,3$. Let $\xi_{0} < \min\{0,E_{0}\}$. We firstly show:
\begin{equation}
\label{identsand}
\mathcal{I}_{j}(b,\omega,\xi) = R_{\infty}(b,\omega,\xi_{0}) A_{j}(b,\omega,\xi) R_{\infty}(b,\omega,\xi_{0}),
\end{equation}
where $A_{j}(b,\omega,\xi)$ consists of a finite linear combination of operator products of type:
\begin{equation}
\label{prodop}
 (\xi-\xi_{0})^{q} R_{\infty}^{r}(b,\omega,\xi)\{P_{k}(b) R_{\infty}(b,\omega,\xi) P_{l}(b)\}^{s} R_{\infty}^{t}(b,\omega,\xi) \quad r+s+t \geq 1,
\end{equation}
where the exponents $q,r,t \in \{0,1,2\}$, $s\in \{0,1\}$, $k,l \in \{1,2\}$ depend on $j$. Notice that Lemma \ref{Tiinftybounded} implies that $\mathbb{P}$-a.s., $\forall \eta>0$ there exists a $\omega$-independent polynomial $p(\cdot\,)$ s.t. $\forall b \in \mathbb{R}$ and $ \forall \xi \in \mathbb{C}$, $d(\xi) \geq \eta$:
\begin{equation}
\label{esprodop}
\Vert A_{j}(b,\omega,\xi) \Vert \leq \vert p(\xi)\vert.
\end{equation}
Obviously \eqref{identsand} holds for $j=0$. If  $j=1,2$, we  use once  the  first resolvent  equation  for the resolvent  appearing in  \eqref{calIj}:
$$ \mathcal{I}_{j}(b,\omega,\xi)  = R_{\infty}(b,\omega,\xi_0) T_{j,\infty}(b,\omega,\xi) +  (\xi-\xi_{0}) R_{\infty}(b,\omega,\xi_0)  \mathcal{I}_{j}(b,\omega,\xi).$$
Then we use   \eqref{T1C}-\eqref{T2C},  and again the  first  resolvent equation  for the last resolvent in the expression  \eqref{T1C}-\eqref{T2C}. This leads to \eqref{identsand} by a straightforward calculation. For  $j=3$, we use  twice the identity  \eqref{T1C} in   \eqref{calI3} and we repeat the same procedure as above. This proves \eqref{identsand}. Furthermore  from  \eqref{constv},  \cite[Sect.A2]{Si1} together with \cite[Eq. (2.40)]{BHL1}, then  $\mathbb{P}$-a.s., $ \forall \xi_0 <\min \{0,E_{0}\}$  and  large enough, there exists a $\omega$-independent constant $c>0$ s.t. $\forall b \in \mathbb R$:
$$ \Vert  R_{\infty}(b, \omega, \xi_0) \Vert_{1,2} = \Vert  R_{\infty}(b, \omega, \xi_0) \Vert_{2,\infty} \leq c, $$
where $ \Vert  \cdot \Vert_{p,q} $ denotes the norm  for operator  from $ L^p(\mathbb R^3)$ to $ L^q(\mathbb R^3 )$, $ 1 \leq p,q\leq \infty$.  Further, let $B(b,\omega,\xi)$ be a bounded operator on $L^{2}(\mathbb{R}^{3})$. Then for any $\varphi, \psi  \in \mathcal{C}_0^\infty(\mathbb R^3 )$:
$$ \vert  \big( \varphi, R_{\infty}(b, \omega, \xi_0) B(b, \omega, \xi ) R_{\infty}(b, \omega, \xi_0)\psi)\vert  \leq c^2  \Vert \varphi  \Vert_{1}  \Vert \psi  \Vert_{1}  \Vert  B(b, \omega, \xi)\Vert. $$
Suppose that the operator $R_{\infty}(b, \omega, \xi_0) B(b, \omega, \xi) R_{\infty}(b, \omega, \xi_0)$ has a jointly continuous integral kernel. Then by using a limiting procedure we conclude that:
\begin{equation}
\label{limpro}
\forall(\bold{x},\bold{y}) \in  \mathbb{R}^{6},\quad \vert (R_{\infty}(b, \omega, \xi_0) B(b, \omega, \xi ) R_{\infty}(b, \omega, \xi_0) )(\bold{x},\bold{y}) \vert  \leq c^2 \Vert B(b, \omega, \xi)\Vert.
\end{equation}
Thus by setting $B(b,\omega,\xi) = A_{j}(b,\omega,\xi)$, we get  \eqref{majpinom} from \eqref{identsand} and   \eqref{esprodop}.\\
Let us prove  $iii)$. As a result of \eqref{covrela} and \eqref{covrelato},  the following covariance relation holds: $\forall\bold{k}\in\mathbb{R}^{3}$, $T_{\bold{k},b}\mathcal{I}_{j}(b,\omega,\xi) T_{-\bold{k},b}=\mathcal{I}_{j}(b,\tau_{\bold{k}}\omega,\xi)$. This implies $ \forall  (\bold{x},\bold{y}) \in  \mathbb{R}^{6}$:
\begin{equation*}
\forall \bold{k} \in \mathbb{R}^{3},\quad \mathcal{I}_{j}(\bold{x},\bold{y};b,\tau_{\bold{k}}\omega,\xi)= \mathrm{e}^{ib \phi(\bold{x},\bold{k})} \mathcal{I}_{j}(\bold{x}-\bold{k},\bold{y}-\bold{k};b,\omega,\xi) \mathrm{e}^{-ib \phi(\bold{y},\bold{k})}.
\end{equation*}
Then $(\omega,\bold{x}) \mapsto \mathcal{I}_{j}(\bold{x},\bold{x};b,\omega,\xi)$ is  well defined and $\mathbb{R}^{3}$-stationary. \eqref{expectfi} follows from the $\omega$-independent estimate \eqref{majpinom}.\qed \\

We deduce  properties of the operators in \eqref{opIinfty0}-\eqref{opIinfty2} from the following.
Let $ \xi_0$ as above. The first resolvent equation  and the Cauchy integral formula  imply that $\mathbb{P}$ a.s.,      $\forall  \beta >0$, $\forall b \in \mathbb{R}$,  $\forall z \in \mathcal{D}_{\epsilon}$ and $\forall K $ compact subset of $ \mathcal{D}_{\epsilon}$, s.t. $z \in K$:
\begin{equation} \label{L0T}
 \mathcal{L}_{\infty,0}^{(\omega)}(\beta,b,z,\epsilon)  = \\ \frac{i}{2\pi} \int_{\Gamma_{K}} \mathrm{d}\xi\, (\xi - \xi_{0})^2 \mathfrak{f}_{\epsilon}(\beta,z;\xi)\mathcal{I}_0(b,\omega,\xi),
\end{equation}
and, from \eqref{calIj}-\eqref{calI3}, 
\begin{gather}
\label{L1T}
\mathcal{L}_{\infty,1}^{(\omega)}(\beta,b,z,\epsilon) = -\frac{i}{2\pi} \int_{\Gamma_{K}} \mathrm{d}\xi\, \mathfrak{f}_{\epsilon}(\beta,z;\xi) \mathcal{I}_{1}(b,\omega,\xi),\\ \mathcal{L}_{\infty,2}^{(\omega)}(\beta,b,z,\epsilon) = \frac{i}{\pi} \int_{\Gamma_{K}} \mathrm{d}\xi\, \mathfrak{f}_{\epsilon}(\beta,z;\xi) \Big(\mathcal{I}_{3}(b,\omega,\xi) - \mathcal{I}_{2}(b,\omega,\xi)\Big). \nonumber
\end{gather}
Then Proposition \ref{ffithm} together with the estimate \eqref{expdecr}  imply:
\begin{corollary}
\label{fithm} $\mathbb{P}$-a.s. on $ \Omega$,   $ \forall \beta >0$, $ \forall b \in \mathbb R$, $ \forall  z \in \mathcal{D}_{\epsilon}$ and for    any compact subset   $K$ of $ \mathcal{D}_{\epsilon}$ s.t. $z \in K$,   one has for $n=0,1,2$:\\
$i)$ For all open and bounded set $U \subset \mathbb{R}^{3}$, $\mathcal{\chi}_{U} \mathcal{L}_{\infty,n}^{(\omega)}(\beta,b,z,\epsilon)\mathcal{\chi}_{U}$ is trace class on $L^{2}(\mathbb{R}^{3})$.\\
$ii)$ $\mathcal{L}_{\infty,n}^{(\omega)}(\beta,b,z,\epsilon)$ has an integral kernel $\mathcal{L}_{\infty,n}^{(\omega)}(\cdot\,,\cdot\,;\beta,b,z,\epsilon)$ jointly continuous on $\mathbb{R}^{6}$.\\
$iii)$ $(\omega,\bold{x}) \mapsto \mathcal{L}_{\infty,n}^{(\omega)}(\bold{x},\bold{x};\beta,b,z,\epsilon)$ is a $\mathbb{R}^{3}$-ergodic random field with a finite expectation.
\end{corollary}

The last two  results of this section  needed to prove our main theorems are the  following.

\begin{proposition} \label{cP1} $\mathbb{P}$-a.s. on $\Omega$, $\forall b \in \mathbb{R}$ and  $j=0,1,2,3$, then the following maps:\\
i)
\begin{equation*}
\xi \in \Gamma_{K} \mapsto  \frac{1}{\vert \Lambda_{L}\vert} \mathrm{Tr}_{L^{2}(\mathbb{R}^{3})}\Big(\chi_{\Lambda_{L}} \mathcal{I}_{j}(b,\omega,\xi) \chi_{\Lambda_{L}}\Big), 
\end{equation*}
is continuous  uniformly in $L \in (0,\infty)$.\\
ii) $\xi \in \Gamma_{K}  \mapsto \mathbb{E}[\mathcal{I}_{j}(\bold{0},\bold{0};b,\omega,\xi)]$ is continuous.
\end{proposition}

\noindent \textit{Proof.} Let $\xi_{1}  \in \Gamma_{K}$ be fixed. Due to Proposition \ref{ffithm}, to prove $i)$ it is sufficient to show  that $\mathbb{P}$-a.s., $\forall b\in \mathbb{R}$,  there exists a constant $c>0$ independent of $L$  s.t.  $\forall\bold{x} \in \mathbb{R}^{3}$, $\forall\xi \in \Gamma_{K}$ with $\vert\delta \xi \vert := \vert \xi - \xi_{1}\vert $ small enough:
\begin{equation}
\label{boinekxi}
\vert \mathcal{I}_j(\bold{x},\bold{x};b,\omega,\xi) - \mathcal{I}_j(\bold{x},\bold{x};b,\omega,\xi_{1})\vert \leq c \vert \delta\xi \vert.
\end{equation}
 Let $j=0$. From   \eqref{calI0} together with the first  resolvent equation, we get for any $\bold{x} \in \mathbb{R}^{3}$:
\begin{multline*}
\mathcal{I}_0(\bold{x},\bold{x};b,\omega,\xi) - \mathcal{I}_0(\bold{x},\bold{x};b,\omega,\xi_{1}) = \\ \big(\xi - \xi_{1}) \big(R_{\infty}(b,\omega,\xi_{0}) R_{\infty}(b,\omega,\xi) R_{\infty}(b,\omega,\xi_{1}) R_{\infty}(b,\omega,\xi_{0})\big)(\bold{x},\bold{x}).
\end{multline*}
Since $ \Vert R_{\infty}(b,\omega,\xi) R_{\infty}(b,\omega,\xi_{1}) \Vert \leq \eta^{-2}$, by applying \eqref{limpro} with $B(b,\omega,\xi) = R_{\infty}(b,\omega,\xi) R_{\infty}(b,\omega,\xi_{1})$, we get \eqref{boinekxi}.  Let $j=1,2,3$.  In view of  \eqref{identsand}, we have to estimate:
\begin{multline}
\label{diefIjmx}
\mathcal{I}_{j}(\bold{x},\bold{x};b,\omega,\xi) - \mathcal{I}_{j}(\bold{x},\bold{x};b,\omega,\xi_{1}) = \\ \Big(R_{\infty}(b,\omega,\xi_{0})\big(A_{j}(b,\omega,\xi) - A_{j}(b,\omega,\xi_{1})\big)R_{\infty}(b,\omega,\xi_{0})\Big)(\bold{x},\bold{x}),
\end{multline} 
where $A_{j}(b,\omega,\cdot)$ consists of a finite linear combination of operators of type \eqref{prodop}. Choose  a generic term appearing in $(A_{j}(b,\omega,\xi) - A_{j}(b,\omega,\xi_{1}))$:
\begin{multline*}
C(b,\omega,\xi,\xi_{1}) :=
(\xi-\xi_{0})^{q} R_{\infty}(b,\omega,\xi) P_{k}(b)R_{\infty}(b,\omega,\xi) P_{l}(b) + \\
 - (\xi_{1}-\xi_{0})^{q} R_{\infty}(b,\omega,\xi_{1}) P_{k}(b)R_{\infty}(b,\omega,\xi_{1}) P_{l}(b) \quad q\in \{0,1,2\}.
\end{multline*}
By using  twice the  resolvent  equation   in the first term of the r.h.s.  of this formula, we get:
\begin{multline*}
C(b,\omega,\xi,\xi_{1}) = ((\xi-\xi_{0})^{q} - (\xi_{1}-\xi_{0})^{q}) R_{\infty}(b,\omega,\xi_{1}) P_{k}(b)R_{\infty}(b,\omega,\xi_{1}) P_{l}(b) +
 (\xi - \xi_{1}) (\xi-\xi_{0})^{q} \times \\
R_{\infty}(b,\omega,\xi_{1})\Big(R_{\infty}(b,\omega,\xi)P_{k}(b) R_{\infty}(b,\omega,\xi) P_{l}(b) + P_{k}(b)  R_{\infty}(b,\omega,\xi)  R_{\infty}(b,\omega,\xi_{1}) P_{l}(b)\Big).
 \end{multline*}
By \eqref{pRp} and \eqref{V2}, $\mathbb{P}$-a.s., $\forall b\in \mathbb{R}$, $ \forall \xi \in \Gamma_{K}$ sufficiently near $\xi_1$, there exists $c>0$ independent of $\xi$  and $L$ s.t.
$\Vert C(b,\omega,\xi,\xi_{1})\Vert \leq  c \vert \delta \xi\vert$. This also  holds for $\Vert A_{j}(b,\omega,\xi) - A_{j}(b,\omega,\xi_{1})\Vert$ which implies \eqref{boinekxi} from  \eqref{limpro} and  \eqref{diefIjmx}. Now $ii)$  follows from the continuity of $\xi \mapsto \mathcal{I}_{j}(\bold{x},\bold{x};b,\omega,\xi)$, $ \forall \bold{x} \in \mathbb{R}^{3}$  together with the $\omega$-independent estimate \eqref{majpinom}. \qed

\begin{proposition} \label{cP2} $\mathbb{P}$-a.s. on $\Omega$,  $\forall \xi \in \Gamma_{K}$ and  $j=0,1$, then the following maps:\\
i)
\begin{equation*}
b \in \mathbb{R} \mapsto \frac{1}{\vert \Lambda_{L}\vert} \mathrm{Tr}_{L^{2}(\mathbb{R}^{3})}\Big(\chi_{\Lambda_{L}} \mathcal{I}_{j}(b,\omega,\xi) \chi_{\Lambda_{L}}\Big),
\end{equation*}
is  continuous uniformly in $L \in (0,\infty)$.\\
ii) $ b \in \mathbb{R}\mapsto \mathbb{E}[\mathcal{I}_{j}(\bold{0},\bold{0};b,\omega,\xi)]$ is continuous.
\end{proposition}

To prove this result we need the following. Introduce on  $L^2( {\mathbb R}^3)$ the operator   $\bold{W}(b,b_{0},\omega,\xi)$ through its integral kernel defined on $\mathbb{R}^{6}\setminus D_{\infty}$ as:
\begin{equation} \label{W}
\bold{W}(\bold{x},\bold{y};b,b_{0},\omega,\xi):=  \mathrm{e}^{i \delta b \phi(\bold{x},\bold{y})} \bold{a}(\bold{x}-\bold{y}) R^{(1)}_{\infty}(\bold{x},\bold{y};b_{0}, \omega, \xi ).
\end{equation}
From \eqref{eskres}, $\mathbb{P}$-a.s.,  $\forall (b,b_{0}) \in \mathbb{R}^{2}$, $\forall \xi \in \Gamma_{K}$, it is bounded and there exists  a polynomial $p(\cdot\,)$ independent of $(b_0,b)$  s.t.:
\begin{equation*}
\Vert \bold{W}(b,b_{0},\omega,\xi) \Vert \leq \vert p(\xi)\vert.
\end{equation*}

\begin{lema} \label{WP}
$\mathbb{P}$-a.s. on $\Omega$,  $\forall (b_0,b) \in \mathbb{R}^2$, $ \forall \xi \in \Gamma_{K}$, $\bold{P}(b)\cdot \bold{W}(b,b_{0},\omega,\xi)$ is bounded and there exists a  polynomial $p(\cdot\,)$ s.t. $\forall b \in \mathbb{R}^2$, $\vert b - b_{0}\vert$ small enough and $\forall \xi \in \Gamma_{K}$:
\begin{equation}
\label{fornd}
\Vert \bold{P}(b) \cdot \bold{W}(b,b_{0},\omega,\xi)\Vert \leq \vert p(\xi)\vert.
\end{equation}
\end{lema}

\noindent \textit{Proof.} Similarly  to  the proof of Lemma \ref{lema3.1},  then $\mathbb{P}$-a.s. , $\forall (b,b_{0}) \in \mathbb{R}^2$ and  $\forall \xi \in \Gamma_{K}$, we have in the bounded operators sense  (see also \cite[Prop. 3.2]{CN3}):
\begin{equation}
\label{approxi}
R_{\infty}(b,\omega,\xi) = \tilde{R}_{\infty}(b,b_{0},\omega,\xi) - R_{\infty}(b,\omega,\xi) \tilde{T}_{\infty}(b,b_{0},\omega,\xi),
\end{equation}
where $\tilde{R}_{\infty}(b,b_{0},\omega,\xi)$ is the operator generated by the kernel defined in \eqref{regkres} with $L=\infty$ and:
\begin{equation}
\label{tildeTinf}
\tilde{T}_{\infty}(b,b_{0},\omega,\xi) := \delta b \tilde{T}_{1,\infty}(b,b_{0},\omega,\xi) + (\delta b)^{2} \tilde{T}_{2,\infty}(b,b_{0},\omega,\xi),
\end{equation}
with $\tilde{T}_{j,\infty}(b,b_{0},\omega,\xi)$ the operator generated by the kernel defined in \eqref{regkTiL} with $L=\infty$. Notice that due to \eqref{eskres}, $\mathbb{P}$-a.s., there exists a polynomial $p(\cdot\,)$ s.t. $\forall (b,b_{0}) \in \mathbb{R}^2$, $\forall \xi \in \Gamma_{K}$:
\begin{equation}
\label{nedes}
\Vert \tilde{R}_{\infty}(b,b_{0},\xi)\Vert, \Vert \tilde{T}_{2,\infty}(b,b_{0},\xi)\Vert \leq \vert p(\xi)\vert.
\end{equation}
Now we remove  the $( \omega,\xi)$-dependence in the notations. From  \eqref{approxi}, we have on $\mathbb{R}^{6}\setminus D_{\infty}$:
\begin{equation*}
(\bold{P}(b)\cdot \bold{W}(b,b_{0}))(\bold{x},\bold{y})=
\bold{a}(\bold{x}-\bold{y})\cdot\bold{P}(b) R_{\infty}^{(1)}(\bold{x},\bold{y};b) + \bold{a}(\bold{x}-\bold{y})\cdot\bold{P}(b) (R_{\infty}(b) \tilde{T}_{\infty}(b,b_{0}))(\bold{x},\bold{y}).
\end{equation*}
The first term in the r.h.s of this expression is the  integral kernel  of  $T_{1,\infty}(b)$ which is bounded, and its norm satisfies \eqref{esopTiinfty}. We want to study the second term. On $\mathbb{R}^{6}\setminus D_{\infty}$, we have:
\begin{multline}
\label{xx}
\bold{a}(\bold{x}-\bold{y})\cdot\bold{P}(b) (R_{\infty}(b) \tilde{T}_{\infty}(b,b_{0}))(\bold{x},\bold{y}) = \int_{\mathbb{R}^{3}} \mathrm{d}\bold{z}\, \bold{a}(\bold{x}-\bold{z})\cdot \bold{P}(b) R_{\infty}^{(1)}(\bold{x},\bold{z};b) \tilde{T}_{\infty}(\bold{z},\bold{y};b,b_{0}) + \\
+ \int_{\mathbb{R}^{3}} \mathrm{d}\bold{z}\, \bold{P}(b) R_{\infty}^{(1)}(\bold{x},\bold{z};b) \bold{a}(\bold{z}-\bold{y}) \tilde{T}_{\infty}(\bold{z},\bold{y};b,b_{0}),
\end{multline}
The first term of the r.h.s. of \eqref{xx} is the integral kernel of
$T_{1,\infty}(b)\tilde{T}_{\infty}(b,b_{0}) = \frac{i}{2}(P_{1}(b)R_{\infty}(b)P_{2} (b)- P_{2}(b)R_{\infty}(b)P_{1}(b))R_{\infty}(b) \tilde{T}_{\infty}(b,b_{0})$ which is bounded due to \eqref{pRp} and \eqref{nedes} knowing \eqref{approxi}, its norm is bounded above by a  $b$-independent quantity. Furthermore in view of \eqref{comute}, on $\mathbb{R}^{6}\setminus D_{\infty}$:
\begin{multline}\label{apres}
\bold{a}(\bold{z}-\bold{y}) \tilde{T}_{\infty}(\bold{z},\bold{y};b,b_{0})
= \mathrm{e}^{i \delta b \phi(\bold{z},\bold{y})} \bold{a}(\bold{z}-\bold{y})\Big(\delta b \bold{P}(b)\cdot \bold{a}(\bold{z}-\bold{y}) + \frac{(\delta b)^{2}}{2} \bold{a}^{2}(\bold{z}-\bold{y})\Big) R_{\infty}^{(1)}(\bold{z},\bold{y};b_{0}) \\
= \delta b \bold{P}(b) \mathrm{e}^{i \delta b \phi(\bold{z},\bold{y})} \bold{a}^{2}(\bold{z}-\bold{y})  R_{\infty}^{(1)}(\bold{z},\bold{y}; b_{0}) -  \frac{1}{2}(\delta b)^{2} \mathrm{e}^{i \delta b \phi(\bold{z},\bold{y})} \bold{a}^{3}(\bold{z}-\bold{y}) R_{\infty}^{(1)}(\bold{z},\bold{y}; b_{0}),
\end{multline}
which is the integral kernel of $2 \delta b \bold{P}(b) \tilde{T}_{2,\infty}(b,b_{0})
- (\delta b)^{2} \bold{Y}(b,b_{0})$, where $\bold{Y}(b,b_{0})=\bold{Y}(b,b_{0},\omega,\xi)$ is the operator defined via its  integral kernel on $\mathbb{R}^{6}\setminus D_{\infty}$:
\begin{equation} \label{Y}
\bold{Y}(\bold{x},\bold{y};b,b_{0},\omega,\xi) := \frac{1}{2} \bold{a}^{2}(\bold{x}-\bold{y}) \bold{W}(\bold{x},\bold{y};b,b_{0},\omega,\xi).
\end{equation}
Notice that from \eqref{eskres} and \eqref{W}, $\Vert \bold{Y}(b,b_{0},\omega,\xi)\Vert \leq \vert p(\xi)\vert$ for some polynomial $p(\cdot\,)$ independent of $(b,b_{0})$.
Since $\bold{Y}(b,b_{0})$ is bounded, then  by Lemmas \ref{Tiinftybounded1}  and \ref{Tiinftybounded} the second term of the r.h.s. of \eqref{xx} is the kernel of the bounded operator $\delta b\bold{P}(b)R_{\infty}(b)\big(2 \bold{P}(b) \tilde{T}_{2,\infty}(b,b_{0})- \delta b \bold{Y}(b,b_{0})\big)$ and its norm is bounded  above by a $b$-independent polynomial   in $\xi$.\qed
\medskip

\noindent{\it Proof of Proposition \ref{cP2}.} Let $j=0$. Define $\tilde{\mathcal{I}}_{0}(b,b_{0},\omega,\xi)$  as in  \eqref{calI0}  but we replace each operator $R_{\infty}(b,.)$ with $\tilde{R}_{\infty}(b,b_{0},.)$. Then from \eqref{approxi} and \eqref{calI0}, $\mathbb{P}$-a.s., $\forall (b,b_0) \in \mathbb{R}^2$, $\forall \xi \in \Gamma_{K}$:
\begin{equation*}
\mathcal{I}_0(b,\omega,\xi) - \mathcal{I}_0(b_{0},\omega,\xi) = \tilde{\mathcal{I}}_{0}(b,b_{0},\omega,\xi) - \mathcal{I}_{0}(b_{0},\omega,\xi) - \tilde{\mathcal{R}}_{0}(b,b_{0},\omega,\xi),
\end{equation*}
where  $\tilde{\mathcal{R}}_{0}(b,b_{0},\omega,\xi) $ is the following bounded operator:
\begin{multline} \label{reste0}
(\tilde R_{\infty}(\cdot,\xi_0)) ^2R_{\infty}(\cdot,\xi)  \tilde{T}_{\infty}(\cdot,\xi) + R_{\infty}(\cdot,\xi_0) \tilde{T}_{\infty}(\cdot,\xi_0)R_{\infty}(\cdot,\xi_0) R_{\infty}(\cdot,\xi) + \\
+ \tilde R_{\infty}(\cdot,\xi_0) R_{\infty}(\cdot, \xi_0)  \tilde{T}_{\infty}(\cdot,\xi_0)R_{\infty}(\cdot,\xi).
\end{multline}
Firstly in the kernel sense, for any $\bold{x} \in \mathbb{R}^{3}$:
\begin{multline*}
\tilde{\mathcal{I}}_0(\bold{x},\bold{x};b,b_{0},\omega,\xi) - \mathcal{I}_0(\bold{x},\bold{x};b_{0},\omega,\xi) = \int_{\mathbb{R}^{6}} \mathrm{d}\bold{z}_{1}\mathrm{d}\bold{z}_{2} \, \{\mathrm{e}^{i \delta b (\phi(\bold{x},\bold{z}_{1})+\phi(\bold{z}_{1},\bold{z}_{2})+\phi(\bold{z}_{2},\bold{x}))} - 1\} \times \\
\times R_{\infty}^{(1)}(\bold{x},\bold{z}_{1};b_{0},\omega,\xi_{0}) R_{\infty}^{(1)}(\bold{z}_{1},\bold{z}_{2};b_{0},\omega,\xi) R_{\infty}^{(1)}(\bold{z}_{2},\bold{x};b_{0},\omega,\xi_{0}).
\end{multline*}
Since $\vert \mathrm{e}^{i \delta b (\phi(\bold{x},\bold{z}_{1})+\phi(\bold{z}_{1},\bold{z}_{2})+\phi(\bold{z}_{2},\bold{x}))} - 1\vert \leq \vert \delta b\vert \vert \bold{x}-\bold{z}_{1}\vert \vert \bold{z}_{1} - \bold{z}_{2}\vert$, then  \eqref{eskres}  implies  that  $\mathbb{P}$-a.s., there exists  a polynomial $p(\cdot\,)$ s.t. $\forall\xi \in \Gamma_{K}$, $\forall\bold{x} \in \mathbb{R}^{3}$, $\forall b \in \mathbb{R}$ with $\vert b - b_{0}\vert $ small enough:
\begin{equation}
\label{cfv}
\vert\tilde{\mathcal{I}}_0(\bold{x},\bold{x};b,b_{0},\omega,\xi) - \mathcal{I}_0(\bold{x},\bold{x};b_{0},\omega,\xi)\vert \leq \vert \delta b\vert \vert p(\xi)\vert.
\end{equation}
Let us  now estimate $\mathrm{Tr}_{L^{2}(\mathbb{R}^{3})}\{\chi_{\Lambda_{L}} \tilde{\mathcal{R}}_{0}(b,b_{0},\omega,\xi)\chi_{\Lambda_{L}}\}$.   Consider   the following  term in  \eqref{reste0}:
\begin{equation*}
{r}_{0}(b,b_{0},\omega,\xi):=  \big(\tilde{R}_{\infty}(b,b_{0},\omega,\xi_{0})\big)^{2}  R_{\infty}(b,\omega,\xi) \tilde{T}_{\infty}(b,b_{0},\omega,\xi).
\end{equation*}
By using  \eqref{regkTiL}, \eqref {regopTL} with $L=\infty$ and  \eqref{comute}, we have on $\mathbb{R}^{6}$:
\begin{multline*}
(R_{\infty}(b,\omega,\xi) \tilde{T}_{\infty}(b,b_{0},\omega,\xi))(\bold{x},\bold{y}) = - (\delta b)^{2} (R_{\infty}(b,\omega,\xi) \tilde{T}_{2,\infty}(b,b_{0},\omega,\xi))(\bold{x},\bold{y}) + \\
+ \delta b \int_{\mathbb{R}^{3}} \mathrm{d}\bold{z}\, R_{\infty}^{(1)}(\bold{x},\bold{z};b,\omega,\xi) (i\nabla_{\bold{z}} + b\bold{a}(\bold{z}))\cdot  \bold{a}(\bold{z}-\bold{y}) \mathrm{e}^{i \delta b \phi(\bold{z},\bold{y})} R_{\infty}^{(1)}(\bold{z},\bold{y};b_{0},\omega,\xi).
\end{multline*}
Then
\begin{equation}
\label{rewRtiT}
R_{\infty}(b,\omega,\xi) \tilde{T}_{\infty}(b,b_{0},\omega,\xi) = \delta b R_{\infty}(b,\omega,\xi) \Big(\bold{P}(b) \cdot {\bold{W}}(b,b_{0},\omega,\xi) - \delta b  \tilde{T}_{2,\infty}(b,b_{0},\omega,\xi) \Big),
\end{equation}
where ${\bold{W}}(b,b_{0},\omega,\xi)$ is defined in \eqref{W}. Hence we have to estimate:
\begin{equation*}
\Big\Vert\chi_{\Lambda_{L}} \big(\tilde{R}_{\infty}(b,b_{0},\omega,\xi_{0})\big)^{2} R_{\infty}(b,\omega,\xi) \big(\bold{P}(b)\cdot {\bold{W}}(b,b_{0},\omega,\xi)  \chi_{\Lambda_{L}} - \tilde{T}_{2,\infty}(b,b_{0},\omega,\xi)\chi_{\Lambda_{L}}\big)\Big\Vert_{\mathfrak{I}_{1}}.
\end{equation*}
In view of  \eqref{W} and  \eqref{eskres},  $\mathbb{P}$-a.s., there exists a polynomial $p(\cdot\,)$ s.t. $\forall \xi \in \Gamma_{K}$, $\forall b \in \mathbb{R}$ and $\forall L\in (0,\infty)$:
\begin{equation}
\label{noI1}
\Vert \chi_{\Lambda_{L}}\tilde{R}_{\infty}(b,b_{0},\omega,\xi_{0})  \Vert_{\mathfrak{I}_{2}},\,\,\Vert {\bold{W}}(b,b_{0},\omega,\xi)  \chi_{\Lambda_{L}}\Vert_{\mathfrak{I}_{2}},\,\,  \Vert \tilde{T}_{2,\infty}(b,b_{0},\omega,\xi) \chi_{\Lambda_{L}}\Vert_{\mathfrak{I}_{2}} \leq \vert p(\xi)\vert L^{\frac{3}{2}}.
\end{equation}
Then from  \eqref{V2} and \eqref{nedes}, $\mathbb{P}$-a.s.,  there exists another polynomial $p(\cdot\,)$ s.t. $\forall \xi \in \Gamma_{K}$, $\forall b \in \mathbb{R}$ with $\vert b-b_{0}\vert$ small enough and $\forall L \in (0,\infty)$:
\begin{equation} \label{noRtiT}
\vert \Lambda_{L}\vert^{-1} \vert \mathrm{Tr}_{L^{2}(\mathbb{R}^{3})} \{\chi_{\Lambda_{L}} {r}_{0}(b,b_{0},\omega,\xi) \chi_{\Lambda_{L}}\}\vert \leq \vert \delta b\vert \vert p(\xi)\vert.
\end{equation}
Now consider the operator $
{r}_{1}(b,b_{0},\omega,\xi):=  R_{\infty}(\cdot,\xi_0) \tilde{T}_{\infty}(\cdot,\xi_0)  R_{\infty}(\cdot,\xi_0)R_{\infty}(\cdot,\xi) $. From \eqref{rewRtiT}:
\begin{equation*}
{r}_{1}(b,b_{0},\omega,\xi)=  \delta b R_{\infty}(\cdot,\xi_0)\Big(\bold{P}(\cdot) \cdot {\bold{W}}(\cdot ,\xi_{0}) - \delta b  \tilde{T}_{2,\infty}(\cdot,\xi_{0}) \Big)R_{\infty}(\cdot,\xi_{0})R_{\infty}(\cdot,\xi).
\end{equation*}
Then by using \eqref{fornd} and the  above arguments, we conclude that
$\mathbb{P}$-a.s., there exists a polynomial $p(\cdot\,)$ s.t. $\forall \xi \in \Gamma_{K}$, $\forall b \in \mathbb{R}$ with $\vert b-b_{0}\vert$ small enough and $\forall L \in (0,\infty)$:
\begin{equation} \label{noRtiT1}
\vert \Lambda_{L}\vert^{-1} \vert \mathrm{Tr}_{L^{2}(\mathbb{R}^{3})} \{\chi_{\Lambda_{L}} {r}_{1}(b,b_{0},\omega,\xi) \chi_{\Lambda_{L}}\}\vert \leq \vert \delta b\vert \vert p(\xi)\vert.
\end{equation}
This also holds for the last term of \eqref{reste0} and then $\vert \Lambda_{L}\vert^{-1} \vert \mathrm{Tr}_{L^{2}(\mathbb{R}^{3})}\{\chi_{\Lambda_{L}} \tilde{\mathcal{R}}_{0}(b,b_{0},\omega,\xi)\chi_{\Lambda_{L}}\}\vert \leq \vert \delta b\vert \vert p(\xi)\vert$.
This together  with \eqref{cfv} prove $i)$ with $j=0$. Let us show $ii)$. From  \eqref{essecbul1}-\eqref{essecbul2}, \eqref{tildeTinf},   \eqref{reste0}  and Lemma \ref{proestim} $ii)$, then $\mathbb{P}$-a.s., there exists a polynomial $p(\cdot\,)$ s.t. $\forall\xi \in \Gamma_{K}$, $\forall b\in \mathbb{R}$ with $\vert b - b_{0}\vert$ small enough:
$$ \forall \bold{x} \in \mathbb{R}^{3},\quad \vert \tilde{\mathcal{R}}_{0}(\bold{x},\bold{x};b,b_{0},\omega,\xi)\vert \leq \vert \delta b\vert\vert p(\xi)\vert (1 + \vert \bold{x}\vert^{\alpha}).$$
This together with \eqref{cfv} imply:
\begin{equation}
\label{difI_0}
\vert \mathcal{I}_{0}(\bold{0},\bold{0};b,\omega,\xi) - \mathcal{I}_{0}(\bold{0},\bold{0};b_{0},\omega,\xi)\vert \leq \vert \delta b\vert \vert p(\xi)\vert.
\end{equation}
Then $ii)$ with $j=0$ follows from \eqref{difI_0} and the $\omega$-independent estimate \eqref{majpinom}.\\
\indent Let $j=1$.  Define the function on  $ \mathbb{R}^{3}$:
\begin{equation*}
\tilde{\mathcal{I}}_{1}(\bold{x},\bold{x};b,b_{0},\omega,\xi) := \int_{\mathbb{R}^{3}} \mathrm{d}\bold{z}\, \tilde{R}_{\infty}^{(1)}(\bold{x},\bold{z};b,b_{0},\omega,\xi) \bold{a}(\bold{z}-\bold{x})\cdot (i\nabla_{\bold{z}} + b \bold{a}(\bold{z})) \tilde{R}_{\infty}^{(1)}(\bold{z},\bold{x};b,b_{0},\omega,\xi).
\end{equation*}
Due to \eqref{essecbul1}-\eqref{essecbul2} and \eqref{phase}, $\mathbb{P}$-a.s., $\forall (b,b_0) \in \mathbb{R}^{2}$, $\forall \xi \in \Gamma_{K}$, it is well-defined since by  \eqref{comute}:
\begin{multline}
\label{hduf}
\tilde{\mathcal{I}}_{1}(\bold{x},\bold{x};b,b_{0},\omega,\xi) =
\int_{\mathbb{R}^{3}} \mathrm{d}\bold{z}\,  R_{\infty}^{(1)}(\bold{x},\bold{z};b_{0},\omega,\xi) T_{1,\infty}(\bold{z},\bold{x};b_{0},\omega,\xi) + \\
+ 2 \delta b \int_{\mathbb{R}^{3}} \mathrm{d}\bold{z}\,  R_{\infty}^{(1)}(\bold{x},\bold{z};b_{0},\omega,\xi) T_{2,\infty}(\bold{z},\bold{x};b_{0},\omega,\xi).
\end{multline}
From the definition \eqref{calIj} of ${\mathcal{I}}_{1}$,  we replace the resolvent on the left and the one in $T_{1,\infty}$  (see \eqref{T1inftyk}) with the r.h.s. of \eqref{approxi}. Then $\forall\bold{x} \in \mathbb{R}^{3}$:
\begin{equation*}
\mathcal{I}_{1}(\bold{x},\bold{x};b,\omega,\xi) - \mathcal{I}_{1}(\bold{x},\bold{x};b_{0},\omega,\xi) =
\tilde{\mathcal{I}}_{1}(\bold{x},\bold{x};b,b_0,\omega,\xi) - \mathcal{I}_{1}(\bold{x},\bold{x};b_{0},\omega,\xi) + \tilde{\mathcal{R}}_{1}(\bold{x},\bold{x};b,b_{0},\omega,\xi),
\end{equation*}
where $\tilde{\mathcal{R}}_{1}(\cdot,\cdot;b,b_{0},\omega,\xi) := {s}_{0}(\cdot,\cdot;b,b_{0},\omega,\xi) + {s}_{1}(\cdot,\cdot;b,b_{0},\omega,\xi)$ and 
\begin{gather}
\label{tilder1}
{s}_{0}(\cdot,\cdot;b,b_{0},\omega,\xi) := \big(R_{\infty}(b,\omega,\xi) \tilde{T}_{\infty}(b,b_{0},\omega,\xi) T_{1,\infty}(b,\omega,\xi)\big)(\cdot,\cdot),\\
\label{tildes1}
{s}_{1}(\cdot ,\cdot;b,b_{0},\omega,\xi):=
\int_{\mathbb{R}^{3}} \mathrm{d}\bold{z}\, \tilde{R}_{\infty}^{(1)}(\cdot,\bold{z};b,b_{0},\omega,\xi) \bold{a}(\bold{z} -\cdot)\cdot\bold{P}(b) \big(R_{\infty}(b,\omega,\xi) \tilde{T}_{\infty}(b,b_{0},\omega,\xi)\big)(\bold{z},\cdot).
\end{gather}
First by  \eqref{hduf}, we have for any $\bold{x} \in \mathbb{R}^{3}$:
\begin{equation*}
\tilde{\mathcal{I}}_{1}(\bold{x},\bold{x};b,b_0,\omega,\xi) - \mathcal{I}_{1}(\bold{x},\bold{x};b_{0},\omega,\xi) =
2 \delta b \big((R_{\infty}(b_{0},\omega,\xi)T_{2,\infty}(b_{0},\omega,\xi)\big)(\bold{x},\bold{x}).
\end{equation*}
Then by \eqref{essecbul1}-\eqref{essecbul2}, $\mathbb{P}$-a.s., there exists a polynomial $p(\cdot\,)$ s.t. $\forall \xi \in \Gamma_{K}$, $\forall b \in \mathbb{R}$, $\forall \bold{x} \in \mathbb{R}^{3}$:
\begin{equation}
\label{fstet}
\vert \tilde{\mathcal{I}}_{1}(\bold{x},\bold{x};b,b_0,\omega,\xi) - \mathcal{I}_{1}(\bold{x},\bold{x};b_{0},\omega,\xi) \vert \leq \vert \delta b\vert \vert p(\xi)\vert.
\end{equation}
From \eqref{rewRtiT}, \eqref{tilder1} is the  diagonal part of the integral kernel of the bounded operator:
\begin{multline*}
{s}_{0}(b,b_{0},\omega,\xi) := \frac{i\delta b}{2}  R_{\infty}(b,\omega,\xi)\Big( \bold{P}(b) \cdot {\bold{W}}(b,b_{0},\omega,\xi) -
\delta b   \tilde{T}_{2,\infty}(b,b_{0},\omega,\xi)\Big) \times \\
\Big(P_{1}(b) R_{\infty}(b,\omega,\xi) P_{2}(b) - P_{2}(b) R_{\infty}(b,\omega,\xi) P_{1}(b)\Big) R_{\infty}(b,\omega,\xi).
\end{multline*}
By using again the same arguments as above together with   \eqref{pRp}, \eqref{fornd} and   \eqref{nedes}, then  $\mathbb{P}$-a.s. there exists a polynomial $p(\cdot\,)$ s.t. $\forall \xi \in \Gamma_{K}$, $\forall b \in \mathbb{R}$ with $\vert b-b_{0}\vert$ small enough and $\forall L \in (0,\infty)$:
\begin{equation}
\label{trrtild1}
\vert \Lambda_{L}\vert^{-1} \vert \mathrm{Tr}_{L^{2}(\mathbb{R}^{3})}\{\chi_{\Lambda_{L}} {s}_{0}(b,b_{0},\omega,\xi)\chi_{\Lambda_{L}}\}\vert \leq \vert \delta b\vert \vert p(\xi)\vert.
\end{equation}
\noindent Similary to  the proof of Lemma \ref{WP}.  Then   \eqref{comute},  \eqref{T1C}, \eqref{apres} and  \eqref{rewRtiT}, show that  \eqref{tildes1} is the  diagonal part of the integral kernel of the bounded operator :
\begin{multline*}
{s}_{1}(b,b_{0},\omega,\xi) = \frac{i}{2}\tilde{R}_{\infty}(b,b_{0},\xi)\Big(P_{1}(b) R_{\infty}(b,\xi) P_{2}(b) - P_{2}(b) R_{\infty}(b,\xi) P_{1}(b)\Big) R_{\infty}(b,\xi) \tilde{T}_{\infty}(b,b_{0},\xi) + \\
+ \delta b \tilde{R}_{\infty}(b,b_{0},\xi)\bold{P}(b)R_{\infty}(b,\xi) \Big(2 \bold{P}(b) \tilde{T}_{2,\infty}(b,b_{0},\xi)- \delta b \bold{Y}(b,b_{0},\xi)\Big),
\end{multline*}
with $\bold{Y}(b,b_{0},\xi)$ the operator defined in  \eqref{Y}.
Due to \eqref{pRp}, the first term  in the above r.h.s. can be treated exactly as the operator $r_0(b,b_{0},\omega,\xi)$ at the beginning of this proof. For the second term, we have \eqref{noI1} and under the same conditions, $\Vert \bold{Y}(b,b_{0},\xi) \chi_{\Lambda_{L}}\Vert_{\mathfrak{I}_{2}} \leq \vert p (\xi)\vert L^{\frac{3}{2}}$.
Therefore  $\mathbb{P}$-a.s., there exists another polynomial $p(\cdot\,)$ s.t. $\forall \xi \in \Gamma_{K}$, $\forall b \in \mathbb{R}$ with $\vert b - b_{0}\vert$ small enough and $\forall L \in (0,\infty)$:
\begin{equation*}
\vert \Lambda_{L}\vert^{-1} \vert \mathrm{Tr}_{L^{2}(\mathbb{R}^{3})} \{\chi_{\Lambda_{L}} {s}_{1}(b,b_{0},\omega,\xi) \chi_{\Lambda_{L}}\}\vert \leq \vert \delta b\vert \vert p(\xi)\vert.
\end{equation*}
Due to \eqref{trrtild1}, this also holds for $\vert \Lambda_{L}\vert^{-1} \vert \mathrm{Tr}_{L^{2}(\mathbb{R}^{3})}\{\chi_{\Lambda_{L}} \tilde{\mathcal{R}}_{1}(b,b_{0},\omega,\xi)\chi_{\Lambda_{L}}\}\vert$. This together  with \eqref{fstet} prove $i)$ with $j=1$. Now consider \eqref{tilder1}-\eqref{tildes1}. From \eqref{essecbul1}-\eqref{essecbul2} and Lemma \ref{proestim} $ii)$, $\mathbb{P}$-a.s., there exists a polynomial $p(\cdot\,)$ s.t. $\forall\xi \in \Gamma_{K}$, $\forall b\in \mathbb{R}$, $\vert b - b_{0}\vert$ small enough:
$$ \forall \bold{x} \in \mathbb{R}^{3},\quad \vert \tilde{\mathcal{R}}_{1}(\bold{x},\bold{x};b,b_{0},\omega,\xi)\vert \leq \vert \delta b\vert\vert p(\xi)\vert (1 + \vert \bold{x}\vert^{\alpha})^{2}.$$
This together with \eqref{fstet} imply:
\begin{equation}
\label{difI_1}
\vert \mathcal{I}_{1}(\bold{0},\bold{0};b,\omega,\xi) - \mathcal{I}_{1}(\bold{0},\bold{0};b_{0},\omega,\xi)\vert \leq \vert \delta b\vert \vert p(\xi)\vert.
\end{equation}
Then $ii)$ follows from \eqref{difI_1} and the $\omega$-independent estimate \eqref{majpinom}.\qed


\section{Proof of Theorem \ref{maintheorem} and Theorem \ref{maintheorem1}}

In the whole of this section, for any compact subset  $K$ of   $  \mathcal{D}_{\epsilon}$ and any $ \beta \in [\beta_1,\beta_2] \in \mathbb R$,  $\Gamma_K$ is defined for $ \beta= \beta_2$. We  will use  $ \eta := \min\{E_0-E_K, \frac{\vartheta_{K}}{2\beta_2}\}$. We  first give  two important results proven subsequently.

\begin{proposition}\label{result1}
i) $\mathbb{P}$-a.s. on $\Omega$,
$\forall b \in \mathbb{R}$, $\forall 0<\beta_1<\beta_2$, $ \forall K$ compact subset of $  \mathcal{D}_{\epsilon}$, one has:
\begin{equation*}
\lim_{L \rightarrow \infty} \frac{1}{\vert \Lambda_{L} \vert} \int_{\Lambda_{L}} \mathrm{d}\bold{x}\, \mathcal{L}_{\infty,n}^{(\omega)}(\bold{x},\bold{x};\beta,b,z,\epsilon) = \mathbb{E}\big[\mathcal{L}_{\infty,n}^{(\omega)}(\bold{0},\bold{0};\beta,b,z,\epsilon)\big]\quad n=0,1,
\end{equation*}
uniformly in $(\beta,z) \in [\beta_1, \beta_2] \times K$.\\
ii) $\mathbb{P}$-a.s. on $\Omega$, $\forall 0<\beta_1<\beta_2$, $\forall K$ compact subset of $\mathcal{D}_{\epsilon}$, one has:
\begin{equation*}
\lim_{L \rightarrow \infty} \frac{1}{\vert \Lambda_{L} \vert} \int_{\Lambda_{L}} \mathrm{d}\bold{x}\, \mathcal{L}_{\infty,2}^{(\omega)}(\bold{x},\bold{x};\beta,0,z,\epsilon) = \mathbb{E}\big[\mathcal{L}_{\infty,2}^{(\omega)}(\bold{0},\bold{0};\beta,0,z,\epsilon)\big],
\end{equation*}
uniformly in $(\beta,z) \in [\beta_1, \beta_2] \times K$.
\end{proposition}

\begin{proposition}
\label{secondresult}  $\mathbb{P}$-a.s. on $\Omega$,
$ \forall b \in \mathbb{R}$, $\forall 0<\beta_1<\beta_2$, $ \forall K$    compact subset of $  \mathcal{D}_{\epsilon}$, one has:
\begin{equation}
\label{ingred1}
\lim_{L \rightarrow \infty} \frac{1}{\vert \Lambda_{L}\vert} \frac{1}{\beta} \bigg\vert  \int_{\Lambda_{L}} \mathrm{d}\bold{x}\, \bigg( \frac{\partial^{n} \mathcal{L}_{L}^{(\omega)}}{\partial b^{n}}(\bold{x},\bold{x};\beta,b,z,\epsilon) - \mathcal{L}_{\infty,n}^{(\omega)}(\bold{x},\bold{x};\beta,b,z,\epsilon)\bigg) \bigg\vert=0 \quad n=0,1,2,
\end{equation}
uniformly in  $(\beta,z) \in [\beta_1, \beta_2] \times K$.
\end{proposition}

\subsection{Proof of Theorem \ref{maintheorem}}

\subsubsection{Proof of $i)$} The results follow  from Propositions \ref{result1} and  \ref{secondresult}, with the simple relation:
\begin{multline}
\label{diffondr}
\mathcal{X}_{L,n}^{(\omega)}(\beta,b,z,\epsilon) = \bigg(\frac{q}{c}\bigg)^{n} \frac{\epsilon}{\beta \vert \Lambda_{L} \vert} \int_{\Lambda_{L}} \mathrm{d}\bold{x}\, \bigg( \frac{\partial^{n} \mathcal{L}_{L}^{(\omega)}}{\partial b^{n}}(\bold{x},\bold{x};\beta,b,z,\epsilon) - \mathcal{L}_{\infty,n}^{(\omega)}(\bold{x},\bold{x};\beta,b,z,\epsilon)\bigg) + \\
+ \bigg(\frac{q}{c}\bigg)^{n} \frac{\epsilon}{\beta \vert \Lambda_{L} \vert} \int_{\Lambda_{L}} \mathrm{d}\bold{x}\, \mathcal{L}_{\infty,n}^{(\omega)}(\bold{x},\bold{x};\beta,b,z,\epsilon).
\end{multline}

\subsubsection{Proof of $ii)$}

Define  the operator on $L^2({\mathbb R}^3)$:
 \begin{equation*}
\mathcal{J}_{\infty}^{(\omega)}(\beta,b,z,\epsilon) := \frac{i}{2\pi} \int_{\Gamma_{K}} \mathrm{d}\xi\, \frac{z \mathrm{e}^{-\beta \xi}}{1+ \epsilon z \mathrm{e}^{-\beta \xi}} R_{\infty}(b,\omega,\xi).
\end{equation*}
Since the function $\xi \mapsto  \frac{z \mathrm{e}^{-\beta \xi}}{1+ \epsilon z \mathrm{e}^{-\beta \xi}}$ is  also  exponentially decreasing on $\Gamma_{K}$ when $\vert\Re \xi\vert \to \infty $, the study of  the operator $\mathcal{J}_{\infty}^{(\omega)}(\beta,b,z,\epsilon)$ is therefore similar to the one of $\mathcal{L}_{\infty,0}^{(\omega)}(\beta,b,z,\epsilon)$. Then we deduce that $\mathbb{P}$-a.s.,
$\forall 0< \beta_1 <\beta_2 $,  $\forall b \in \mathbb R$ and   for any compact subset $ K $ of  $\mathcal{D}_{\epsilon}$:
\begin{equation}
\label{limro}
\rho_{\infty}(\beta,b,z,\epsilon):= \lim_{L \rightarrow \infty} \rho_{L}^{(\omega)}(\beta,b,z,\epsilon) = \mathbb{E}[\mathcal{J}_{\infty}^{(\omega)}(\bold{0},\bold{0};\beta,b,z,\epsilon)],
\end{equation}
uniformly in $(\beta, z) \in [\beta_1, \beta_2]\times K$. Here  $\mathcal{J}_{\infty}^{(\omega)}(\cdot\,,\cdot\,;\beta,b,z,\epsilon)$ is the integral kernel of $\mathcal{J}_{\infty}^{(\omega)}(\beta,b,z,\epsilon)$.

Let $ K'$ be the interior of $K$ which is supposed nonempty. Fix $z_{0} \in K'$. Let $z \in K $ near $z_{0}$. Since  $\mathbb{P}$-a.s., $\forall \beta>0$ and $ \forall b \in \mathbb{R}$, $P_{L}^{(\omega)}(\beta,b,\cdot\,,\epsilon)$ is an analytic function on $\mathcal{D}_{\epsilon}$,  then:
\begin{equation} \label{111}
P_{L}^{(\omega)}(\beta,b,z,\epsilon) = P_{L}^{(\omega)}(\beta,b,z_{0},\epsilon) + (z-z_{0}) \frac{1}{\beta z_{0}}\rho_{L}^{(\omega)}(\beta,b,z_{0},\epsilon) + o((z-z_{0})).
\end{equation}
Notice that since $\Vert R_L(b,\omega, \xi) \Vert \leq c, \forall \xi \in \Gamma_K$ and $L \in (0, \infty] $, the third term in   the r.h.s. of \eqref{111}  is  $L$-independent.
It follows from Theorem \ref{maintheorem} $i)$, $\mathbb{P}$-a.s., $\forall \beta>0$, $ \forall b \in \mathbb{R}$ and for $z \in K$  near $z_0$, the following  identity between non-random limits:
\begin{equation*}
P_{\infty}(\beta,b,z,\epsilon) - P_{\infty}(\beta,b,z_{0},\epsilon)  = (z-z_{0}) \frac{1}{\beta z_{0}} \mathbb{E}[\mathcal{J}_{\infty}^{(\omega)}(\bold{0},\bold{0};\beta,b,z_{0},\epsilon)] + o(z-z_{0}).
\end{equation*}
So $P_{\infty}(\beta,b,\cdot\,,\epsilon)$ is  analytic in $z_{0}$, and $\beta z_{0} \frac{\partial  P_{\infty}}{\partial z}(\beta,b,z_{0},\epsilon) = \mathbb{E}[\mathcal{J}_{\infty}^{(\omega)}(\bold{0},\bold{0};\beta,b,z_{0},\epsilon)]$.

\subsubsection{Proof of $ iii)$}
Let $\alpha \in (0,\frac{1}{3})$ and $b_{0}\in \mathbb{R}$ be fixed. Since $\mathbb{P}$-a.s., $\forall \beta >0$ and $\forall z \in \mathcal{D}_{\epsilon}$, $b \mapsto P_{L}^{(\omega)}(\beta,b,z,\epsilon)$ is a $\mathcal{C}^{\infty}$-function, then for $b \in \mathbb{R}$  near $b_{0}$:
\begin{equation*}
P_{L}^{(\omega)}(\beta,b,z,\epsilon) = P_{L}^{(\omega)}(\beta,b_{0},z,\epsilon) +  (b-b_{0}) \frac{c}{q} \mathcal{X}_{L,1}^{(\omega)}(\beta,b_{0},z,\epsilon) + o((b-b_{0})).
\end{equation*}
In  virtue of Remark  \ref{pour la fin2} below,  the  third term  in the r.h.s. of the above equality is  uniformly bounded  in $L \in (0,\infty)$. It follows from Theorem \ref{maintheorem} $i)$, $\mathbb{P}$-a.s., $\forall \beta >0$, $\forall z \in \mathcal{D}_{\epsilon}$ and for $b\in \mathbb{R}$ sufficiently close to $b_{0}$, the following relation:
\begin{equation*}
P_{\infty}(\beta,b,z,\epsilon) - P_{\infty}(\beta,b_{0},z,\epsilon) = (b-b_{0}) \frac{c}{q}\mathcal{X}_{\infty,1}(\beta,b_{0},z,\epsilon) +   o((b-b_{0})).
\end{equation*}
Then $P_{\infty}(\beta,\cdot\,,z,\epsilon)$ is differentiable at  $b_{0}$ and $
\frac{\partial P_{\infty}}{\partial b}(\beta,b_{0},z,\epsilon) =  (\frac{c}{q})\mathcal{X}_{\infty,1}(\beta,b_{0},z,\epsilon)$. 


\subsection{Proof of Theorem \ref{maintheorem1}} $i)$ follows from  \eqref{diffondr} for $n=2$ and $b=0$, Propositions \ref {result1} $ii)$ and  \ref{secondresult}. Let us prove $ii)$. Let $\alpha \in (0,\frac{1}{4})$. Since $\mathbb{P}$-a.s. $\forall\beta>0$ and $\forall z \in \mathcal{D}_{\epsilon}$, $b \mapsto \mathcal{X}_{L,1}^{(\omega)}(\beta,b,z,\epsilon)$ is a $\mathcal{C}^{\infty}$-function near $b_{0}=0$, then for real $b$ sufficiently small:
\begin{equation*}
\mathcal{X}_{L,1}^{(\omega)}(\beta,b,z,\epsilon) =
\mathcal{X}_{L,1}^{(\omega)}(\beta,0,z,\epsilon) +  b \frac{c}{q} \mathcal{X}_{L,2}^{(\omega)}(\beta,0,z,\epsilon) + o(b).
\end{equation*}
From Remarks \ref{pour la fin1} and \ref{pour la fin2},  the last term  in the r.h.s. of this equality is  uniformly bounded in $L\in (0,\infty)$.  It follows from Theorems \ref{maintheorem} $i)$ and \ref{maintheorem1} $i)$:
\begin{equation*}
\mathcal{X}_{\infty,1}(\beta,b,z,\epsilon) - \mathcal{X}_{\infty,1}(\beta,0,z,\epsilon) = b \frac{c}{q}\mathcal{X}_{\infty,2}(\beta,0,z,\epsilon) +   o(b).
\end{equation*}
This together with  Theorem \ref{maintheorem} $iii)$,   imply  that $P_{\infty}(\beta,\cdot\,,z,\epsilon)$ is twice differentiable near $b=0$ and $\frac{\partial^2 P_{\infty}}{\partial b^{2}}(\beta,0,z,\epsilon) =  (\frac{c}{q})^{2} \mathcal{X}_{\infty,2}(\beta,0,z,\epsilon)$.

\subsection{ Proof of Proposition \ref{result1}}

By Corollary \ref{fithm},  we can apply the Birkhoff-Khintchine theorem \cite[Prop. 1.13]{PF}  to $(\omega,\bold{x}) \mapsto \mathcal{L}_{\infty,n}^{(\omega)}(\bold{x},\bold{x};\beta,b,z,\epsilon), n=0,1,2$. This  implies that $\mathbb{P}$-a.s., for any $\beta >0$ and $z \in K$:
\begin{equation*}
\label{limiti}
\lim_{L \rightarrow \infty} \frac{1}{\vert \Lambda_{L} \vert} \int_{\Lambda_{L}} \mathrm{d}\bold{x}\, \mathcal{L}_{\infty,n}^{(\omega)}(\bold{x},\bold{x};\beta,b,z,\epsilon) = \mathbb{E}\big[\mathcal{L}_{\infty,n}^{(\omega)}(\bold{0},\bold{0};\beta,b,z,\epsilon)\big] \quad n=0,1,2.
\end{equation*}
But this is not sufficient to prove Proposition \ref{result1}. Consider the case of $n=0$. In view of \eqref{L0T} and  Proposition \ref{ffithm},  we  can use   the Birkhoff-Khintchine theorem for 
 $(\bold{x}, \omega) \mapsto  {\mathcal I}_0(\bold{x},\bold{x};b, \omega, \xi)$.  Then 
$\forall b \in \mathbb R$ and $\forall \xi \in \Gamma_K$, there exists $ \Omega_{\xi,b} \subset \Omega$ with ${\mathbb P}(\Omega_{\xi,b})=1$ s.t.  $\forall \omega  \in \Omega_{\xi,b}$:
\begin{equation} \label {besoin}
 \lim_{L \rightarrow \infty} \frac{1}{\vert \Lambda_{L} \vert} \int_{\Lambda_{L}} \mathrm{d}\bold{x}\, {\mathcal I}_0(\bold{x},\bold{x}; b, \omega, \xi) = \mathbb{E}\big[{\mathcal I}_0(\bold{0},\bold{0}; b, \omega, \xi)\big].
 \end{equation}
Now choose a countable  dense subset of $\Gamma_K$, $ \Gamma_c:=\{\xi_i,  i \in \mathbb N\}$.  Then  $\forall b \in \mathbb R$, there exists $ \Omega_{b} \subset \Omega$ with ${\mathbb P}(\Omega_{b})=1$  s.t. $ \forall \omega \in \Omega_{b}$ and $\forall \xi_i \in \Gamma_c$,  \eqref{besoin} holds. Next we use Proposition \ref{cP1} $i)$:
$ \forall b \in \mathbb R$,  there exists  $  \Omega'_{b} \subset \Omega_{b}$ with ${\mathbb P}(\Omega'_{b})=1$ s.t. $ \forall \omega \in \Omega'_{b}$, the map $ \xi \in \Gamma_K \mapsto \varsigma_L(b, \omega, \xi) := \frac{1}{\vert \Lambda_{L} \vert} \int_{\Lambda_{L}} \mathrm{d}\bold{x}\, {\mathcal I}_0(\bold{x},\bold{x}; b, \omega, \xi) $ is continuous uniformly in $ L \in (0,\infty)$. Let $\xi \in \Gamma_K$. Then $ \forall \varepsilon >0$ there exists $ \xi_i \in \Gamma_c$ s.t. $ \forall L \in (0,\infty)$,
$\vert \varsigma_L(b, \omega, \xi)-\varsigma_L(b, \omega, \xi_i)\vert \leq \varepsilon$.
On the other hand by Proposition \ref{cP1} $ii)$, $\xi_i$ can also be chosen s.t.  $\vert \mathbb{E}\big[{\mathcal I}_0(\bold{0},\bold{0}; b, \omega, \xi)\big]- \mathbb{E}\big[{\mathcal I}_0(\bold{0},\bold{0}; b, \omega, \xi_i)\big]\vert \leq \varepsilon$. Then taking the real part we get:
\begin{multline*}\Re \{ \mathbb{E}\big[{\mathcal I}_0(\bold{0},\bold{0}; b, \omega, \xi)\big]\} -2 \varepsilon \leq \liminf _L \Re  \{\varsigma_L(b, \omega, \xi)\} \leq \\ \limsup_L  \Re \{\varsigma_L(b, \omega, \xi)\} \leq  \Re \{ \mathbb{E}\big[{\mathcal I}_0(\bold{0},\bold{0}; b, \omega, \xi)\big]\}+2\varepsilon. \end{multline*}
Obviously, this also holds true for the imaginary part. Consequently $ \forall b \in \mathbb R$, $ \forall \omega \in \Omega'_{b}$, \eqref{besoin} holds  $\forall
\xi \in \Gamma_K$. We can repeat the same arguments as above to remove the $b$-dependance but with the use of  Proposition \ref{cP2}. Then we conclude that $\mathbb{P}$-a.s., $\forall b \in \mathbb R$ and $\forall \xi \in \Gamma_K$ \eqref{besoin} holds true.
Notice  that  since $\mathbb{P}$-a.s., $\forall b \in \mathbb R$ and $\forall \xi \in \Gamma_K$ the integral kernel of ${\mathcal I}_0(b,\omega, \xi)$ is uniformly bounded by a constant independent of $\omega$ and $\xi$, therefore we have:
$$ \frac{i}{2\pi} \int_{\Gamma_{K}} \mathrm{d}\xi\, (\xi - \xi_{0})^2 \mathfrak{f}_{\epsilon}(\beta,z;\xi) \mathbb{E}\big[{\mathcal I}_0(\bold{0},\bold{0}; b, \omega, \xi)\big]=  \mathbb{E}\big[\mathcal{L}_{\infty,0}^{(\omega)}(\bold{0},\bold{0};\beta,b,z,\epsilon)\big].$$
Afterwards consider the quantity:
$$\mathcal{Q}_{0}(\beta, b,z, \epsilon):=  \frac{i}{2\pi} \int_{\Gamma_{K}} \mathrm{d}\xi\, (\xi - \xi_{0})^2 \mathfrak{f}_{\epsilon}(\beta,z;\xi)\bigg( \frac{1}{\vert \Lambda_{L} \vert} \int_{\Lambda_{L}} \mathrm{d}\bold{x}\, {\mathcal I}_0(\bold{x},\bold{x}; b, \omega, \xi)  - \mathbb {E}\big[{\mathcal I}_0(\bold{0},\bold{0}; b, \omega, \xi)\big]\bigg).$$
Then by using \eqref{expdecr}  there exists a constant $c= c(\beta_{1},K)>0$ s.t. $\forall\,\beta \in [\beta_{1},\beta_{2}]$ and $\forall\,z \in K$:
$$\vert \mathcal{Q}_{0}(\beta, b,z, \epsilon) \vert \leq c   \int_{\Gamma_{K}} \vert \mathrm{d}\xi \vert\, \vert\xi - \xi_{0}\vert^2e^{-\beta_1 \Re \xi}\bigg\vert  \frac{1}{\vert \Lambda_{L} \vert} \int_{\Lambda_{L}} \mathrm{d}\bold{x}\, {\mathcal I}_0(\bold{x},\bold{x}; b, \omega, \xi)  - \mathbb{E}\big[{\mathcal I}_0(\bold{0},\bold{0}; b, \omega, \xi)\big]\bigg\vert. $$
In view of \eqref{besoin}, this proves the proposition  for $n=0$. 
  
  Similarly we consider the case of $n=1$ from  \eqref{L1T}.  In view of  Propositions  \ref{ffithm},  \ref{cP1} and \ref{cP2},  $\mathbb{P}$-a.s., $\forall b \in \mathbb R$ and $\forall \xi \in \Gamma_K$,  \eqref{besoin}  also holds  true if we consider  now the kernel ${\mathcal I}_1(\bold{x},\bold{x}; b, \omega, \xi)$ instead  of ${\mathcal I}_0(\bold{x},\bold{x}; b, \omega, \xi)$. Hence,  following  the proof for the case of $n=0$ step by step,  we conclude the  proof for the case of $n=1$. This proves $i)$. $ii)$ also follows by the same arguments but we disregard the $b$-dependence since we only treat the zero-field case.

\subsection{Proof of Proposition  \ref{secondresult}}

Let  $L >0$ and  $\kappa>0$. We use the decomposition:
\begin{equation}
\label{68.3}
\Lambda_{L} = (\Lambda_{L}\setminus \Lambda_{\kappa}) \cup \Lambda_{\kappa}\quad \textrm{with $\Lambda_{\kappa}:= \{\bold{x} \in \Lambda_{L}\,:\, \mathrm{d}(\bold{x}) < \kappa\}$},
\end{equation}
where we set $\mathrm{d}(\bold{x}):= \mathrm{dist}(\bold{x},\partial \Lambda_{L})$.
Note that the definition \eqref{68.3} implies:
\begin{equation}
\label{68.98}
\vert \Lambda_{\kappa}\vert = \mathcal{O}(L^{2}) \quad \textrm{when $L \rightarrow \infty$}.
\end{equation}
Hereafter we denote by $\chi_{\Lambda_{\kappa}}$ the characteristic function of  $\Lambda_{\kappa}$.

\begin{lema}
\label{diffcont}
$\mathbb{P}$-a.s. on $\Omega$, $\forall b \in \mathbb{R}$, $\forall 0<L<\infty$ and $\forall \xi \in \Gamma_{K}$, then:\\
i) $\bold{x} \mapsto \big(R_{L}^{(1)}(\bold{x},\bold{y};b,\omega,\xi) - R_{\infty}^{(1)}(\bold{x},\bold{y};b,\omega,\xi)\big)\big\vert_{\bold{y}=\bold{x}}$ and  $\bold{x} \mapsto \big((i\nabla_{\bold{x}} + b \bold{a}(\bold{x}))(R_{L}^{(1)}(\bold{x},\bold{y};b,\omega,\xi) - R_{\infty}^{(1)}(\bold{x},\bold{y};b,\omega,\xi))\big)\big\vert_{\bold{y}=\bold{x}}$ are continuous functions on $\Lambda_{L}\setminus \Lambda_{\kappa}$.\\
ii) There exists a constant $\gamma>0$ and a polynomial $p(\cdot\,)$ both $L$-independent s.t. on $\Lambda_{L}^{2}\setminus D_{L}$:
\begin{multline}
\label{esdifres}
\vert R_{L}^{(1)}(\bold{x},\bold{y};b,\omega,\xi) - R_{\infty}^{(1)}(\bold{x},\bold{y};b,\omega,\xi)\vert \\
\leq \vert p(\xi)\vert  (1+\vert \bold{x}\vert^{\alpha} + \vert \bold{y}\vert^{\alpha}) \mathrm{e}^{- \frac{\gamma}{1+\vert \xi\vert} \vert \bold{x}-\bold{y}\vert} \bigg(\frac{\chi_{\Lambda_{\kappa}}(\bold{x})}{\vert \bold{x}-\bold{y}\vert} + \frac{\chi_{\Lambda_{\kappa}}(\bold{y})}{\vert \bold{x}-\bold{y}\vert} + \mathrm{e}^{- \frac{\gamma}{1+\vert \xi\vert}(\mathrm{d}(\bold{x}) + \mathrm{d}(\bold{y}))}\bigg),
\end{multline}
\begin{multline}
\label{esdifdres}
\vert (i\nabla_{\bold{x}} + b \bold{a}(\bold{x}))(R_{L}^{(1)}(\bold{x},\bold{y};b,\omega,\xi) - R_{\infty}^{(1)}(\bold{x},\bold{y};b,\omega,\xi))\vert \\
\leq \vert p(\xi)\vert
(1+\vert \bold{x}\vert^{\alpha}+ \vert \bold{y}\vert^{\alpha})^{2} \mathrm{e}^{- \frac{\gamma}{1+\vert \xi\vert} \vert \bold{x}-\bold{y}\vert} \bigg(\frac{\chi_{\Lambda_{\kappa}}(\bold{x})}{\vert \bold{x}-\bold{y}\vert^{2}} + \frac{\chi_{\Lambda_{\kappa}}(\bold{y})}{\vert \bold{x}-\bold{y}\vert^{2}} + \mathrm{e}^{- \frac{\gamma}{1+\vert \xi\vert} (\mathrm{d}(\bold{x}) + \mathrm{d}(\bold{y}))}\bigg).
\end{multline}
\end{lema}

\noindent \textit{Proof.} Under the conditions of Lemma \ref{diffcont} and  from the Green's identity, we have on $\Lambda_{L}^{2}\setminus D_{L}$:
\begin{equation*}
R_{L}^{(1)}(\bold{x},\bold{y};b,\omega,\xi) - R_{\infty}^{(1)}(\bold{x},\bold{y};b,\omega,\xi) = \\
- \frac{1}{2} \int_{\partial \Lambda_{L}} \mathrm{d}\sigma(\bold{z})\, R_{\infty}^{(1)}(\bold{x},\bold{z};b,\omega,\xi) [\bold{n}_{\bold{z}} \cdot \nabla_{\bold{z}} R_{L}^{(1)}(\bold{z},\bold{y};b,\omega,\xi)],
\end{equation*}
where $\mathrm{d}\sigma(\bold{z})$  is  the measure on $\partial\Lambda_{L}$ and $\bold{n}_{\bold{z}}$ the outer normal to $\partial\Lambda_{L}$.  From \eqref{eskres}, Lemma \ref{prokdres} and since ${\bf z } \in \partial \Lambda_{L}$, then  $\mathbb{P}$-a.s., $\forall b \in \mathbb{R}$ there exists a constant $\gamma>0$ and a polynomial $p(\cdot\,)$ s.t. $\forall \xi \in \Gamma_{K}$ the integrand in the r.h.s. is a continuous function on $  (\Lambda_{L} \setminus \Lambda_{\kappa})^2$ satisfying
\begin{multline}
\label{intermes}
\int_{\partial \Lambda_{L}} \mathrm{d}\sigma(\bold{z})\,\Big\vert R_{\infty}^{(1)}(\bold{x},\bold{z};b,\omega,\xi) [\bold{n}_{\bold{z}} \cdot \nabla_{\bold{z}} R_{L}^{(1)}(\bold{z},\bold{y};b,\omega,\xi)]\Big\vert   \\
\leq \vert p(\xi )\vert (1+\vert \bold{x}\vert^{\alpha}+\vert \bold{y}\vert^{\alpha}) \mathrm{e}^{-\frac{\gamma}{1 + \vert \xi \vert} \vert \bold{x} - \bold{y} \vert} \mathrm{e}^{- \frac{\gamma}{1 + \vert \xi\vert} (\mathrm{d}(\bold{x}) + \mathrm{d}(\bold{y}))}.
\end{multline}
Then 
\eqref{intermes} together with \eqref{eskres} imply \eqref{esdifres}.
We also have on $\Lambda_{L}^{2}\setminus D_{L}$:
\begin{multline*}
(i\nabla_{\bold{x}} + b \bold{a}(\bold{x}))(R_{L}^{(1)}(\bold{x},\bold{y};b,\omega,\xi) - R_{\infty}^{(1)}(\bold{x},\bold{y};b,\omega,\xi)) = \\
- \frac{1}{2} \int_{\partial \Lambda_{L}} \mathrm{d}\sigma(\bold{z})\, (i\nabla_{\bold{x}} + b \bold{a}(\bold{x})) R_{\infty}^{(1)}(\bold{x},\bold{z};b,\omega,\xi) [\bold{n}_{\bold{z}} \cdot \nabla_{\bold{z}} R_{L}^{(1)}(\bold{z},\bold{y};b,\omega,\xi)],
\end{multline*}
Then Lemma \ref {prokdres} together with the above arguments show that  $(i\nabla_{\bold{x}} + b \bold{a}(\bold{x}))(R_{L}^{(1)}(\cdot\,,\cdot\,;b,\omega,\xi) - R_{\infty}^{(1)}(\cdot\,,\cdot\,;b,\omega,\xi))$ is also a continuous function on $(\Lambda_{L} \setminus \Lambda_{\kappa})^2 $ and \eqref{esdifdres} holds.\qed

\remark \label{63.8}   Definitions \eqref{kT1L} and \eqref{kT2L}  then  imply the following estimates.  Under  the same conditions as  in Lemma \ref{diffcont},  $\forall (\bold{x},\bold{y}) \in \Lambda_{L}^2 \setminus D_{L}$ and for $j=1,2$:
\begin{multline}
\label{63.9}
\vert T_{j,L}(\bold{x},\bold{y};b,\omega,\xi) - T_{j,\infty}(\bold{x},\bold{y};b,\omega,\xi) \vert  \\  \leq \vert p(\xi)\vert  (1+\vert \bold{x}\vert^{\alpha} + \vert \bold{y}\vert^{\alpha})^{2} \mathrm{e}^{-\frac{\gamma}{1 + \vert \xi \vert} \vert \bold{x} - \bold{y}\vert}\bigg(\frac{\chi_{\Lambda_{\kappa}}(\bold{x})}{\vert \bold{x}-\bold{y}\vert} + \frac{\chi_{\Lambda_{\kappa}}(\bold{y})}{\vert \bold{x}-\bold{y}\vert} + \mathrm{e}^{- \frac{\gamma}{1 + \vert \xi \vert} (\mathrm{d}(\bold{x}) + \mathrm{d}(\bold{y}))}\bigg),
\end{multline}
for another constant $\gamma>0$ and polynomial $p(\cdot\,)$.

\subsubsection {The case of n=0} Under the conditions of Proposition \ref{secondresult}, we have:
\begin{multline*}
\frac{1}{\vert \Lambda_{L}\vert} \frac{1}{\beta} \int_{\Lambda_{L}} \mathrm{d}\bold{x}\, \Big( \mathcal{L}_{L}^{(\omega)}(\bold{x},\bold{x};\beta,b,z,\epsilon) - \mathcal{L}_{\infty,0}^{(\omega)}(\bold{x},\bold{x};\beta,b,z,\epsilon)\Big)=\\
\frac{1}{\beta} \frac{1}{L^{3}}\frac{i}{2\pi} \bigg(\int_{\Gamma_{K}} \mathrm{d}\xi\, \mathfrak{f}_{\epsilon}(\beta,z;\xi) \int_{\Lambda_{L} \setminus \Lambda_{\kappa}} \mathrm{d}\bold{x}\, \big(R_{L}^{(1)}(\bold{x},\bold{y};b,\omega,\xi) - R_{\infty}^{(1)}(\bold{x},\bold{y};b,\omega,\xi)\big)\big\vert_{\bold{y} = \bold{x}} + \\
 \int_{\Gamma_{K}} \mathrm{d}\xi\, (\xi - \xi_{0}) \mathfrak{f}_{\epsilon}(\beta,z;\xi) \int_{\Lambda_{\kappa}} \mathrm{d}\bold{x}\, (R_{L}(b,\omega,\xi_{0})R_{L}(b,\omega,\xi))(\bold{x},\bold{x}) - (R_{\infty}(b,\omega,\xi_{0})R_{\infty}(b,\omega,\xi))(\bold{x},\bold{x}) \bigg).
\end{multline*}
Here we  have used the first resolvent equation  and  the integral Cauchy formula to rewrite the second term of the r.h.s. of this expression. From Remark \ref{2.2}  and  \eqref{expdecr} this  second term is bounded above by $c\times \vert \Lambda_{\kappa}\vert$ for some constant $c=c(\beta_{1},K)>0$.
On the other hand, due to \eqref{esdifres}, a straightforward calculus leads to:
$$ \bigg\vert \int_{\Lambda_{L} \setminus \Lambda_{\kappa}} \mathrm{d}\bold{x}\, \big(R_{L}^{(1)}(\bold{x},\bold{y};b,\omega,\xi) - R_{\infty}^{(1)}(\bold{x},\bold{y};b,\omega,\xi)\big)\big\vert_{\bold{y}=\bold{x}} \bigg\vert  \leq \vert p(\xi)\vert L^{2+\alpha},$$
for another polynomial $p(\cdot\,)$.  Hence  as  in the proof of Proposition \ref{result1}, then $\mathbb{P}$-a.s., $\forall \beta \in [\beta_1, \beta_2]$, $\forall b \in \mathbb{R}$, $\forall z \in K$  and  for $L$ sufficiently large:
$$ \bigg\vert \frac{1}{\vert \Lambda_{L}\vert}  \int_{\Lambda_{L}} \mathrm{d}\bold{x}\, \Big( \mathcal{L}_{L}^{(\omega)}(\bold{x},\bold{x};\beta,b,z,\epsilon) - \mathcal{L}_{\infty,0}^{(\omega)}(\bold{x},\bold{x};\beta,b,z,\epsilon)\Big) \bigg\vert \leq c \frac{1}{{L^{1- \alpha}}},$$
for some constant $c=c(\beta_1,b,K)>0$. This proves the case of $n=0$.


\subsubsection{The cases of n=1,2}
For any $b \in \mathbb{R}$, $L \in (0,\infty]$ and $\xi \in \Gamma_{K}$, introduce  the notations:
 $$\mathcal{K}_{L,j}(\bold{x},\bold{z})=\mathcal{K}_{L,j}(\bold{x},\bold{z};b,\omega,\xi) := (-1)^{j+1} R_{L}^{(1)}(\bold{x},\bold{z};b,\omega,\xi) T_{j,L}(\bold{z},\bold{x};b,\omega,\xi)\quad j=1,2,
$$
where $(\bold{x}, \bold{z}) \in \Lambda_L ^2$ and $\bold{x} \not=\bold{z}$. Furthermore for $j=1,2$ set:
\begin{gather}
\label{uLn}
u_{L,j}^{(\omega)}(b,\xi) :=   \int_{\Lambda_{L} } \mathrm{d}\bold{x} \int_{\Lambda_{L}} \mathrm{d}\bold{z}\, \mathcal{K}_{L,j}(\bold{x},\bold{z}), \quad
u_{\infty,j}^{(\omega)}(b,\xi) :=  \int_{\Lambda_{L}} \mathrm{d}\bold{x} \int_{\Lambda_{L}} \mathrm{d}\bold{z}\, \mathcal{K}_{\infty,j}(\bold{x},\bold{z}), \\
v_{\infty,j}^{(\omega)}(b,\xi) :=   \int_{\Lambda_{L} } \mathrm{d}\bold{x} \int_{\mathbb{R}^{3} \setminus \Lambda_{L}} \mathrm{d}\bold{z}\, \mathcal{K}_{\infty,j}(\bold{x},\bold{z}).\nonumber
\end{gather}
Obviously  we have the following estimate:
\begin{equation*}
\bigg\vert  \int_{\Lambda_{L}} \mathrm{d}\bold{x}\, \bigg(  \int_{\Lambda_{L}} \mathrm{d}\bold{z}\, \mathcal{K}_{L,j}(\bold{x},\bold{z})-  \int_{\mathbb{R}^{3} } \mathrm{d}\bold{z}\, \mathcal{K}_{\infty,j}(\bold{x},\bold{z}) \bigg)\bigg\vert \\
\leq \vert u_{L,j}^{(\omega)}(b,\xi) - u_{\infty,j}^{(\omega)}(b,\xi)\vert + \vert v_{\infty,j}^{(\omega)}(b,\xi) \vert.
\end{equation*}
We want to estimate each term in the above r.h.s. Firstly let us prove the following result.  $\mathbb{P}$-a.s., $\forall b \in  \mathbb{R}$, there exists a polynomial $p(\cdot\,)$ s.t. $\forall \xi \in \Gamma_K$ and  for large $L$:
\begin{equation}
\label{64.12P}
\vert u_{L,j}^{(\omega)}(b,\xi) - u_{\infty,j}^{(\omega)}(b,\xi)\vert \leq \vert p( \xi) \vert L^{2+ 2\alpha}.
\end{equation}
With our  previous notations we have:
\begin{equation}
\label{diffu}
u_{L,j}^{(\omega)}(b,\xi) - u_{\infty,j}^{(\omega)}(b,\xi) =
 \int_{\Lambda_{L} } \mathrm{d}\bold{x} \int_{\Lambda_{L}} \mathrm{d}\bold{z}\, \Big(\mathcal{K}_{L,j}(\bold{x},\bold{z}) - \mathcal{K}_{\infty,j}(\bold{x},\bold{z})\Big).
\end{equation}
By leaving out the dependence on $b$, $\omega$ and $\xi$ for the kernels, we have for $ j=1,2$:
\begin{equation}\label{main11}
\vert \mathcal{K}_{L,j}(\bold{x},\bold{z}) - \mathcal{K}_{\infty,j}(\bold{x},\bold{z})\vert \leq  \vert (R_{L}^{(1)}- R_{\infty}^{(1)})(\bold{x},\bold{z}) T_{j,L}(\bold{z},\bold{x}) \vert +  \vert R_{\infty}^{(1)}(\bold{x},\bold{z})
 (T_{j,L} - T_{j,\infty})(\bold{z},\bold{x})\vert .
\end{equation}
In view of \eqref{main11}, let us  estimate the r.h.s of \eqref{diffu}.
By \eqref{essecbul2}-\eqref{essecbul1}  for $L = \infty$, \eqref{esdifres} and \eqref{63.9}, then $\mathbb{P}$-a.s., $\forall b \in \mathbb{R}$ there exists a constant $\gamma > 0$ and a polynomial $p(\cdot\,)$ s.t. $ \forall \xi \in \Gamma_K$, for $L$ large  and $\forall (\bold{x},  \bold{z}) \in\Lambda_{L}^{2}\setminus D_{L}$:
\begin{equation}
\label{supestimm}
 \vert \mathcal{K}_{L,j}(\bold{x},\bold{z}) - \mathcal{K}_{\infty,j}(\bold{x},\bold{z})\vert \leq  \vert p( \xi)\vert    L^{2\alpha}
(\frac{ \chi_{\Lambda_{\kappa}}(\bold{x})}{\vert \bold{x} - \bold{z}\vert}  + \frac{ \chi_{\Lambda_{\kappa}}(\bold{z})} {\vert \bold{x} - \bold{z}\vert }+  \mathrm{e}^{- \frac{\gamma}{1+\vert \xi\vert}( \mathrm{d}(\bold{z}) +\mathrm{d}(\bold{x} ))})   \frac{\mathrm{e}^{-2 \frac{\gamma}{1 + \vert \xi \vert} \vert \bold{x} - \bold{z}\vert}}{\vert \bold{x} - \bold{z}\vert}.
\end{equation}
Put the estimate \eqref{supestimm} in the r.h.s. of  \eqref{diffu} and a straightforward  computation leads to \eqref{64.12P}.
On the other hand from  \eqref{essecbul2}-\eqref{essecbul1},  we have:
\begin{equation*}
\forall\, \xi \in \Gamma_{K},\quad \vert \mathcal{K}_{\infty,j}(\bold{x},\bold{z})\vert \leq \vert p(\xi) \vert  (1 + \vert \bold{z}\vert^{\alpha})  \frac{\mathrm{e}^{-\frac{\gamma}{1 + \vert \xi\vert} \vert \bold{x} - \bold{z} \vert}}{\vert \bold{x} - \bold{z}\vert^2},
\end{equation*}
for another $\gamma>0$ and polynomial $p(\cdot\,)$.  We  now use  that
there exists a constant $c > 0$ s.t.:
\begin{equation*}
\forall \gamma>0,\,\forall L \in (0,\infty),\quad \int_{\Lambda_{L}} \mathrm{d}\bold{x}\int_{\mathbb{R}^{3} \setminus \Lambda_{L}} \mathrm{d}\bold{z}\, \frac{\mathrm{e}^{- \gamma \vert \bold{x} - \bold{z}\vert}}{\vert \bold{x} - \bold{z} \vert^{k}} \leq c \gamma^{-(2+k)} L^{2} \quad k=1,2.
\end{equation*}
Then $\mathbb{P}$-a.s., $\forall b \in \mathbb{R}$, there exists another polynomial $p(\cdot\,)$ s.t. $\forall \xi \in \Gamma_{K}$ and for large $L$:
\begin{equation}
\label{64.13P}
\vert v_{\infty,j}^{(\omega)}(b,\xi)\vert \leq  \vert p( \xi) \vert L^{2+ \alpha}.
\end{equation}
 Then   \eqref{64.12P}-\eqref{64.13P} imply:
\begin{equation}
\label{main12}
\bigg\vert \int_{\Lambda_{L}} \mathrm{d}\bold{x} \bigg(\int_{\Lambda_{L}} \mathrm{d}\bold{z}\, \mathcal{K}_{L,j}(\bold{x},\bold{z}) - \int_{\mathbb{R}^{3}} \mathrm{d}\bold{z}\, \mathcal{K}_{\infty,j}(\bold{x},\bold{z})\bigg)\bigg\vert \leq \vert p(\xi)\vert L^{2+2 \alpha}.
\end{equation}
From this analysis we conclude the case of $n=1$. Indeed from
\eqref{64.12P}-\eqref{64.13P},  $\mathbb{P}$-a.s., $\forall b \in \mathbb{R}$ there exists a constant $c= c(\beta_1,K,b)>0$ s.t. $\forall \beta \in [\beta_{1},\beta_{2}]$, $\forall z \in K$ and large $L$:
\begin{multline*}
\bigg\vert \frac{1}{\vert \Lambda_{L}\vert} \int_{\Lambda_{L}} \mathrm{d}\bold{x}\, \bigg( \frac{\partial  \mathcal{L}_{L}^{(\omega)}}{\partial b}(\bold{x},\bold{x};\beta,b,z,\epsilon) - \mathcal{L}_{\infty,1}^{(\omega)}(\bold{x},\bold{x};\beta,b,z,\epsilon)\bigg)\bigg\vert \\
\leq \frac{1}{\vert \Lambda_{L}\vert} \int_{\Gamma_{K}} \vert \mathrm{d}\xi\vert \vert \mathfrak{f}_{\epsilon}(\beta,z;\xi)\vert  \bigg\vert  \int_{\Lambda_{L}} \mathrm{d}\bold{x}\, \bigg(  \int_{\Lambda_{L}} \mathrm{d}\bold{z}\, \mathcal{K}_{L,1}(\bold{x},\bold{z})-  \int_{\mathbb{R}^{3} } \mathrm{d}\bold{z}\, \mathcal{K}_{\infty,1}(\bold{x},\bold{z}) \bigg)\bigg\vert  \leq c \frac{1}{L^{1 -2\alpha}}.
\end{multline*}

We now  complete the proof of the proposition. Define for any $b \in \mathbb{R}$, $L \in (0,\infty]$ and $\xi \in \Gamma_{K}$:
$$\mathcal{K}_{L,3}(\bold{x},\bold{Z}) :=R_{L}^{(1)}(\bold{x},\bold{z}_{1};b,\omega,\xi) T_{1,L}(\bold{z}_{1},\bold{z}_{2};b,\omega,\xi) T_{1,L}(\bold{z}_{2},\bold{x} ;b,\omega,\xi),$$
where $\bold{Z}:=(\bold{z}_1,\bold{z}_2)$, $(\bold{x}, \bold{z}_1,\bold{z}_2) \in \Lambda_L ^3$ and
$\bold{x} \not=\bold{z}_1 \not=\bold{z}_2$. Introduce the quantities $u_{L,3}^{(\omega)}(b,\xi), u_{\infty,3}^{(\omega)}(b,\xi)$ as in \eqref{uLn} but with $\mathcal{K}_{L,3}(\bold{x},\bold{Z})$ instead of  $\mathcal{K}_{L,j}(\bold{x},\bold{Z}), j=1,2$ and integrating w.r.t. the measure $\mathrm{d}\bold{x} \mathrm{d}\bold{Z}$. Moreover set:
\begin{equation}
\label{winftyn3}
v_{\infty,3}^{(\omega)}(b,\xi) :=    \int_{\Lambda_{L}} \mathrm{d}\bold{x}\, \bigg\{\int_{\mathbb{R}^{3} \setminus \Lambda_{L}  } \mathrm{d}\bold{z}_{1}\int_{\mathbb{R}^{3}} \mathrm{d}\bold{z}_{2}\, \mathcal{K}_{\infty,3}(\bold{x},\bold{Z}) + \\ \int_{ \Lambda_{L}} \mathrm{d}\bold{z}_{1}\int_{ \mathbb{R}^{3} \setminus \Lambda_{L}} \mathrm{d}\bold{z}_{2}\,\mathcal{K}_{\infty,3}(\bold{x},\bold{Z})\bigg\}.
\end{equation}
Let us estimate $u_{L,3}^{(\omega)}(b,\xi) - u_{\infty,3}^{(\omega)}(b,\xi) $ with the same method as above.  \eqref{main11} is replaced with:
\begin{multline*}
\vert \mathcal{K}_{L,3}(\bold{x},\bold{Z}) - \mathcal{K}_{\infty,3}(\bold{x},\bold{Z})\vert \leq  \vert (R_{L}^{(1)}- R_{\infty}^{(1)})(\bold{x},\bold{z}_1) T_{1,L}(\bold{z}_1,\bold{z}_2) T_{1,L}(\bold{z}_2,\bold{x})\vert + \\ \vert R_{\infty}^{(1)}(\bold{x},\bold{z}_1)
 (T_{1,L} - T_{1,\infty})(\bold{z}_1,\bold{z}_2)T_{1,L}(\bold{z}_2,\bold{x})\vert +   \vert R_{\infty}^{(1)}(\bold{x},\bold{z}_1)
T_{1,\infty}(\bold{z}_1,\bold{z}_2) (T_{1,L} - T_{1,\infty})(\bold{z}_2,\bold{x})\vert.
\end{multline*}
Here it is convenient to set $ \bold{z}_0=\bold{z}_3= \bold{x} $. By using \eqref{essecbul2}, \eqref{essecbul1}  together with \eqref{esdifres}, \eqref{esdifdres}, then $\mathbb{P}$-a.s., $\forall b \in \mathbb{R}$ there exists a constant $\gamma>0$ and a polynomial $p(\cdot\,)$ s.t. $\forall \xi \in \Gamma_{K}$, for large $L$ and  $\forall(\bold{x},\bold{Z}) \in \Lambda_{L}^{3}$ with $\bold{x}\neq \bold{z}_{1}\neq\bold{z}_{2}$:
 \begin{multline}
\label{supestimm2}
 \vert \mathcal{K}_{L,3}(\bold{x},\bold{Z}) - \mathcal{K}_{\infty,3}(\bold{x},\bold{Z})\vert \leq  \vert p( \xi)\vert L^{3\alpha} \mathrm{e}^{- \frac{\gamma}{1 + \vert \xi \vert} \sum^2_{l=0}\vert \bold{z}_l - \bold{z}_{l+1}\vert} \times \\
 \times (\prod_{l=0}^2 \frac{1}{\vert \bold{z}_{l+1} - \bold{z}_l \vert} )\sum^2_{l=0}( \chi_{\Lambda_{\kappa}}(\bold{z}_l) +  \vert \bold{z}_{l+1} - \bold{z}_l \vert \mathrm{e}^{- \frac{\gamma}{1+\vert \xi \vert}( \mathrm{d}(\bold{z}_{l+1} )+ \mathrm{d}(\bold{z}_{l}))}).
\end{multline}
So  from \eqref{supestimm2} by a tedious computation, we obtain  that $\mathbb{P}$-a.s., $\forall b \in \mathbb{R}$, there exists another polynomial $p(\cdot\,)$ s.t. $\forall \xi \in \Gamma_{K}$ and for  large $L$:
\begin{equation}
\label{64.123P}
 \vert u_{L,3}^{(\omega)}(b,\xi) - u_{\infty,3}^{(\omega)}(b,\xi)\vert \leq \vert p( \xi) \vert L^{2+ 3\alpha}.
\end{equation}
We also have the estimate:
\begin{equation*}
\vert \mathcal{K}_{\infty,3}(\bold{x},\bold{Z})\vert \leq \vert p(\xi) \vert  (1 + \vert \bold{z}_1\vert^{\alpha})  (1 + \vert \bold{z}_2\vert^{\alpha}) (\prod_{l=0}^2 \frac{1}{\vert \bold{z}_{l+1} - \bold{z}_l \vert} )\mathrm{e}^{- \frac{\gamma}{2(1 + \vert \xi \vert)} \sum^2_{l=0}\vert \bold{z}_l - \bold{z}_{l+1}\vert},
\end{equation*}
for another constant $\gamma>0$ and polynomial $p(\cdot\,)$. In view of \eqref{winftyn3}, by some straightforward estimates, $\mathbb{P}$-a.s., $\forall b \in \mathbb{R}$, there exists another  polynomial $p(\cdot\,)$ s.t. $\forall\, \xi \in \Gamma_{K}$ and large $L$:
\begin{equation}
\label{64.133P}
\vert v_{\infty, 3}^{(\omega)}(b,\xi)\vert \leq  \vert p( \xi) \vert L^{2+ 2\alpha}.
\end{equation}
Consequently,
\begin{equation} \label{main3}
\bigg\vert  \int_{\Lambda_{L}} \mathrm{d}\bold{x}\, \bigg(  \int_{\Lambda_{L}^2} \mathrm{d}\bold{Z}\, \mathcal{K}_{L,3}(\bold{x},\bold{Z})-  \int_{\mathbb{R}^{3} } \mathrm{d}\bold{Z}\, \mathcal{K}_{\infty,3}(\bold{x},\bold{Z}) \bigg)\bigg\vert
 \leq \vert p( \xi) \vert L^{2+ 3\alpha}.
\end{equation}
Let us prove the case $n=2$. From
\eqref{main12} and \eqref{main3},  $\mathbb{P}$-a.s., $\forall b \in \mathbb{R}$ there exists a constant $c= c(\beta_1,K,b)>0$ s.t. $\forall \beta \in [\beta_{1},\beta_{2}]$, $\forall z \in K$ and for  large $L$:
\begin{multline} \label{main}
\bigg\vert \frac{1}{\vert \Lambda_{L}\vert} \int_{\Lambda_{L}} \mathrm{d}\bold{x}\, \bigg(\frac{\partial^{2} \mathcal{L}_{L}^{(\omega)}}{\partial b^{2}}(\bold{x},\bold{x};\beta,b,z,\epsilon) - \mathcal{L}_{\infty,2}^{(\omega)}(\bold{x},\bold{x};\beta,b,z,\epsilon)\bigg)\bigg\vert \\
\leq \frac{1}{\vert \Lambda_{L}\vert} \int_{\Gamma_{K}} \vert \mathrm{d}\xi\vert \vert \mathfrak{f}_{\epsilon}(\beta,z;\xi)\vert  \bigg\{  \bigg\vert  \int_{\Lambda_{L}} \mathrm{d}\bold{x}\, \bigg(  \int_{\Lambda_{L}^2} \mathrm{d}\bold{Z}\, \mathcal{K}_{L,3}(\bold{x},\bold{Z})-  \int_{\mathbb{R}^{6} } \mathrm{d}\bold{Z}\, \mathcal{K}_{\infty,3}(\bold{x},\bold{Z})\bigg)\bigg\vert  + \\
 \bigg\vert  \int_{\Lambda_{L}} \mathrm{d}\bold{x}\, \bigg(  \int_{\Lambda_{L}} \mathrm{d}\bold{z}\, \mathcal{K}_{L,2}(\bold{x},\bold{z})-  \int_{\mathbb{R}^{3} } \mathrm{d}\bold{z}\, \mathcal{K}_{\infty,2}(\bold{x},\bold{z})  \bigg)\bigg\vert \bigg\}\leq c \frac{1}{L^{1 - 3\alpha}}.
 \end{multline}
Since we suppose that $ 0< \alpha < \frac{1}{3}$, the proposition follows. \qed

\remark \label{pour la fin2}   \eqref{main12}, \eqref{main3} and \eqref{opochiLn}  show that the finite-volume magnetization and susceptibility,  if they  are defined, are uniformly bounded in $L$ provided that
$\alpha \in (0,\frac{1}{3})$. In fact
 these  considerations   can be extended to estimate    $\mathcal{X}_{L,3}^{(\omega)}(\beta,b,z,\epsilon) $ in \eqref{suscep3}. Indeed  let:
\begin{equation}
\label{d2}
 \sum_{j=1}^2\bigg( \int_{\Lambda_{L}} \mathrm{d}\bold{x}\, \mathcal{U}^{(\omega)}_{L,j}(\bold{x},\bold{x};b,\xi)  - \int_{\Lambda_{L}} \mathrm{d}\bold{x}\, \mathcal{U}^{(\omega)}_{\infty,j}(\bold{x},\bold{x};b,\xi)  \bigg),
 \end{equation}
where $\mathcal{U}^{(\omega)}_{L,j}(\cdot\,,\cdot\,;b,\xi)$ are defined in \eqref{Ujk} and $\mathcal{U}^{(\omega)}_{\infty ,j}(\cdot\,,\cdot\,;b,\xi)$ are defined in the same way but with $L=\infty$. \eqref{d2}
can be estimated    as above.   In that case,  handling heavy technicalities, we can  show that  $\mathbb{P}$-a.s., $\forall b \in \mathbb{R}$ there exists a polynomial $p(\cdot\,)$  s.t. $\forall\xi \in \Gamma_{K}$, the quantity in \eqref{d2} is bounded above by $\vert p(\xi)\vert L^{2 +4 \alpha}$. It follows that if $\alpha \in (0,\frac{1}{4})$, then \eqref{suscep3} and  \eqref {expdecr}  imply that
$\mathbb{P}$-a.s., $\forall \beta>0$, $\forall b \in \mathbb{R}$ and $\forall z \in K$,  $\mathcal{X}_{L,3}^{(\omega)}(\beta,b,z,\epsilon) $ is also  bounded uniformly in $L \in (0,\infty)$.


\section {Appendix}
\subsection{Proof of Lemma \ref{prokdres}} For all $L \in (0,\infty]$ and $b \in \mathbb{R}$ denote by $G_{L}(\cdot\,,\cdot\,;t,b)$ the integral kernel  of the strongly continuous semigroup $ \{\mathrm{e}^{- t H_{0,L}(b)}, t>0\}$; here $H_{0,L}(b), L<\infty, L=\infty $ is the free  operator defined  respectively  in \eqref{HL} and \eqref{Hinfini}  with $ V^{(\omega)}= 0$. It is known that $G_{L}(\cdot\,,\cdot\,;t,b)$  is smooth and obeys, see \cite[Eq. (2.31) \& (4.13)]{BrCoLo1}:
\begin{equation}
\label{BCLes}
\forall (\bold{x},\bold{y}) \in \Lambda_{L}^{2}, \quad \vert (i\nabla_{\bold{x}} + b \bold{a}(\bold{x}))G_{L}(\bold{x},\bold{y};t,b)\vert \leq c (1+\vert b\vert)^{3} (1+t)^{5} t^{-2} \mathrm{e}^{- \frac{\vert \bold{x}-\bold{y}\vert^{2}}{16t}},
\end{equation}
where $c>0$ is  a $L$-independent constant.
Let $\lambda < \min\{0, E_{0}\}$. By using the Laplace transform \cite[Eq. (2.5)]{If} in the kernels sense, we get on $\Lambda_{L}^{2}\setminus D_{L}$:
\begin{equation}
\label{kdLaTr}
(i\nabla_{\bold{x}} + b \bold{a}(\bold{x})) R_{0,L}^{(1)}(\bold{x},\bold{y};b,\lambda) = \int_{0}^{+\infty} \mathrm{d}t\, \mathrm{e}^{\lambda t} (i\nabla_{\bold{x}} + b \bold{a}(\bold{x})) G_{L}(\bold{x},\bold{y};t,b),
\end{equation}
where $R_{0,L}^{(1)}(\cdot\,, \cdot\,;b,\lambda)$ denotes the integral kernel of $R_{0,L}(b,\lambda) := (H_{0,L}(b)- \lambda)^{-1}$, $L \in (0,\infty]$. Due to \eqref{BCLes}, \eqref{kdLaTr} is well-defined and the function $(\bold{x},\bold{y}) \mapsto (i\nabla_{\bold{x}} + b \bold{a}(\bold{x}))  R_{0,L}^{(1)}(\bold{x},\bold{y};b,\lambda)$ is jointly continuous on $\Lambda_{L}^{2}\setminus D_{L}$. Besides, there exists a constant $c=c(\lambda)>0$ s.t.:
\begin{equation}
\label{eskdRLl0}
\forall (\bold{x},\bold{y}) \in  \Lambda_{L}^{2}\setminus D_{L},\quad \vert (i\nabla_{\bold{x}} + b \bold{a}(\bold{x})) R_{0,L}^{(1)}(\bold{x},\bold{y};b,\lambda)\vert \leq c (1+\vert b\vert)^{3} \frac{\mathrm{e}^{- \frac{\sqrt{-\lambda}}{2 \sqrt{2}} \vert \bold{x}-\bold{y}\vert}}{\vert \bold{x}-\bold{y}\vert^{2}}.
\end{equation}
Now $\mathbb{P}$-a.s., $\forall b \in \mathbb{R}$, $\forall \lambda < \min\{0,E_{0}\}$ and $\forall L \in (0,\infty]$, the second resolvent equation:
\begin{equation}
\label{eqres}
R_{L}(b,\omega,\lambda) = R_{0,L}(b,\lambda) - R_{0,L}(b,\lambda)V^{(\omega)}R_{L}(b,\omega,\lambda),
\end{equation}
holds in the bounded operators sense. Indeed,   $\mathbb{P}$-a.s., $\forall b \in \mathbb R$,  $V^{(\omega)}R_{L}(b,\omega,\lambda)$ is bounded if $L < \infty$. When $L=\infty$, $R_{0, \infty}(b,\lambda)V^{(\omega)}R_{\infty}(b,\omega,\lambda)$ is defined  as the closure of the same  operator defined  on the domain $(H_{\infty}(b,\omega)- \lambda)\mathcal{C}_{0}^{\infty}(\mathbb{R}^{3})$ which is  dense in $L^{2}(\mathbb{R}^{3})$.  Hence,  we have 
\begin{gather}
\label{k2eqad}
(i\nabla_{\bold{x}} + b \bold{a}(\bold{x})) R_{L}^{(1)}(\bold{x},\bold{y};b,\omega,\lambda) = (i\nabla_{\bold{x}} + b \bold{a}(\bold{x})) R_{0,L}^{(1)}(\bold{x},\bold{y};b,\lambda) -  \mathcal{M}_{L}(\bold{x},\bold{y};b,\omega,\lambda),\\
 \mathcal{M}_{L}(\bold{x},\bold{y};b,\omega,\lambda) := \int_{\Lambda_{L}} \mathrm{d}\bold{z}\, (i\nabla_{\bold{x}} + b \bold{a}(\bold{x})) R_{0,L}^{(1)}(\bold{x},\bold{z};b,\lambda) V^{(\omega)}(\bold{z}) R_{L}^{(1)}(\bold{z},\bold{y};b,\omega,\lambda). \nonumber
\end{gather}
According  to the decomposition $V^{(\omega)}= V_{1}^{(\omega)} + V_{2}^{(\omega)}$, set on $\Lambda_{L}^{2}\setminus D_{L}$:
$$\mathcal{M}_{L}(\bold{x},\bold{y};b,\omega,\lambda)= \mathcal{M}_{L,1}(\bold{x},\bold{y};b,\omega,\lambda) + \mathcal{M}_{L,2}(\bold{x},\bold{y};b,\omega,\lambda).$$
From   \eqref{eskres}, \eqref{eskdRLl0},  Lemma \ref{Vcontinuity} with (R1) if $l=1$ and Lemma \ref{continuity} $i)$ with \eqref{V2cond1} if  $l=2$,  together with  \eqref{kdLaTr} and \eqref{eqres}, then  $\mathbb{P}$-a.s., $\forall b \in \mathbb{R}$, $\forall \lambda < \min\{0,E_{0}\}$, $\forall L \in (0,\infty]$ and $\forall(\bold{x},\bold{y}) \in \Lambda_{L}^{2}\setminus D_{L}$, $(i\nabla_{\bold{x}} + b \bold{a}(\bold{x}))  R_{0,L}^{(1)}(\bold{x},\cdot\,;b,\lambda) V_{l}^{(\omega)}(\cdot\,) R_{L}^{(1)}(\cdot\,,\bold{y};b,\omega,\lambda ) \in L^{1}(\Lambda_{L})$ $\forall(\bold{x},\bold{y}) \in \Lambda_{L}^{2}\setminus D_{L}$. From Lemmas \ref{continuity} and \ref{Vcontinuity}  then
$(\bold{x},\bold{y}) \mapsto \mathcal{M}_{L,l}(\bold{x},\bold{y};b,\omega,\lambda)$ are  jointly continuous on $\Lambda_{L}^{2}\setminus D_{L}$. Therefore this also holds for $ (\bold{x},\bold{y}) \mapsto (i\nabla_{\bold{x}} + b \bold{a}(\bold{x}))R_{L}^{(1)}(\bold{x},\bold{y};b,\omega,\lambda)$.\\
On the other hand, from Lemmas \ref{proestim} and \ref{propLuloc} together with \eqref{eskres}, \eqref{eskdRLl0} and the inequality:
\begin{equation} \label{inq1}
\vert \bold{x} - \bold{z}\vert^{-1} \vert \bold{z} - \bold{y}\vert^{-1} \leq \vert \bold{x} - \bold{y}\vert^{-1}( \vert \bold{x} - \bold{z}\vert^{-1} + \vert \bold{z} - \bold{y}\vert^{-1}) \quad \bold{x}\neq\bold{y}\neq\bold{z},
\end{equation}
then there exists a constant $c>0$ s.t.:
\begin{equation}
\label{1111}
\forall(\bold{x},\bold{y}) \in \Lambda_{L}^{2}\setminus D_{L},\quad \vert \mathcal{M}_{L,1}(\bold{x},\bold{y};b,\omega,\lambda)\vert \leq
c(1+\vert b\vert)^{3} \frac{\mathrm{e}^{-c^{'}(\lambda) \vert \bold{x}-\bold{y}\vert}}{\vert \bold{x}-\bold{y}\vert},
\end{equation}
where $c^{'}(\lambda):=  \frac{1}{2}\min\{\frac{\sqrt{-\lambda}}{2 \sqrt{2}}, \frac{\gamma}{1 + \vert \lambda \vert} \} = \frac{1}{2}\frac{\gamma}{1 + \vert \lambda \vert}$ if  $ \lambda <0$ is chosen large enough.
From \eqref{V2cond1}, \eqref{eskres}, \eqref{eskdRLl0} together with \eqref{inq1}, then  there exists $c>0$ s.t.:
\begin{equation}
\label{1112}
\forall (\bold{x},\bold{y}) \in \Lambda_{L}^{2}\setminus D_{L},\quad \vert \mathcal{M}_{L,2}(\bold{x},\bold{y};b,\omega,\lambda) \vert \leq
c  (1+\vert b\vert)^{3} (1+\vert \bold{x}\vert^{\alpha} + \vert \bold{y}\vert^{\alpha}) \frac{\mathrm{e}^{- \frac{c^{'}(\lambda)}{4} \vert \bold{x}-\bold{y}\vert}}{\vert \bold{x}-\bold{y}\vert}.
\end{equation}
Therefore in view of \eqref{k2eqad}, then \eqref{eskdRLl0} together with \eqref{1111} and \eqref{1112} imply that $\mathbb{P}$-a.s.,  $\forall \lambda < \min\{0,E_{0}\}$, there exists  $c>0$ s.t. $\forall L \in (0,\infty]$, $\forall b \in \mathbb{R}$, we have on $\Lambda_{L}^{2}\setminus D_{L}$:
\begin{equation}
\label{eskdRLlV}
\vert (i\nabla_{\bold{x}} + b \bold{a}(\bold{x})) R_{L}^{(1)}(\bold{x},\bold{y};b,\omega,\lambda)\vert \leq c  (1+\vert b\vert)^{3} (1+\vert \bold{x}\vert^{\alpha}+\vert \bold{y}\vert^{\alpha}) \frac{\mathrm{e}^{- \frac{c^{'}(\lambda)}{8}\vert \bold{x}-\bold{y}\vert}}{\vert \bold{x}-\bold{y}\vert^{2}}.
\end{equation}
Now the first resolvent equation allows us to write $ \forall (\bold{x},\bold{y}) \in \Lambda_{L}^{2}\setminus D_{L}$:
\begin{multline}
\label{k1eq}
(i\nabla_{\bold{x}} + b \bold{a}(\bold{x})) R_{L}^{(1)}(\bold{x},\bold{y};b,\omega,\xi) = (i\nabla_{\bold{x}} + b \bold{a}(\bold{x})) R_{L}^{(1)}(\bold{x},\bold{y};b,\omega,\lambda) + \mathcal{N}_{L}(\bold{x},\bold{y};b,\omega,\xi,\lambda),\\
 \mathcal{N}_{L}(\bold{x},\bold{y};b,\omega,\xi,\lambda) :=  (\xi-\lambda) \int_{\Lambda_{L}} \mathrm{d}\bold{z}\, (i\nabla_{\bold{x}} + b \bold{a}(\bold{x})) R_{L}^{(1)}(\bold{x},\bold{z};b,\omega,\lambda) R_{L}^{(1)}(\bold{z},\bold{y};b,\omega,\xi).
\end{multline}
Due to \eqref{eskres}, \eqref{eskdRLlV} and  Lemma \ref{continuity} $i)$, $\mathbb{P}$-a.s., $\forall b \in \mathbb{R}$, $\forall L \in (0,\infty]$, $\forall \eta>0$, $\forall \lambda < \min \{0,E_{0}\}$, and $\forall \xi \in \mathbb{C}$ s.t. $d(\xi) \geq \eta$, $(\bold{x},\bold{y}) \mapsto  \mathcal{N}_{L}(\bold{x},\bold{y};b,\omega,\xi,\lambda)$ is jointly continuous on $\Lambda_{L}^{2}\setminus D_{L}$. This proves Lemma \ref{prokdres} $i)$. $ii)$ follows from \eqref{eskres}, \eqref{eskdRLlV} and the fact that for $\vert \lambda\vert$ large enough, there exists $ \gamma'>0$ s.t. $ \forall \xi \in  \mathbb{C}$, $d(\xi) \geq \eta$, $\min \{\frac{c^{'}(\lambda)}{16}, \frac{ \gamma}{1 + \vert  \xi\vert}\} \geq \frac{ \gamma'}{1 + \vert  \xi\vert} $. \qed

\subsection{Some kernel estimates}

Here we give  some useful estimates needed in this paper.

\begin{lema}
\label{continuity}
Let $U \subseteq \mathbb{R}^{3}$ be an open set and $D:=\{(\bold{x},\bold{y})\in U^{2} : \bold{x}=\bold{y}\}$.\\  Let $K_{l}(\cdot\,,\cdot\,):U^{2}\setminus D\rightarrow \mathbb{C}$, $l=1,2$ be  integral kernels satisfying:\\
(h1) $K_{1}(\cdot\,,\bold{z})$ and $K_{2}(\bold{z},\cdot\,)$ are continuous on $U \setminus \{\bold{z}\}$ for almost all $\bold{z}\in U$.\\
(h2) There exist real numbers $c_{l},\gamma_{l}>0$ and $\nu_{l},\mu_{l} \geq 0$ as well as $\delta_{l}\in[0,3)$ s.t.:
\begin{equation}
\label{eskjut}
\forall(\bold{x},\bold{y}) \in U^{2}\setminus D,\quad \vert K_{l}(\bold{x},\bold{y})\vert \leq c_{l}(\vert \bold{x}\vert^{\nu_{l}} + \vert \bold{y}\vert^{\mu_{l}}) \frac{\mathrm{e}^{-\gamma_{l} \vert \bold{x} - \bold{y}\vert}}{\vert \bold{x} - \bold{y}\vert^{\delta_{l}}} \quad l=1,2.
\end{equation}
Then:\\
i) $K_{1}(\bold{x},\cdot\,)K_{2}(\cdot\,,\bold{y}) \in L^{1}(U)$ for all $(\bold{x},\bold{y}) \in U^{2}\setminus D$. Furthermore the map:\\ $(\bold{x},\bold{y})\mapsto \mathcal{K}(\bold{x},\bold{y}) := \int_{U} \mathrm{d}\bold{z}\, K_{1}(\bold{x},\bold{z}) K_{2}(\bold{z},\bold{y})$
    is jointly continuous on $U^{2}\setminus D$.\\
ii) Under the additional assumption $\delta_{1} + \delta_{2} \in [0,3)$, $K_{1}(\bold{x},\cdot\,)K_{2}(\cdot\,,\bold{x}) \in L^{1}(U)$ for all $\bold{x} \in U$. Moreover  $\bold{x}\mapsto \mathcal{K}(\bold{x},\bold{x})$ is continuous on $U$.
\end{lema}

This result is obtained by using standard arguments, see e.g. \cite[Sect. 3]{BGP}.

\begin{lema}
\label{proestim}
Consider the assumptions (h1)-(h2) of Lemma \ref{continuity} but with $\mu_{l},\nu_{l}=0$.\\
i) Let $\delta_{l}=1,2$ and $\gamma:= \min\{\gamma_{1},\gamma_{2}\}$. Then there exists a constant $c >0$ s.t. on $U^{2}$:
\begin{equation}
\label{esintk}
\int_{U} \mathrm{d}\bold{z}\, \vert K_{1}(\bold{x},\bold{z}) K_{2}(\bold{z},\bold{y})\vert \leq \frac{c} {\gamma} \mathrm{e}^{-\frac{\gamma}{2} \vert \bold{x} - \bold{y}\vert} \times \left\{\begin{array}{ll}
\frac{1}{\vert \bold{x} - \bold{y}\vert^{\min\{\delta_{1},\delta_{2}\}}} &\textrm{if $\delta_{1},\delta_{2} \neq 1$ and $\bold{x}\neq\bold{y}$} \\
1  &\textrm{if $\delta_{1},\delta_{2}=1$}
\end{array}\right..
\end{equation}
ii)  Let $\gamma_{l}=\gamma$ and $\delta_{l} =l$, $l=1,2$. Then $\forall k\geq 2$ there exists a constant $c>0$ s.t. on $U^{2}$:
\begin{gather}
\label{fiestie}
 \int_{U^{k}} \mathrm{d}\bold{z}_{1} \dotsb \mathrm{d}\bold{z}_{k}\, \vert K_{1}(\bold{x},\bold{z}_{1}) K_{1}(\bold{z}_{1},\bold{z}_{2}) \dotsb K_{1}(\bold{z}_{k},\bold{y})\vert \leq \frac{c}{\gamma^{2k -1}} \mathrm{e}^{-\frac{\gamma}{2^{k}} \vert \bold{x} - \bold{y}\vert}, \\
\label{secestie}
\int_{U^{k}} \mathrm{d}\bold{z}_{1} \dotsb \mathrm{d}\bold{z}_{k}\, \vert K_{1}(\bold{x},\bold{z}_{1}) K_{2}(\bold{z}_{1},\bold{z}_{2}) \dotsb K_{2}(\bold{z}_{k},\bold{y})\vert \leq \frac{c}{\gamma^{k}} \frac{\mathrm{e}^{-\frac{\gamma}{2^{k}} \vert \bold{x} - \bold{y}\vert}}{\vert \bold{x} - \bold{y}\vert}\quad \textrm{if $\bold{x} \neq \bold{y}$}.
\end{gather}
\end{lema}

\noindent \textit{Proof.} The main ingredient is the following estimate.  For all $\gamma>0$ and $\delta \in [0,3)$,
\begin{equation}
\label{estimcle}
\sup_{\bold{x} \in U} \int_{\mathbb{R}^{3}} \mathrm{d}\bold{y}\, \frac{\mathrm{e}^{-\gamma \vert \bold{x} - \bold{y}\vert}}{\vert \bold{x} - \bold{y}\vert^{\delta}} = \int_{0}^{\infty} \mathrm{d}r\, r^{2-\delta} \mathrm{e}^{-\gamma r} = \frac{\Gamma(3-\delta)}{\gamma^{3 - \delta}},
\end{equation}
where $\Gamma(\cdot\,)$ denotes the usual Gamma Euler function. When $\delta_{1}=\delta_{2}=1$, we have:
\begin{equation*}
\forall (\bold{x},\bold{y}) \in U^2,\quad \int_{U} \mathrm{d}\bold{z}\, \vert K_{1}(\bold{x},\bold{z}) K_{2}(\bold{z},\bold{y})\vert \leq c_{1}c_{2} \mathrm{e}^{-\frac{\gamma}{2} \vert \bold{x} - \bold{y}\vert}  \int_{\mathbb{R}^{3}} \mathrm{d}\bold{z}\, \frac{\mathrm{e}^{-\frac{\gamma_{1}}{2} \vert \bold{x} - \bold{z}\vert}}{\vert \bold{x} - \bold{z}\vert}\frac{\mathrm{e}^{-\frac{\gamma_{2}}{2} \vert \bold{z} - \bold{y}\vert}}{\vert \bold{z} - \bold{y}\vert}.
\end{equation*}
Then by the Cauchy-Schwarz inequality  and \eqref{estimcle}, \eqref{esintk} follows. Similarly  we get the cases of $\delta_{1}=2, \delta_{2}=1$ and $\delta_{1}  = \delta_{2}= 2$ from \eqref{inq1} combined with \eqref{estimcle}.\\
The estimates \eqref{fiestie} and \eqref{secestie} are obtained by induction from the above arguments. \qed

\begin{lema}
\label{propLuloc}
Let $\gamma>0$, $\delta \in (0,3)$ and $p > \frac{3}{(3 - \delta)}$. Suppose that $V \in L^{p}_{\mathrm{uloc}}(\mathbb{R}^{3})$.
Then $\forall\bold{x} \in \mathbb{R}^{3}$, $V(\cdot) \mathrm{e}^{- \gamma \vert \bold{x} - \cdot\,\vert} \vert \bold{x} - \cdot\,\vert^{-\delta} \in L^{1}(\mathbb{R}^{3})$, and there exists a constant $c = c(\gamma, \delta, \Vert V\Vert_{p,\mathrm{uloc}})>0$ s.t.:
\begin{equation*}
\sup_{\bold{x} \in \mathbb{R}^{3}} \Vert V(\cdot\,) \mathrm{e}^{-\gamma \vert \bold{x} - \cdot\,\vert} \vert \bold{x} - \cdot\,\vert^{-\delta} \Vert_{1} \leq c.
\end{equation*}
\end{lema}

\noindent \textit{Proof.} Let $ \frac {1}{p} + \frac {1}{q} =1$ and $p > \frac{3}{(3 - \delta)}$. By the H\"older inequality, we get:
$$ \int_{\vert \bold{x} -  \bold{y} \vert <1} \mathrm{d}\bold{y}\, \vert V(\bold{y})\vert \frac{\mathrm{e}^{- \gamma  \vert \bold{x} - \bold{y} \vert}}{\vert \bold{x} -\bold{y}  \vert^{\delta}} \leq   \Vert  V\Vert_{p,\mathrm{uloc}}  \bigg(\int_{\vert \bold{x} -  \bold{y} \vert <1} \mathrm{d}\bold{y}\, \frac{\mathrm{e}^{- q\gamma \vert \bold{x} - \bold{y} \vert}}{ \vert \bold{x} -\bold{y}  \vert^{q\delta}}\bigg)^{\frac{1}{q}} < \infty.$$
On the other hand, with same $p,q$, we have:
\begin{equation*}
\int_{\vert \bold{x} -  \bold{y} \vert \geq 1} \mathrm{d}\bold{y}\, \vert V(\bold{y}) \vert \frac{\mathrm{e}^{- \gamma  \vert \bold{x} - \bold{y} \vert}}{\vert \bold{x} -\bold{y}  \vert^{\delta}}  \leq
\bigg( \sum_{k=1}^\infty   \int_{k<\vert  \bold{x} -  \bold{y} \vert \leq k+1} \mathrm{d}\bold{y}\, \vert V(\bold{y}) \vert^p \mathrm{e}^{- \frac{p\gamma}{2}  \vert \bold{x} - \bold{y} \vert}\bigg)^{ \frac {1}{p}}\bigg(  \int_{\vert  \bold{x} -  \bold{y} \vert \geq 1} \mathrm{d}\bold{y}\,  \mathrm{e}^{- \frac{q \gamma}{2}  \vert \bold{x} - \bold{y}\vert} \bigg)^{ \frac {1}{q}}.
\end{equation*}
Since the domain $k<\vert  \bold{x} -  \bold{y} \vert \leq k+1 $ is covered by $cste\times k^2$ unit balls, then the above r.h.s. is bounded from above by $c   \Vert  V\Vert_{p,\mathrm{uloc}}\big( \sum_{k=1}^\infty k^2  \mathrm{e}^{- \frac{p\gamma}{2}k} \big)^{ \frac {1}{p}} < \infty $ for some constant $c>0$.  \qed\\

From Lemma \ref{continuity} together with Lemma \ref{propLuloc}, we finally prove:

\begin{lema}
\label{Vcontinuity}
Consider the assumptions (h1)-(h2) of Lemma \ref{continuity}. Let  $V\in L^{p}_{\mathrm{uloc}}(\mathbb{R}^{3})$ with  $p> \frac{3}{3 - \max\{\delta_{1},\delta_{2}\}}$ if $\delta_{1}+ \delta_{2} \neq 0$, elsewhere $p \geq 1$. Then $K_{1}(\bold{x},\cdot\,)V(\cdot\,)K_{2}(\cdot\,,\bold{y}) \in L^{1}(U)$ for all $\bold{x}\neq\bold{y}$. Furthermore,
$(\bold{x},\bold{y}) \mapsto \int_{U} \mathrm{d}\bold{z}\, K_{1}(\bold{x},\bold{z})V(\bold{z})K_{2}(\bold{z},\bold{y})$
 is jointly continuous on $U^{2}\setminus D$.
\end{lema}

\noindent \textit{Proof.}
Consider only the 'most tricky' case which occurs when $U = \mathbb{R}^{3}$.
Note first that the estimate in $(h2)$ can be rewritten on $U^{2} \setminus D$ as:
\begin{equation}
\label{gooest}
\vert K_{1}(\bold{x},\bold{y}) \vert \leq c_{1} (1+\vert \bold{x}\vert^{\nu_{1}+\mu_{1}}) \frac{\mathrm{e}^{-\frac{\gamma_{1}}{2} \vert \bold{x} - \bold{y}\vert}}{\vert \bold{x} - \bold{y}\vert^{\delta_{1}}},\quad \vert K_{2}(\bold{x},\bold{y}) \vert \leq c_{2} (1+\vert \bold{y}\vert^{\nu_{2}+\mu_{2}}) \frac{\mathrm{e}^{-\frac{\gamma_{2}}{2} \vert \bold{x} - \bold{y}\vert}}{\vert \bold{x} - \bold{y}\vert^{\delta_{2}}},
\end{equation}
for another constants $c_{1},c_{2}>0$. Set $J(\bold{x},\bold{y};\cdot\,):= K_{1}(\bold{x},\cdot\,) V(\cdot\,) K_{2}(\cdot\,,\bold{y})$. Let $0<\varsigma < \frac{1}{2} \vert \bold{x}-\bold{y}\vert$ and denote by $\mathcal{B}(\cdot\,,\varsigma)$ the open ball having the radius $\varsigma>0$. From \eqref{gooest}, there exists a constant $c=c(\vert\bold{x}\vert,\vert \bold{y}\vert)>0$ s.t. $\forall\bold{z} \in \mathcal{B}(\bold{x},\varsigma)$, $\vert J(\bold{x},\bold{y};\bold{z})\vert \leq c \vert V(\bold{z})\vert \mathrm{e}^{-\frac{\gamma_{1}}{2} \vert \bold{x} - \bold{z}\vert} \vert \bold{x} - \bold{z}\vert^{-\delta_{1}}$. Then by Lemma \ref{propLuloc}, $J(\bold{x},\bold{y};\cdot\,)\in L^{1}(\mathcal{B}(\bold{x},\varsigma))$ as soon as $p> 3/(3 - \delta_{1})$. On the same way $J(\bold{x},\bold{y};\cdot\,)\in L^{1}(\mathcal{B}(\bold{y},\varsigma))$ as soon as $p> 3/(3 - \delta_{2})$. Besides there exists a constant $c^{'}=c^{'}(\vert\bold{x}\vert,\vert \bold{y}\vert)>0$ s.t. $\forall\bold{z} \in \mathbb{R}^{3} \setminus (\mathcal{B}(\bold{x},\varsigma) \cup \mathcal{B}(\bold{y},\varsigma))$, $\vert J(\bold{x},\bold{y};\bold{z})\vert \leq c \mathrm{e}^{-\frac{\gamma_{1}}{2} \vert\bold{x} - \bold{z}\vert}\vert V(\bold{z})\vert \mathrm{e}^{-\frac{\gamma_{2}}{2} \vert\bold{z} - \bold{y}\vert}$. As $\sup_{\bold{x} \in \mathbb{R}^{3}} \Vert V(\cdot\,) \mathrm{e}^{-\frac{\gamma_{1}}{2} \vert\bold{x} - \cdot\,\vert}\Vert_{p}, \sup_{\bold{y} \in \mathbb{R}^{3}} \Vert \mathrm{e}^{-\frac{\gamma_{2}}{2} \vert \cdot\, - \bold{y} \vert}\Vert_{q}< \infty$ whenever $p,q\geq 1$, then by the H\"older inequality, $J(\bold{x},\bold{y};\cdot\,) \in L^{1}(\mathbb{R}^{3} \setminus (\mathcal{B}(\bold{x},\varsigma) \cup \mathcal{B}(\bold{y},\varsigma)))$. Therefore $J(\bold{x},\bold{y};\cdot\,) \in L^{1}(\mathbb{R}^{3})$ provided that $p> 3/(3 - \max(\delta_{1},\delta_{2}))$. By standard arguments, the continuity property follows. \qed

\section{Acknowledgments}

The authors  thank Horia Cornean for fruitful discussions about this problem and  his kind invitations to department of mathematics  of Aalborg university, where a part of this work was done.

\end{document}